\documentclass[%
reprint,
superscriptaddress,
 amsmath,amssymb,
aps,
prb,
]{revtex4-2}
\usepackage{soul}
\usepackage{pifont}
\usepackage{verbatim}
\usepackage{subcaption}
\usepackage{graphicx}
\usepackage{dcolumn}
\usepackage{bm}
\usepackage{xcolor}
\usepackage[hidelinks]{hyperref}
\usepackage{url}
\usepackage[normalem]{ulem} 

\usepackage{physics}
\usepackage{amsthm,amsbsy,amsfonts}
\usepackage{wrapfig}

\newcommand{\I}{\mathbb{I}}

\begin{document}

\title{Magic angle conditions for twisted 3D topological insulators}

\author{Aaron Dunbrack}
\affiliation{Department of Physics and Astronomy, Stony Brook University, Stony Brook, New York 11974, USA}

\author{Jennifer Cano}
\affiliation{Department of Physics and Astronomy, Stony Brook University, Stony Brook, New York 11974, USA}
\affiliation{Center for Computational Quantum Physics, Flatiron Institute, New York, New York 10010, USA}

\date{\today}

\begin{abstract}

    We derive a general low-energy theory for twisted moir\'e heterostructures comprised of Dirac materials.
    We apply our theory to heterostructures on the surface of a three dimensional topological insulator (3D TI).
    First, we consider the interface between two 3D TIs arranged with a relative twist angle.
    We prove that if the two TIs are identical, then a necessary condition for a perturbative magic angle where the Dirac cone velocity vanishes is to have an interlayer spin-flipping hopping term.
    Without this term, the Dirac cone velocities can still be significantly renormalized, decreasing to less than half of their original values, but they will not vanish.
    Second, we consider graphene on the surface of a TI arranged with a small twist angle.
    Again, a magic angle is unachievable with only a spin-flipping hopping term; under such conditions, the Dirac cone is renormalized, but only moderately.
    In both cases, our perturbative results are verified by computing the band structure of the continuum model.
    The enhanced density of states that results from decreasing the surface Dirac cone velocity provides a tunable route to realizing interacting topological phases.
    
\end{abstract}

\maketitle

\section{Introduction\label{Sec:Intro}}

Magic angle twisted bilayer graphene exhibits a variety of interacting phases such as superconductivity and the quantum anomalous Hall effect \cite{cao2018correlated,cao2018unconventional,sharpe2019emergent,serlin2020intrinsic}.
The relative twist angle between the two layers acts as a tuning knob that can dramatically change their physical properties.
Specifically, at specific ``magic'' twist angles \cite{Bistritzer12233,lopes2012continuum,suarez2010flat}, the renormalized Dirac cone velocity vanishes, leaving behind gapped flatbands that are susceptible to interacting instabilities.

In this manuscript, we ask whether the physics observed in twisted graphene heterostructures can be realized in other Dirac materials. We are particularly motivated by renormalizing the velocity of the Dirac cone on the surface of a topological insulator (TI), where an interacting gap will yield either a quantum anomalous Hall insulator or a topological insulator \cite{hasan2010colloquium}. Recently, the effect of a slowly spatially-varying ``moir\'e superlattice'' potential on the surface of a topological insulator has been studied \cite{cano2021moire,wang2021moire,guerci2022designer}. Such a potential will increase the surface density of states at the Dirac point by renormalizing the Dirac cone. It will also generate van Hove singularities at an energy slightly above and below the Dirac point. Both of these effects will enhance the instabilities towards interacting topological phases \cite{fu2008superconducting,santos2010,baum2012magnetic,baum2012density,marchand2012lattice,schmidt2012strong,sitte2013interaction,mendler2015magnetic,wang2021moire,guerci2022designer}.
A potential that breaks time-reversal symmetry may produce flat Chern bands \cite{liu2021magnetic}. 
The theory can be generalized to modulate an interacting gap on the edge of a 2D TI \cite{chou2021band}.

Here, we consider a different route to manipulating the surface Dirac cone velocity, which is to incorporate a 3D TI into a twisted heterostructure with another Dirac material. The interlayer coupling between the two Dirac cones offers a new tuning knob compared to the superlattice potential and, as we show, can lead to perfectly flat bands under certain ideal (but not fine-tuned) conditions.

To this end, we derive a general theory of a twisted moir\'e heterostructure between two Dirac materials. The two layers need not be identical, but we require them to feature Dirac cones which, upon being arranged with a small twist angle, are separated by a small momentum difference. We derive an analytical expression for the low-energy theory as a function of the interlayer coupling and discuss the effect of rotation, time-reversal, and interlayer mirror symmetries.

We then apply our theory to two types of twisted heterostructures on the surface of a topological insulator. First, we consider the two-dimensional (2D) interface between two three-dimensional (3D) TIs arranged with a small relative twist angle. To meet our criterion that the two Dirac cones must be separated by  small momentum after twisting, we require that the Dirac cones are not at the center of the Brillouin zone (BZ), but rather at the $(\pi,\pi)$ point. We then derive conditions for gapless flat bands. Our most significant result is that if the two 3D TIs are identical, a gapless flat band requires an interlayer spin-flipping hopping term. We prove that this condition is mathematically equivalent to the known condition in twisted bilayer graphene that an interlayer sublattice-swapping is necessary to achieve a magic angle \cite{bernevig2020tbg}. However, while the latter is natural in twisted bilayer graphene, the origin of a spin-flipping term is not clear, although it is symmetry-allowed. Nonetheless, if the spin-flipping term is not present, the Dirac cone velocity can still be significantly renormalized, reaching a minimum around half of its original value. We back our analytical results with a numerical simulation.

The second heterostructure we examine in detail is graphene on the surface of a topological insulator, arranged with a small relative twist angle. Graphene on a 3D TI surface has been studied extensively without a twist angle, both theoretically \cite{Rossi_2019,jin2012multiple,jin2013proximity,RossiTriolaOld,cao2016heavy,debeule2017transmission,rodriguez2017giant,song2018spin, GRBISE5, PhysRevB.100.165141} and experimentally \cite{dang2010epitaxial,song2010topological,lee2015proximity,steinberg2015tunneling,zhang2016gate,bian2016experimental,vaklinova2016current,rajput2016indirect,zhang2017electronic,GRBISE1,khokhriakov2018tailoring,GRBISE3}.  A twist angle was considered in Refs.~\onlinecite{RossiTriolaOld,GRBISE1}, but flat bands were not considered. The primary goal of the previous literature was to enhance spin-orbit coupling in graphene.
Our motivation is in the opposite direction: we hope the interlayer coupling from graphene will renormalize the Dirac cone on the 3D TI. 
Similar to the interface between two topological insulators, we find that a magic angle condition is not achieved without spin-flipping interlayer hopping, although the 3D TI Dirac cone can still be slightly renormalized. We verify our analytical results with a numerical model.

\section{Magic Angle Condition for Dirac Cones at TRIM Points\label{Sec:Approach}}

We are interested in twisted bilayer heterostructures of two Dirac materials. Although our theory is applicable to any Dirac material, we focus on the case where one layer has a single time-reversal-invariant Dirac cone at a TRIM point, which implies it is the surface of a strong 3D TI.
We write down the most general interlayer coupling Hamiltonian and compute the self-energy of each Dirac cone perturbatively in the coupling. From the self-energy, we extract both an energy shift and a renormalized Fermi velocity for the Dirac cone, the latter of which gives a magic angle condition. We then discuss the physical conditions necessary to achieve the magic angle. Although we use the language specific to a twisted moir\'e heterostructure, the theory presented applies equally well to a moir\'e heterostructure created by stacking two layers with a small lattice mismatch.

\subsection{Model for Dirac materials\label{Sec:Assumps}}

A Dirac Hamiltonian for an isotropic gapless Dirac cone at charge neutrality can be written in the form:
\begin{equation}\label{eq:DiracHam}
    H_D(k)=vk\cdot\sigma,
\end{equation}
where $\sigma$ is some here-unspecified degree of freedom such as spin, sublattice, or orbital. (The proof that Eq.~(\ref{eq:DiracHam}) is fully general is given in Appendix~\ref{Apx:CanHamForm}.)
When the Dirac cone is not at the origin, we will denote the momentum difference from the Dirac point at $k_0$ by $\bar k \equiv k-k_0$. 
In each layer, there may be multiple Dirac cones of the form of Eq.~\eqref{eq:DiracHam}, which are labelled by valley, spin, or other indices. 
Unless symmetry-related, each Dirac cone will have a different Fermi velocity and energy, as well as anisotropy; for simplicity here we do not vary these parameters.

The Hamiltonian for each layer $L=1,2$ is written as $H(\bar k)=H_D(\bar k)\delta_{ij}$ (where $i$, $j$ run over the different Dirac cones, e.g., in graphene, over spin and valley):
\begin{equation}\label{eq:H0opdef}
    \hat H_0=\int d^2\bar k \psi_{L,\bar k}^\dagger H(\bar k)\psi_{L,\bar k}
\end{equation}
Interlayer coupling is then included as:
\begin{equation}\label{eq:Topdef}
    \hat T=\sum_Q \hat T_Q=\sum_Q\int d^2\bar k \hat\psi_{1,\bar k}^\dagger T_{Q}\hat\psi_{2,\bar k+Q}+h.c.
\end{equation}
Here, $\hat\psi_{L,\bar k}$ are the electron field operators on layer $L$ at crystal momentum $k_0+\bar k$ with spin, orbital, valley, etc. indices implicit; note $k_0$ may also implicitly depend on valley and layer.

The vectors $Q$ are the displacements in momentum space of the Dirac cones that result from twisting (see, e.g., Fig.~\ref{fig:TITIBZs}). The collection of vectors $\{Q_i-Q_j\}$ from each valley form the reciprocal lattice of the moir\`e pattern and thereby define the moir\`e BZ. A derivation of which $Q$ enter and how to compute the corresponding interlayer coupling matrix $T_Q$ is given in Appendix~\ref{Apx:CoupDer}.
Under the assumption that the lattices are nearly matched and arranged with a small twist angle, and, in addition, that the interlayer spacing is appreciably larger than the lattice constant of either material, $T_Q$ is only non-negligible for a finite set of $Q$. 

The low-energy continuum model in Eqs.~(\ref{eq:H0opdef}) and (\ref{eq:Topdef}) is valid when $\bar k$ is small (i.e., close enough to a Dirac cone that the linearization is valid) and when the valleys are well-separated (i.e., the distance between distinct valleys is much larger than $Q$).

To study the effect of the interlayer coupling term in Eq.~(\ref{eq:Topdef}) on the Dirac Hamiltonian in Eq.~(\ref{eq:H0opdef}), we assume that the interlayer coupling is small, in the sense that the energy scale of interlayer coupling, $t$, is much less than $v|Q|$ for the smallest value of $|Q|$.
(This implicitly assumes that there is no $Q=0$ term; the possibility of such a term is discussed further in Sec.~\ref{Sec:TITI}.) We then work perturbatively in the parameter $t/v|Q|$.

We further assume the presence of both time-reversal symmetry, $\mathcal{T}$, with $\mathcal{T}^2=-1$, and an in-plane rotational symmetry by $2\pi/n$, denoted $C_{n,z}$, with $n>2$. 
The rotation symmetry ensures that the Dirac cone is isotropic, which is a convenient and physically relevant simplification, but is not necessary.
If the two layers are the same, there may be additional symmetries that exchange the layers. The action of the symmetry operators is derived in Appendix~\ref{Apx:CanHamForm}.

\subsection{Self-energy of a single Dirac cone\label{Sec:Methods}}

We now treat the interlayer coupling in Eq.~\eqref{eq:Topdef} as a perturbation to the Dirac Hamiltonian in Eq.~\eqref{eq:H0opdef} and compute the self-energy to lowest order in $t/v|Q|$. From the self-energy, we will extract the renormalized velocity and energy shift of the original Dirac cone due to interlayer coupling. We compute the self-energy of a single Dirac cone located at a time-reversal-invariant-momentum (TRIM); in Appendix~\ref{Apx:SelfEnCalcDetails} we discuss the generalization for a system with multiple Dirac cones not necessarily at TRIM. Without loss of generality, we take the layer in which we are computing the self-energy to be layer $1$, with Fermi velocity $v_1$, while layer 2 has Fermi velocity $v_2$.

In this single Dirac cone case, the combination of time-reversal and rotational symmetry constrain the self-energy to take the form:
\begin{equation}\label{eq:1DCSelfEn}
    \Sigma^\text{single-cone}(\omega,\bar k)=-[A\omega+B+C(\bar{k}\cdot\sigma)+D(\bar{k}\cross\sigma)]
\end{equation}
The parameter $A$ determines the quasiparticle weight via $Z^{-1} = 1+A$, while the parameters $C$ and $D$ are direct corrections to the renormalized Fermi velocity, which takes the form:
\begin{equation}\label{eq:1DCFermiVel}
    v^*_1=\frac{\sqrt{(v_1-C)^2+D^2}}{1+A},
\end{equation}
where $v_1$ is the original velocity before the interlayer coupling.
The parameter $B$ in Eq.~\eqref{eq:1DCFermiVel} determines an overall energy shift of the renormalized Dirac cone, given by:
\begin{equation}\label{eq:1DCEnShift}
    \Delta E=-\frac{B}{1+A}
\end{equation}

Eq.~\eqref{eq:1DCFermiVel} is an important result: it imposes a condition to find a ``magic angle'' where the renormalized Fermi velocity vanishes, specifically, $v^*_1 = 0$ when $D=0$ and $C=v_1$. In contrast, previous work \cite{cano2021moire} showed that a superlattice potential on the surface of a 3D TI can renormalize the Dirac cone velocity, but does not result in a magic angle condition; in the language of Eq.~\eqref{eq:1DCFermiVel}, it yields $C=D=0$.

We now derive the physical conditions that must be satisfied to realize the magic angle by expressing the abstract parameters $A,B,C,D$ in terms of the physical interlayer coupling terms, $T_Q$, introduced in Eq.~\eqref{eq:Topdef}.
 
For simplicity, we first assume the original Dirac cone is coupled to a single Dirac cone in an adjacent layer, so that $T_Q$ is a $2\times 2$ matrix. For each $Q$, we decompose $T_Q$ as follows:
\begin{equation}\label{eq:TQdef}
    T_Q=t_{0,Q}\I+t_{z,Q}\sigma_z+t_{||,Q}(\hat{Q}\cdot\sigma)+t_{\perp,Q}(\hat{Q}\times\sigma),
\end{equation}
where the parameters $t_{0,z,||,\perp}$ are complex numbers. Note that since $\hat Q$ is ill-defined at $Q=0$, $t_{||}$ and $t_{\perp}$ are not continuous through $\theta=0$; specifically, the behavior of $\hat{Q}$ implies that $t_{||,\perp}\sim \text{sgn}(\theta)$, to lowest order in $\theta$.

In terms of these coefficients,
\begin{subequations}\label{eq:1DCCoeffs}
\begin{align}
    A&=\sum_Q\frac{1}{v_2^2Q^2}(|t_{0,Q}|^2+|t_{z,Q}|^2+|t_{||,Q}|^2+|t_{\perp,Q}|^2)\\
    B&=\sum_Q\frac{2}{v_2|Q|}(\Re[t_{0,Q}t_{||,Q}^*]+\Im[t_{\perp,Q}t_{z,Q}^*])\\
    C&=\sum_Q \frac{1}{v_2Q^2}(|t_{\perp,Q}|^2-|t_{||,Q}|^2)\\
    D&=\sum_Q \frac{-2}{v_2Q^2}\Re[t_{||,Q}t_{\perp,Q}^*]
\end{align}
\end{subequations}
These equations follow directly from the more general results presented in Appendix~\ref{Apx:SelfEnCalc}.

More generally, if the original Dirac cone is coupled to multiple Dirac cones in the second layer, labelled by another (valley, spin, etc.) index, then the decomposition in Eq.~\eqref{eq:TQdef} should be applied to each valley, and the parameters $A,B,C,D$ are found by summing Eq.~\eqref{eq:1DCCoeffs} over all valleys.

In each of the sums in Eq.~\eqref{eq:1DCCoeffs}, all symmetry-related $Q$ values will contribute the same summand, due to the definition of the coefficients in Eq.~\eqref{eq:TQdef} and the symmetry properties discused in Appendix~\ref{Apx:SymmTq}. Accordingly, if only the smallest $Q$ values are non-negligible, it suffices to consider the couplings for one choice of $Q$.

\section{Interface between two 3D topological insulators\label{Sec:TITI}}

The first case we will consider is the interface between two 3D TIs that are stacked on top of each other with a small relative twist angle.
This is also the simplest case, since each surface hosts only a single time-reversal-invariant Dirac cone.

If both layers have a Dirac cone at $\Gamma$, there will be a zero-momentum ($Q=0$) interlayer coupling term. Physically, this term appears because even after twisting, the Dirac cones at $\Gamma$ have no momentum separation. (Mathematically, the Q=0 term arises from our derivation in Appendix~\ref{Apx:CoupDer}.) The presence of a $Q=0$ term invalidates our results in the previous section, which were derived by expanding perturbatively in $t/v|Q|$. Instead of a renormalized gapless Dirac cone, a $Q=0$ term will open a gap at the interface, similar to the interface between two untwisted 3D TIs, but modulated by a superlattice potential. Therefore, to realize a gapless Dirac cone at an interface between two twisted 3D TIs, we require that at least one layer has a Dirac cone at a TRIM that is not $\Gamma$.

\begin{figure*}
    \centering
    \begin{subfigure}[t]{0.45\textwidth}
        \includegraphics[width=0.6\textwidth]{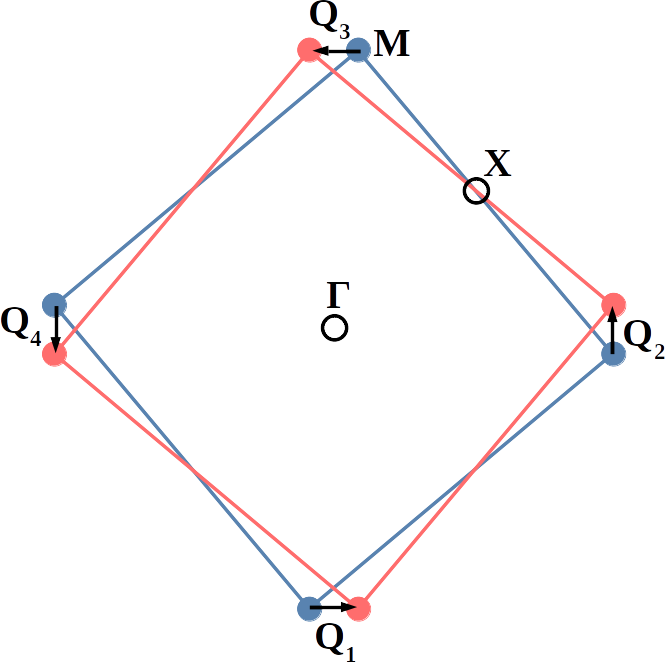}
        \caption{Surface Dirac cones at $M$ in both layers, with identical lattice constants.}
        \label{fig:TITISameBZ}
    \end{subfigure}
    \begin{subfigure}[t]{0.45\textwidth}
        \includegraphics[width=0.6\textwidth]{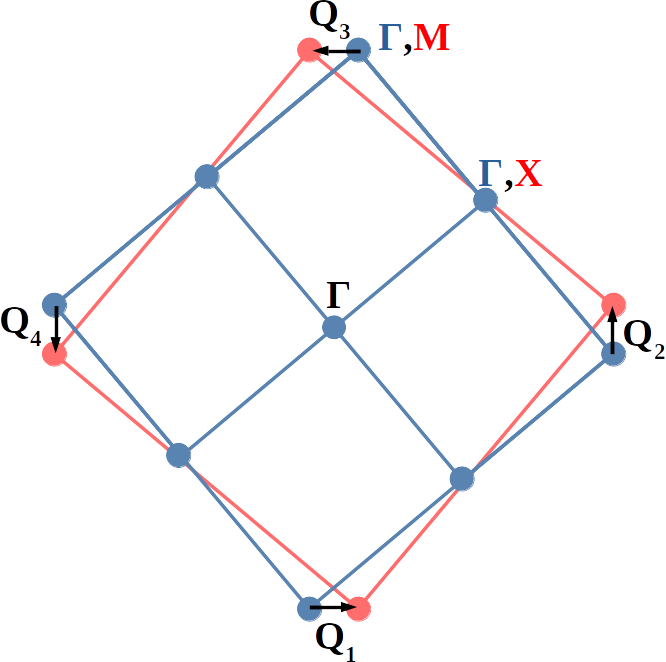}
        \caption{Dirac cone at $M$ on one surface and $\Gamma$ on the other, with lattice constants that differ by a factor of two.}
        \label{fig:TITI2FoldBZ}
    \end{subfigure}
    \caption{BZ at the interface between two 3D TIs arranged with a small relative twist for two scenarios described in Sec.~\ref{Sec:TITI}. Filled circles indicate Dirac cones; empty circles indicate other TRIM in the original BZ. Vectors $Q$ indicate the momentum difference between nearest Dirac cones, which is exactly the effective quasimomentum transfer in the continuum model, per Eq.~\eqref{eq:Topdef}.}
    \label{fig:TITIBZs}
\end{figure*}

This motivates us to consider the hypothetical situation of an interface between two $C_4$-symmetric 3D TIs with the same lattice constant and Dirac cones at the surface $M\equiv (\pi,\pi)$ point. When the two materials are twisted, the surface BZ is illustrated in Fig.~\ref{fig:TITISameBZ}.
Since the Dirac cones are at $(\pi,\pi)$, they are $C_4$-invariant and thus isotropic to linear order (any rotation of order greater than two will yield an isotropic Dirac cone.)
Without such a symmetry, the Dirac cones can flatten anisotropically; if the velocity vanishes in only one direction, an effective 1D system results \cite{kariyado2019flat,kennes2020one}. 
The other TRIM on the surface of a four-fold-symmetric TI are not $C_4$-symmetric: the $(\pi,0)$ and $(0,\pi)$ points mix under a four-fold rotation.
On the surface of a three- or six-fold symmetric TI, the TRIM also mix under the three- or six-fold rotational symmetry, such that there are no TRIM points (besides $\Gamma$) that are three- or six-fold invariant.
Thus, if we seek a Dirac cone at a TRIM that will be renormalized isotropically at a twisted interface between two 3D TIs, our only option is to consider a Dirac cone at $(\pi,\pi)$ on the surface of a four-fold symmetric topological insulator.

We may also consider interfaces between two 3D TIs with different lattice constants. 
For example, consider an interface between one 3D TI with a Dirac cone at $M$ and a second 3D TI with a Dirac cone at $\Gamma$, such that the two materials have a 2:1 lattice vector mismatch (their unit cell areas differ by a factor of four). In this setup, the cone at $M$ from the first layer is folded onto $\Gamma$ from the second when the layers are aligned. When the two materials are twisted, the surface BZ is illustrated in Fig.~\ref{fig:TITI2FoldBZ}. The low-energy theory of the interface is similar to the case where the two layers have the same lattice constant and both have Dirac cones at $M$, but subject to different symmetries. Further, the global topology, i.e., where the Dirac points are located on the surface BZ, may differ, which gives rise to different mechanisms for the topological protection of gapless surface states.
One can generalize to construct interfaces with more complicated supercells describing any integer ratio of lattice constants.

\begin{figure}
    \centering
    \includegraphics[width=0.3\textwidth]{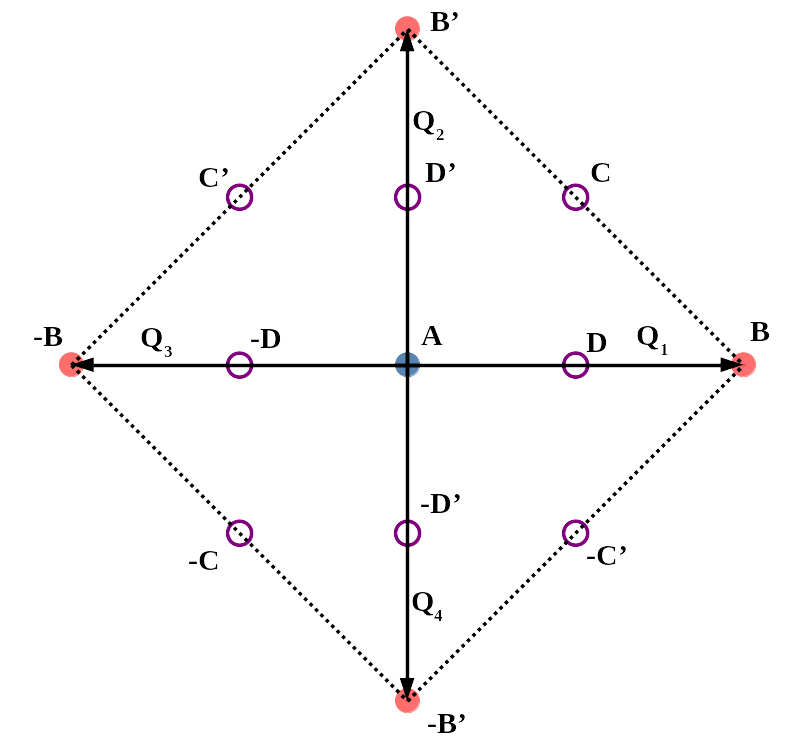}
    \caption{Moir\'e BZ. Blue and red filled circles are the locations of Dirac cones from each layer. Empty circles indicate other points along the BZ path (which may be TRIM points depending on the configuration.)}
    \label{fig:TITIPlotPathArea}
\end{figure}

\subsection{Identical TI surfaces\label{Sec:TITISame}}

We start by considering the first scenario, where both TIs are identical and have Dirac cones at $M$.
Then $C_2\mathcal{T}$ symmetry leaves each $Q$ invariant (as we derive in Eqs~\eqref{eq:TRSonT} and \eqref{eq:RotonT}, it acts as $-\sigma_x\mathcal{K}$, implying $\sigma_xT_Q^*\sigma_x=T_Q$.) 
This constrains the coefficients $t_0$, $t_{||}$ and $t_{\perp}$ to be real, whereas $t_z$ is purely imaginary.
Accordingly, let us define $t_z=i\hat t_z$.

We assume that only the smallest symmetry-related set of four $Q$ contribute nonnegligibly. Without loss of generality, we may then choose these $Q$s along the $x$ and $y$-axes, as illustrated in Fig.~\ref{fig:TITISameBZ}. This choice of $Q$s is still valid in the presence of a small lattice mismatch, although the decomposition into $t_{||}$ and $t_\perp$ will change.
Eq.~\eqref{eq:1DCCoeffs} simplifies to:
\begin{subequations}\label{eq:TITICoeffs}
\begin{align}
A&=\frac{4(t_{0,Q}^2+\hat t_{z,Q}^2+t_{||,Q}^2+t_{\perp,Q}^2)}{v^2Q^2}\\
B&=\frac{8(t_{0,Q}t_{||,Q}+t_{\perp,Q}\hat t_{z,Q})}{v|Q|}\label{eq:TITICoeffB}\\
C&=\frac{4(t_{\perp,Q}^2-t_{||,Q}^2)}{vQ^2}\label{eq:TITICoeffC}\\
D&=\frac{-8t_{||,Q}t_{\perp,Q}}{vQ^2}
\end{align}
\end{subequations}

One of the most important consequences of these formulas is that if $t_{||}$ and $t_{\perp}$ vanish, then the only non-zero term is $A$. 
However, according to the formula for the renormalized velocity in Eq.~\eqref{eq:1DCFermiVel}, a magic angle requires $C> 0$.
Thus, to lowest order in perturbation theory, a magic angle in this system requires $t_{\perp}$ to be non-zero.
Since, by its definition in Eq.~\eqref{eq:TQdef}, $t_{\perp}$ is a coefficient of off-diagonal Pauli matrices,
we conclude that magic angles are only possible in the presence of inter-layer spin-flipping hopping terms.

This result can be understood by analogy to twisted bilayer graphene, where instead of spin, the $\sigma$ matrices act on the sublattice degree of freedom. In Ref.~\cite{bernevig2020tbg}, the authors showed that magic angles do not appear in the ``second chiral limit'' of TBLG where inter-sublattice interlayer hopping vanishes.
When sublattice is replaced by spin, this is exactly our result that magic angles do not appear at the interface between two identical 3D TIs without spin-flipping interlayer hopping.
Although in TBLG the Dirac cones are in sublattice space and protected by spinless $C_2\mathcal{T}$, while on the surface of a 3D TI, the Dirac cone is in spin space and protected by $\mathcal{T}$, we show in Appendix~\ref{Apx:TC2SelfEn} that this analogy is mathematically rigorous.

We now consider the magic angle constraint in more detail.
If the two 3D TI layers are identical, they have an additional layer-interchanging symmetry $C_{2y}$ (which, in combination with $C_4$, generates three other layer-interchanging symmetries). Using the orientation of $Q$s defined in Fig.~\ref{fig:TITIBZs} (along with the action of the rotation operator defined in Eq.~\ref{eq:ILRonT}), this symmetry requires:
\begin{equation}
    \sigma_yT_{Q_1}^\dagger\sigma_y=T_{Q_1}
\end{equation}
Combining the above equation with the decomposition of $T_Q$ given in Eq.~\eqref{eq:TQdef} (and the aforementioned $C_2\mathcal{T}$ constraints) yields the single additional constraint $t_{||}=0$, which, from Eq.~\eqref{eq:TITICoeffs}, implies $D=0$, $C=4t_{\perp,Q}^2/vQ^2$.
According to the expression for the renormalized Fermi velocity in Eq.~\eqref{eq:1DCFermiVel}, a magic angle will result when $D=0$, $C=v_1$. Since varying the twist angle changes the magnitude $Q^2$ that appears in the denominator of $C$, we conclude that as long as $t_{\perp,Q}\neq 0$, there will exist some magic twist angle where $C=v_1$.
Importantly, this magic twist angle requires no fine-tuning.

On the other hand, as mentioned below Eq.~\eqref{eq:TITICoeffs}, $t_{\perp,Q}$ corresponds to a spin-flipping interlayer hopping term.
While this term is symmetry-allowed, it is not obvious how such a term would naturally arise, and we might expect it to be small. 
Since the magic angle occurs when $C=v_1$, which implies $t_{\perp,Q} \sim vQ$, then if $t_{\perp,Q}$ is small, the magic angle requires that $Q$ also be small, which corresponds to a small twist angle (see Fig.~\ref{fig:TITISameBZ}). 
At such small twist angles, there are two places where our perturbative results may fail. First: if spin-preserving couplings dominate, then we require small $|Q|$ for $t_{\perp}/v|Q|$ to be sufficiently large. However, since our expansion in $t$ includes a leading order approximation in $t_0/v|Q|$ as well, the smallness of $|Q|$ may demand working to a higher order in perturbation theory in $t_0$. Second: physically, our assumption of rigid layers may break down at small twist angle if disorder or lattice relaxation effects dominate.

In the absence of a layer-interchanging symmetry, generically $t_{||}\neq 0$. Since this implies $D\neq 0$ whenever $C>0$, by Eq.~\eqref{eq:1DCFermiVel}, lowest-order perturbation theory indicates magic angles will not arise. To higher order in perturbation theory, fine-tuned scenarios may arise where a magic angle exists, but we do not expect it to be generic.

In Sec.~\ref{Sec:TITINum} we will compare our perturbative predictions for the magic angle to a global band structure calculation.

\begin{figure*}
    \centering
    \begin{subfigure}[t]{0.24\textwidth}
        \includegraphics[width=.6\textwidth]{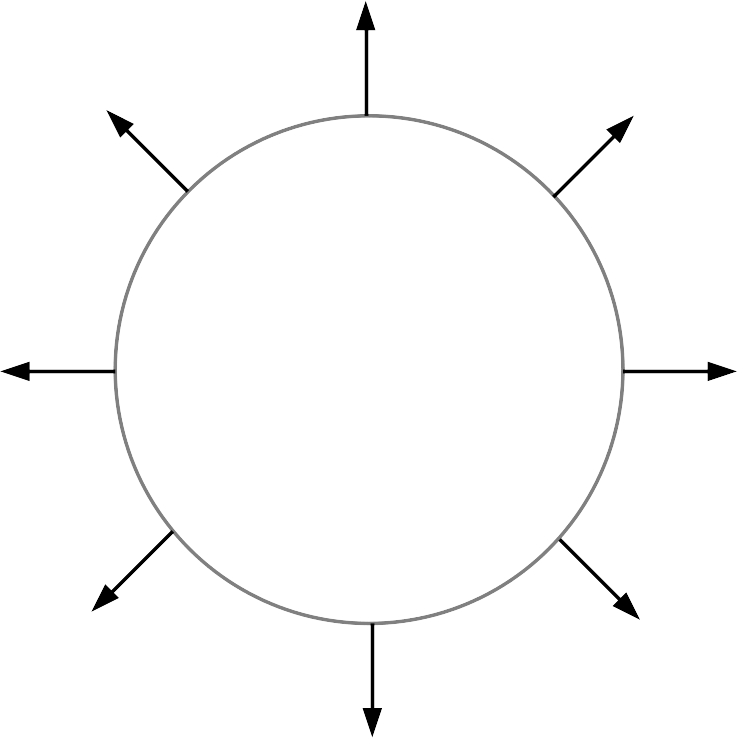}
        \caption{$H=vk\cdot\sigma$}
        \label{fig:dotspin}
    \end{subfigure}
    \begin{subfigure}[t]{0.24\textwidth}
        \includegraphics[width=.6\textwidth]{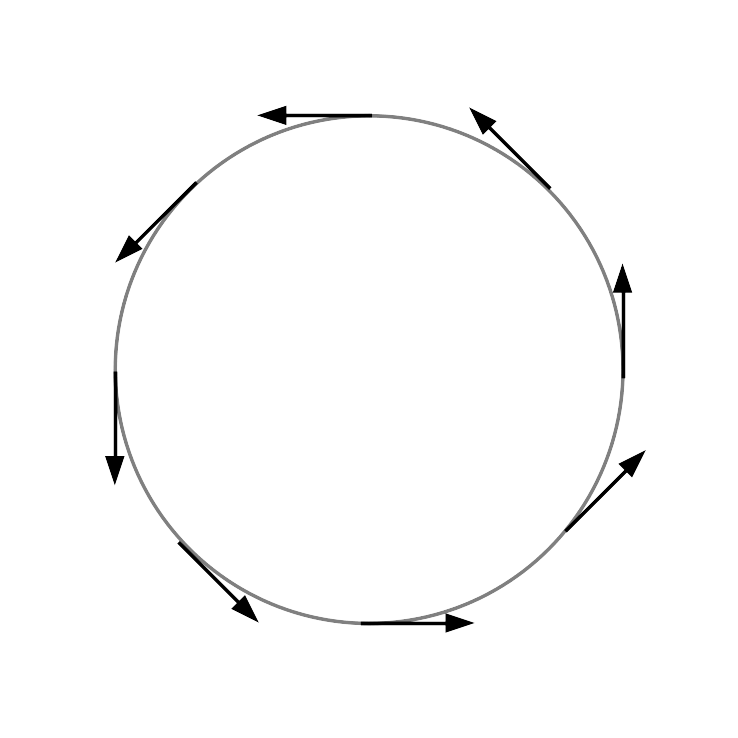}
        \caption{$H=vk\times\sigma$}
        \label{fig:crossspin}
    \end{subfigure}
    \begin{subfigure}[t]{0.24\textwidth}
        \includegraphics[width=.6\textwidth]{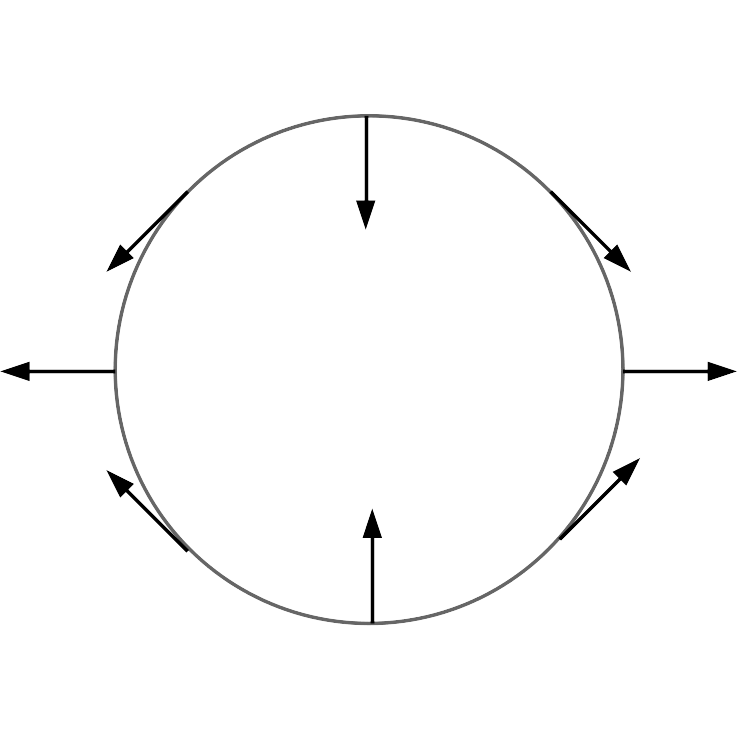}
        \caption{$H=vk\cdot\sigma^*$}
        \label{fig:dotstarspin}
    \end{subfigure}
    \begin{subfigure}[t]{0.24\textwidth}
        \includegraphics[width=.6\textwidth]{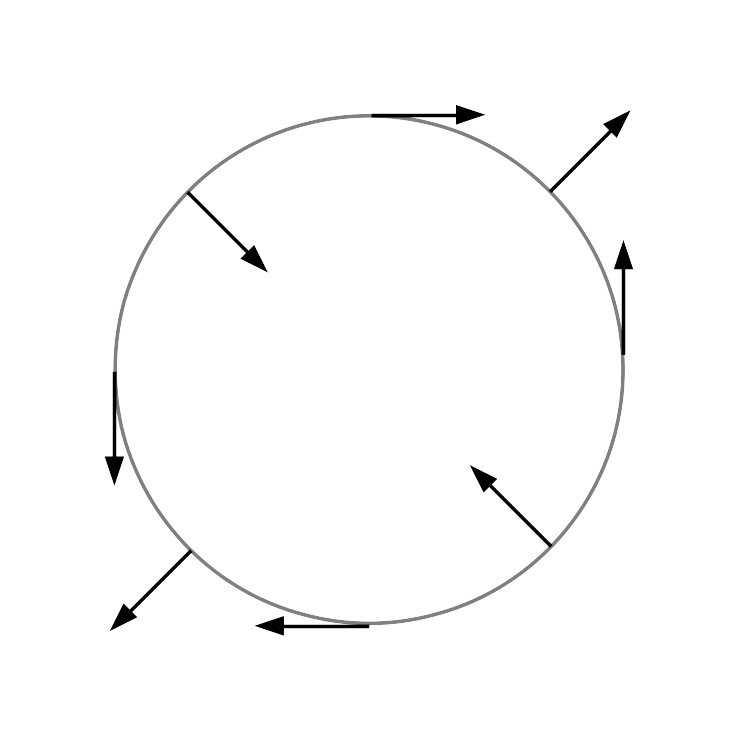}
        \caption{$H=vk\times\sigma^*$}
        \label{fig:crossstarspin}
    \end{subfigure}
    \begin{subfigure}[t]{0.24\textwidth}
        \includegraphics[width=.6\textwidth]{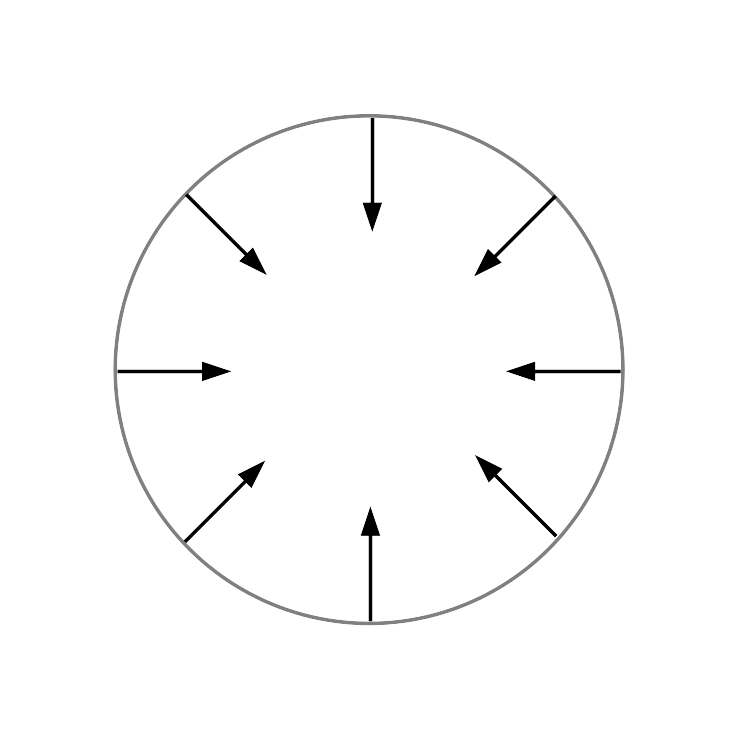}
        \caption{$H=-vk\cdot\sigma$}
        \label{fig:mdotspin}
    \end{subfigure}
    \begin{subfigure}[t]{0.24\textwidth}
        \includegraphics[width=.6\textwidth]{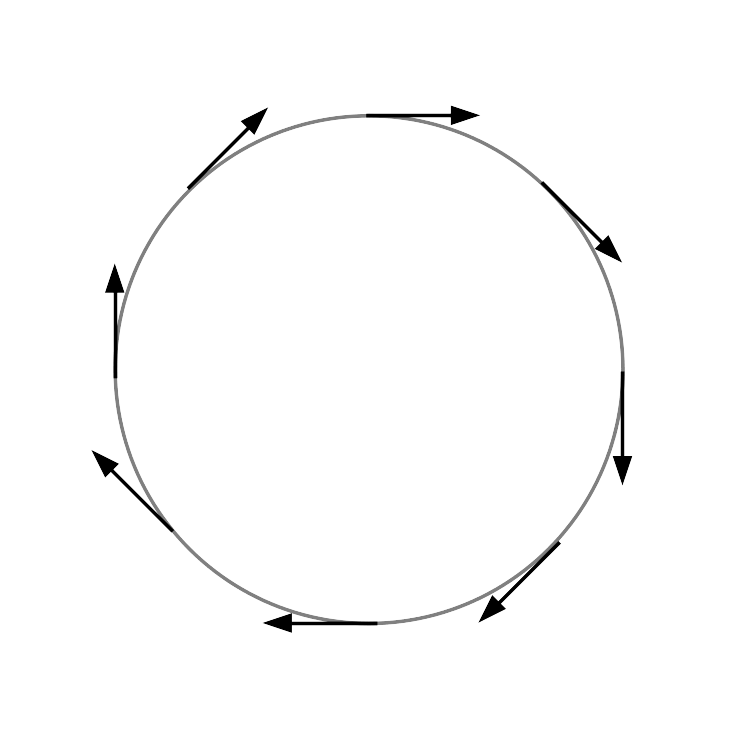}
        \caption{$H=-vk\times\sigma$}
        \label{fig:mcrossspin}
    \end{subfigure}
    \begin{subfigure}[t]{0.24\textwidth}
        \includegraphics[width=.6\textwidth]{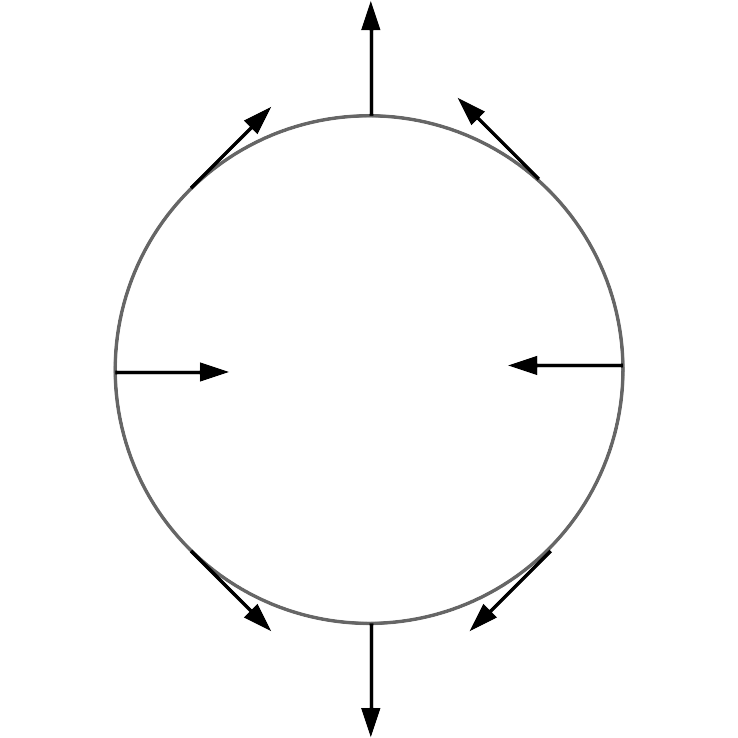}
        \caption{$H=-vk\cdot\sigma^*$}
        \label{fig:mdotstarspin}
    \end{subfigure}
    \begin{subfigure}[t]{0.24\textwidth}
        \includegraphics[width=.6\textwidth]{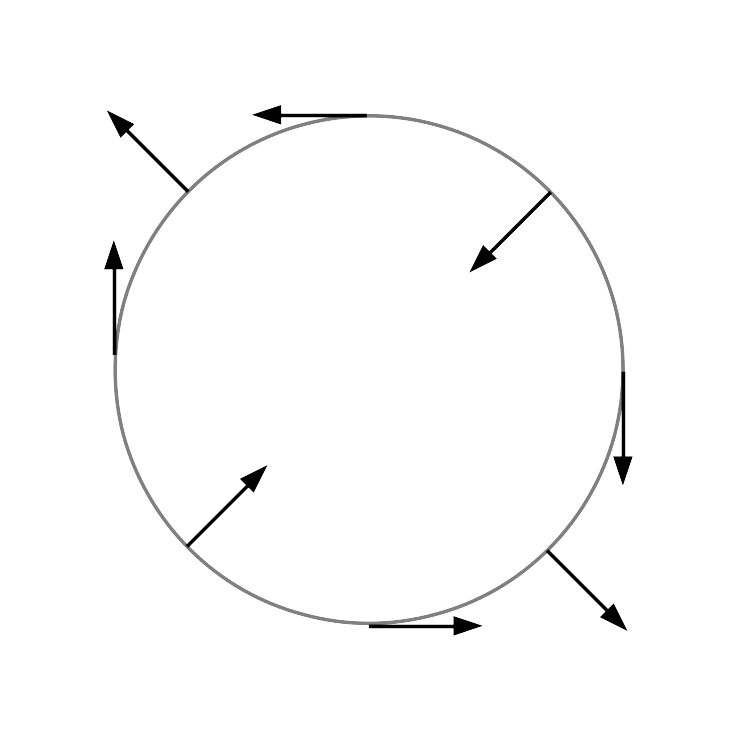}
        \caption{$H=-vk\times\sigma^*$}
        \label{fig:mcrossstarspin}
    \end{subfigure}
    \caption{Spin textures of the conduction bands of different Dirac cones. 
    For Dirac cones where $\sigma$ appears in the Hamiltonian, spin and momentum rotate the same way, whereas for Dirac cones where $\sigma^*$ appears, spin and momentum rotate in opposite directions.
    This winding number also results in different (unitarily equivalent) angular momentum representations.}
    \label{fig:SpinTextures}
\end{figure*}\subsection{Spin vs. pseudospin\label{Sec:TITI-*}}

The low-energy Hamiltonian $v(\bar{k}\cdot \sigma)$ is not the unique description of a Dirac cone invariant under time-reversal and rotation symmetries: specifically, as derived in Appendix~\ref{Apx:vkdotsigma}, any linear combination of $\bar{k}\cdot\sigma$ and $\bar k\times\sigma$ is invariant. Such linear combinations can be brought into the form $v(\bar{k}\cdot \sigma)$ by a unitary transformation, although the transformation also changes the meaning of the spin basis. As a first example, if the Hamiltonian in one layer is described by $v(\bar{k}\cdot \sigma)$ and the Hamiltonian in the other layer is described by $v(\bar k\times\sigma)$ or $-v(\bar k\cdot\sigma)$, then we reach the same conclusion as in the previous section that magic angles exist, but they require a different mixture of $t_{\perp}$ and $t_{||}$. Importantly, spin-flipping hopping is still required.

However, we get a different result if one layer has the low-energy Hamiltonian $-\bar k\cdot\sigma^*$. The unitary transformation that transforms this Hamiltonian to the form $\bar{k}\cdot\tilde\sigma$ exchanges spin-flipping and spin-preserving interactions (following the procedure described in Appendix~\ref{Apx:vkdotsigma}). Thus, the interlayer hopping term that results in a magic angle to leading order in perturbation theory is now a spin-preserving term.  For other $\sigma^*$-type Hamiltonians, the hopping required to achieve a magic angle preserves the projection of spin in the $z$-direction, but not in all directions, i.e., the requirement to achieve a magic angle is a linear combination of $t_0$ and $t_z$.

From a physical perspective, this set-up provides a physically reasonable (spin-preserving) interlayer hopping term to achieve flat bands at the interface between two topological insulators. However, this physical motivation for the interlayer hopping term comes at a cost: it requires two \textit{different} topological insulators, one with effective Hamiltonian $\bar k\cdot\sigma$ and the other with $-\bar k\cdot\sigma^*$.

Physically, materials with a $\bar k\cdot\sigma^*$ Hamiltonian can be distinguished from materials with a $\bar k\cdot\sigma$ Hamiltonian by their spin texture: their spins rotate in opposite directions around the Dirac cone, as illustrated in Fig.~\ref{fig:SpinTextures}. 
Specifically, writing the linearized Hamiltonian for layer $L$ as $k\cdot M_L\cdot\sigma$, the spin-winding of the Dirac cone is classified by the parameter:
\begin{equation}\label{eq:chiLdef}
    \chi_L=\text{sgn}(\det(M_L)) = \pm 1,
\end{equation}
where $\chi_L(k\cdot\sigma)=1$ and $\chi_L(-k\cdot\sigma^*)=-1$. 
To determine whether spin-flipping inter-layer hopping is necessary to realize a magic angle,
the relevant quantity is $\chi:=\chi_1\chi_2$, the product of the $\chi$ of each layer: if the product is $+1$, $z$-spin-preserving interlayer hopping is required, whereas if the product is $-1$, then $z$-spin-flipping interlayer hopping is required.

Note that $\chi_L$ does not completely classify the Hamiltonian, and as a consequence, the mixture of $t_{||}$ and $t_\perp$ required to achieve a magic angle depends on the low-energy Hamiltonian. For example, the ratio of $t_\parallel$ to $t_\perp$ required for a magic angle at the interface between two 3D TIs with $k\cdot \sigma$ Dirac cones will be different than that required at the interface between two 3D TIs with $k\times \sigma$ Dirac cones, even though both cases have $\chi_1=\chi_2 = +1$.

\subsection{Global band structure\label{Sec:TITINum}}

So far we have derived the existence/absence of a magic angle using the renormalized Dirac cone velocity derived in Eq.~\eqref{eq:1DCFermiVel} to leading order. We now study the validity of our conclusions to higher order in both $k$ and $t_0$ by computing the full band structure. We assume spin-preserving coupling and consider two cases: one in which both TIs have effective Hamiltonians of the form $\bar k\cdot\sigma$ (as considered in Sec.~\ref{Sec:TITISame}), and one in which the second TI instead has an effective Hamiltonian of the form $-\bar k\cdot\sigma^*$ (as discussed in Sec.~\ref{Sec:TITI-*}), where $\sigma$ always references the spin degree of freedom, rather than a rotated pseudospin degree of freedom. We will call these the $\chi=+1$ and $\chi=-1$ cases. Note that the band structure for the $\chi=-1$ case with spin-preserving hopping is identical to the $\chi = +1$ case with spin-flipping hopping, i.e., $t_0 \rightarrow t_\perp$.

\begin{figure*}
    \centering
    \begin{subfigure}[b]{0.29\textwidth}
        \includegraphics[width=\textwidth]{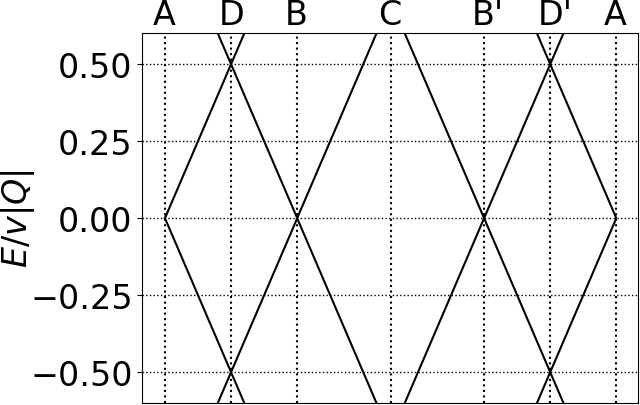}
    \end{subfigure}
    \begin{subfigure}[b]{0.226631\textwidth}
        \includegraphics[width=\textwidth]{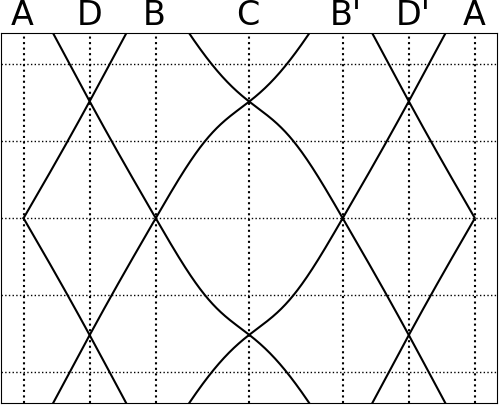}
    \end{subfigure}
    \begin{subfigure}[b]{0.226631\textwidth}
        \includegraphics[width=\textwidth]{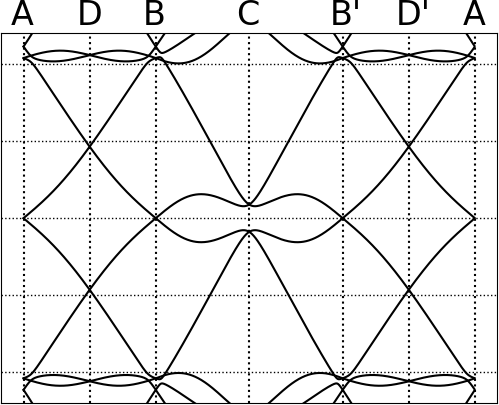}
    \end{subfigure}
    \begin{subfigure}[b]{0.226631\textwidth}
        \includegraphics[width=\textwidth]{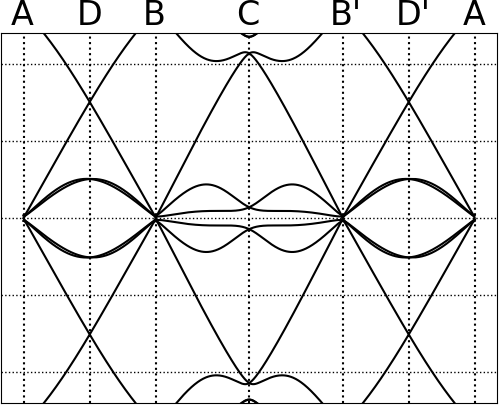}
    \end{subfigure}
    \begin{subfigure}[b]{0.29\textwidth}
        \includegraphics[width=\textwidth]{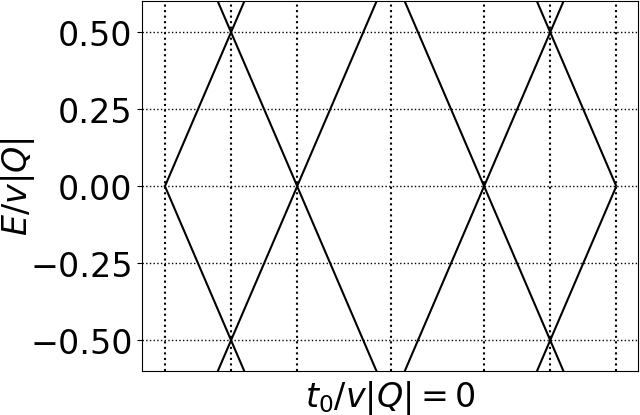}
    \end{subfigure}
    \begin{subfigure}[b]{0.226631\textwidth}
        \includegraphics[width=\textwidth]{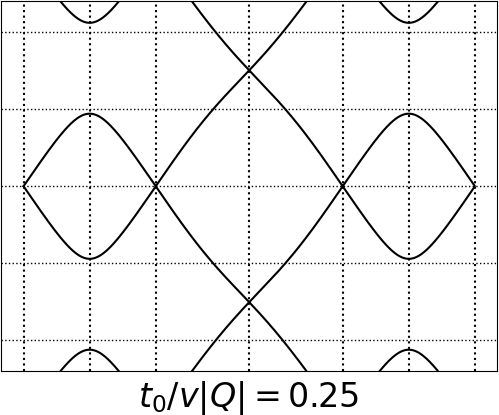}
    \end{subfigure}
    \begin{subfigure}[b]{0.226631\textwidth}
        \includegraphics[width=\textwidth]{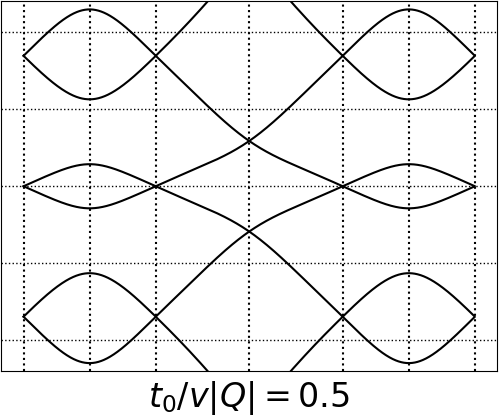}
    \end{subfigure}
    \begin{subfigure}[b]{0.226631\textwidth}
        \includegraphics[width=\textwidth]{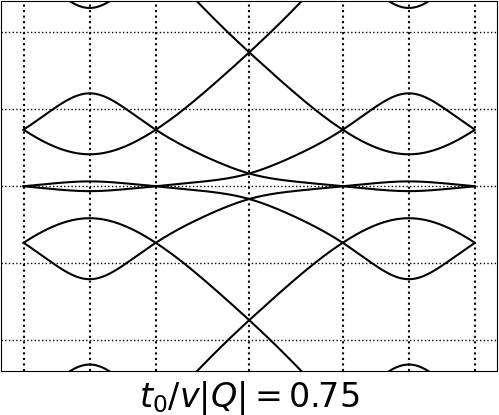}
    \end{subfigure}
    \caption{Moir\'e band structures for the interface between two 3D TIs. The path ABB'A refers to the moir\'e BZ shown in Fig.~\ref{fig:TITIPlotPathArea}. In the top row, the low-energy Hamiltonian for both materials is $vk\cdot\sigma$ ($\chi=+1$), while in the bottom row, the low-energy Hamiltonian for one TI is $vk\cdot\sigma$ and for the other $-vk\cdot\sigma^*$ ($\chi=-1$). In all cases, only a single spin-preserving interlayer coupling term is included. 
    The band structures are identical to those that result from a spin-flipping term by making the substitutions $t_0 \rightarrow t_\perp$ and $\chi \rightarrow -\chi$.
    }
    \label{fig:TITISpectrumPlots}
\end{figure*}

Fig.~\ref{fig:TITISpectrumPlots} shows the spectrum plotted along the ABB'A slice of the moir\'e BZ, which is labelled in Fig.~\ref{fig:TITIPlotPathArea}. The top row of plots are spectra for the $\chi=+1$ case, and the bottom row of plots are spectra for the $\chi=-1$ case, with varying $t_0/v|Q|$ in different columns.

\begin{figure}
    \centering
    \includegraphics[width=0.35\textwidth]{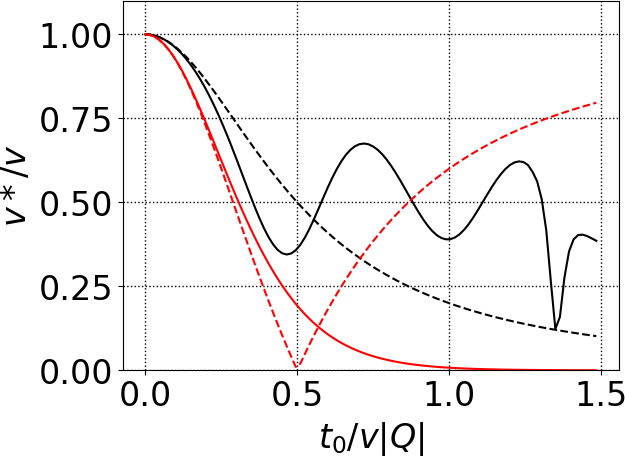}
    \caption{Fermi velocity vs. $t_0$ for spin-preserving couplings in the $\chi=+1$ (black) and $\chi=-1$ (red) cases. Dashed lines are first-order perturbation theory, solid lines are extracted from numerically-computed band structures. The $\chi=-1$ case consistently produces lower Fermi velocities.}
    \label{fig:TITIvt}
\end{figure}

We expect in the $\chi=-1$ case (from Eqs~\eqref{eq:TITICoeffC} and \eqref{eq:1DCFermiVel}, taking $t_0\rightarrow t_\perp$ according to the basis change described in the previous section and derived in Appendix~\ref{Apx:vkdotsigma}) that a magic angle will arise for $t_0/v|Q|=0.5$. Examining the $t_0/v|Q|=0.5$ plot, we find bands that are appreciably flatter, but not yet ``magic." However, we observe that at higher coupling, $t_0/v|Q|=0.75$ (Fig.~\ref{fig:TITISpectrumPlots}, bottom-right) produces appreciably flatter bands near a magic angle. We attribute this mismatch between our perturbative and numerical results to the fact that at the predicted magic angle, our perturbative parameter $t_0/v|Q| =0.5$ is not very small. In Fig.~\ref{fig:TITIvt}, we compare the numerical and perturbative calculations of the Fermi velocity and see that they begin to deviate around $t_0/v|Q|\sim 0.2$.

Nonetheless, the band structures plotted in Fig.~\ref{fig:TITISpectrumPlots} confirm the qualitative intuition (discussed in Sec.~\ref{Sec:TITI-*}) that, while both the $\chi=+1$ and $\chi=-1$ cases produce some band flattening (compare the leftmost column, in which the layers are uncoupled, to the other columns), the $\chi=-1$ case produces much flatter bands with spin-preserving interactions than the $\chi=+1$ case. This can be seen by comparing the spectra in the two rows of Fig.~\ref{fig:TITISpectrumPlots}: the bandwidth along the AB line, and in particular at D, is consistently narrower in the $\chi=-1$ case.

The same trend appears in Fig.~\ref{fig:TITIvt}, where the Fermi velocity is plotted against $t_0/v|Q|$.
The two cases are notably closer in Fermi velocity than first-order perturbation theory would predict for intermediate values of $t_0/v|Q|$ (between $0.2$ and $0.5$), but the $\chi=-1$ case nevertheless consistently yields lower Fermi velocities, and for large $t_0/v|Q|\gtrsim 0.6$, the $\chi=-1$ Fermi velocity becomes smaller by an order of magnitude than the $\chi=+1$ Fermi velocity. In addition, while the Fermi velocity oscillates in amplitude in the $\chi=+1$ case, the Fermi velocity in the $\chi=-1$ case decays monotonically, well past the validity of the perturbation theory.

\begin{figure}
    \centering
    \includegraphics[width=0.45\textwidth]{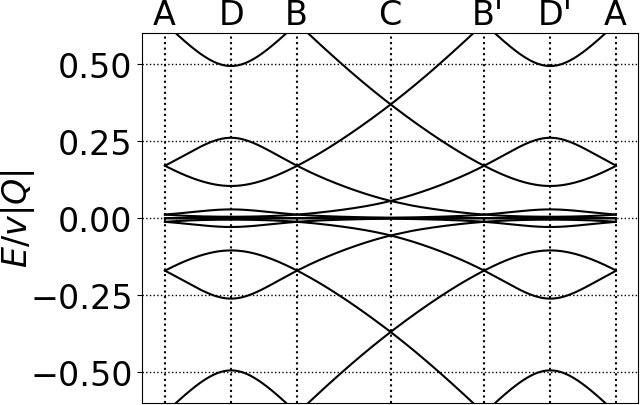}
    \caption{Spectrum for spin-preserving coupling in the $\chi=-1$ case with $t_0/v|Q|=1.3$. 
    The original Dirac cone becomes extremely flat and nearby bands collapse in energy, creating a large density of states.}
    \label{fig:TITIFlatSpectrum}
\end{figure}

In Fig.~\ref{fig:TITIFlatSpectrum}, we illustrate the extremely flat low-energy bands that result in the $\chi=-1$ case with $t_0/v|Q|=1.3$.
The figure shows that not only do the lowest energy bands become flat, but the adjacent bands also collapse onto the flat bands, creating a larger density of states than would result from the flat bands alone. Both the accumulation of flat bands and the exponential suppression of Fermi velocity were also discussed for twisted square lattices in Ref.~\onlinecite{TwistedSquares}.

Returning to the spectra in Fig.~\ref{fig:TITISpectrumPlots}, it is also notable that the spectra are consistently gapless at all energies.
This is not required by the strong $\mathbb{Z}_2$ topological invariant: time-reversal symmetry allows gaps to open at the interface between two 3D TIs. Instead, the gapless surface states are protected as a weak TI: the cones at $A$ and $B$ are separated in the moir\'e BZ (as illustrated in Fig.~\ref{fig:TITIPlotPathArea}), and consequently are protected by the approximate translation symmetry of the moir\'e lattice \cite{Lopes_dos_Santos_2012}, which also protects the cones at $C$ and $C'$. 
By ``moir\'e BZ" we refer to the BZ defined below Eq.~(\ref{eq:Topdef}), not the BZ of any particular commensurate lattice;
hence, the translation symmetries are not sensitive to any particular commensurate structure. 
However, the protection of Dirac cones by moir\'e translation symmetry implies that disorder is sufficiently small on the moir\'e length scale.
(The simplicity of this argument is in contrast to the analogous circumstances in the second chiral limit of TBLG \cite{BernevigII} and in twisted square lattices \cite{TwistedSquares} in which a more nuanced argument is required.)

Dirac cones also appear at $D$ and $D'$ in the top row in Fig.~\ref{fig:TITISpectrumPlots}. These crossings are not topologically protected:
there exist non-spin-conserving interlayer couplings that are symmetry-allowed and would open a gap at $D$.
In particular, in the lower row in Fig.~\ref{fig:TITISpectrumPlots}, the Dirac cones have the opposite spin-momentum locking; hence, the spins are aligned where the Dirac cones overlap at $D$ and a large gap opens.

Instead, these crossings' gaplessness is a consequence of the specific parameters used in our model. In particular: since $D$ lies on the $A-B$ line (see Fig.~\ref{fig:TITIPlotPathArea}), when the Dirac cones have the same spin-momentum locking (i.e., both are of the form $\bar k \cdot \sigma$), they have opposite spin at $D$.
Since the spectra in Fig.~\ref{fig:TITISpectrumPlots} only include spin-conserving hopping, there is no term to open a gap at $D$.
When the two Dirac cones have different velocities or chemical potentials, the crossing still remains gapless, but moves along the $A-B$ line.

\begin{figure*}
    \centering
    \begin{subfigure}[b]{0.29\textwidth}
        \includegraphics[width=\textwidth]{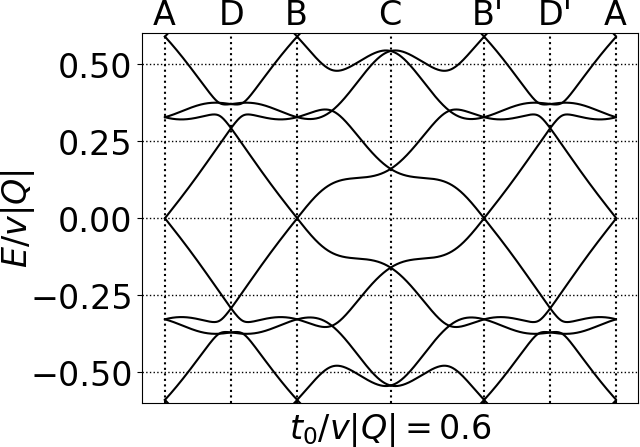}
    \end{subfigure}
    \begin{subfigure}[b]{0.226631\textwidth}
        \includegraphics[width=\textwidth]{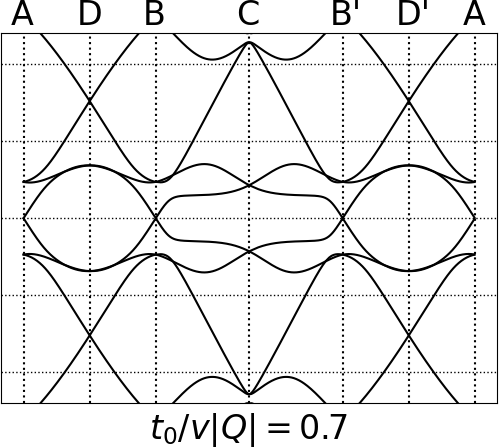}
    \end{subfigure}
    \begin{subfigure}[b]{0.226631\textwidth}
        \includegraphics[width=\textwidth]{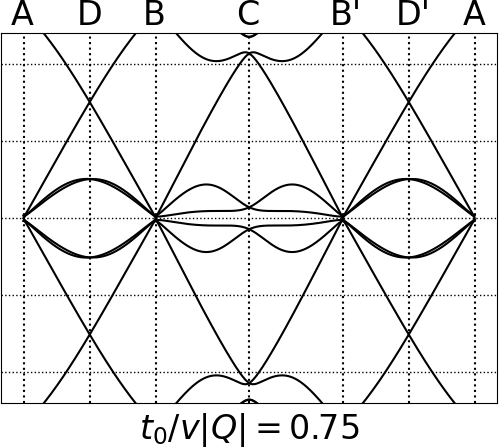}
    \end{subfigure}
    \begin{subfigure}[b]{0.226631\textwidth}
        \includegraphics[width=\textwidth]{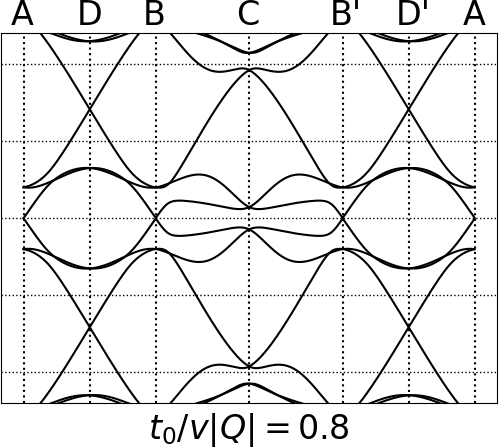}
    \end{subfigure}
    \caption{Band structures for the interface between two identical twisted 3D TIs with spin-preserving coupling (BZ path defined in Fig.~\ref{fig:TITIPlotPathArea}). At $t_0/v|Q|=0.75$, the bands nearest the Fermi level become nearly degenerate along AB.}
    \label{fig:TITIMoreCone}
\end{figure*}

There is one additional interesting feature to examine. In the $\chi=+1$ case, there are two nearly-degenerate bands at $t_0/v|Q|=0.75$. Fig.~\ref{fig:TITIMoreCone} shows a finer range of hopping parameters, which reveals the lowest conduction bands descend to produce additional zero-energy states at a critical value of $t_0/v|Q|=0.75$, then separate again as the coupling increases. A similar critical point where the lowest conduction band dips down and touches the flat band also appears in TBLG, both in the usual Bistritzer-MacDonald model (Fig. 3A of \cite{Bistritzer12233}) and in the ``second chiral model'' with vanishing AB coupling (Fig. 2c of \cite{bernevig2020tbg}), although the nearly-degenerate bands that result at this crossing are specific to our model.

The touching point appears to be distinct in the two cases: in the TI-TI case, the touching point is at the Dirac cones, whereas in TBLG the touching point is at $\Gamma$ (away from the Dirac cones). However, this can be understood by interpreting the touching point as occurring $-Q$ away from a Dirac point, which in TBLG is the $\Gamma$ point and in our TI-TI model is at a Dirac cone.

\section{TI on Graphene\label{Sec:TIGR}}

We have shown that stacking two 3D TIs with a small twist angle can significantly renormalize the Dirac cones at the interface, with nearly flat bands appearing at a specific magic twist angle under certain conditions. In the simplest case, the surface Dirac cone of each TI must not be at the center of the surface BZ.
However, none of the 3D TIs in the Bi$_2$Se$_3$ family meet this condition: their surface Dirac cones are always at $\Gamma$ \cite{zhang2009topological,xia2009observation}.

This motivates us to consider a second example of a twisted heterostructure with multiple Dirac cones: the interface between a topological insulator and graphene. Such interfaces have been considered in previous work
\cite{Rossi_2019,jin2012multiple,jin2013proximity,RossiTriolaOld,cao2016heavy,debeule2017transmission,rodriguez2017giant,song2018spin, GRBISE5,dang2010epitaxial,song2010topological,lee2015proximity,steinberg2015tunneling,zhang2016gate,bian2016experimental,vaklinova2016current,rajput2016indirect,zhang2017electronic,GRBISE1,khokhriakov2018tailoring,GRBISE3, PhysRevB.100.165141}, but the effect of a small relative twist angle was not considered.

\begin{figure}
    \centering
    \includegraphics[height=1.75in]{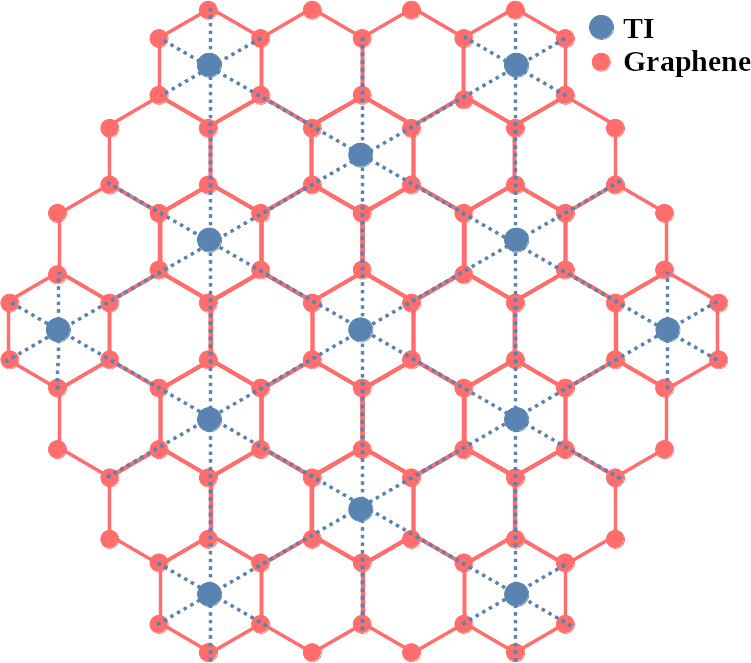}
    \caption{Graphene on Bi$_2$Se$_3$. Blue circles indicate Se atoms on the Bi$_2$Se$_3$ surface; red circles indicate carbon atoms in graphene. The superlattice matching is accurate to $\sim 3$\% \cite{RossiTriolaOld}.}
    \label{fig:TIGRUntwisted}
\end{figure}

\begin{figure}
    \centering
    \includegraphics[height=1.75in]{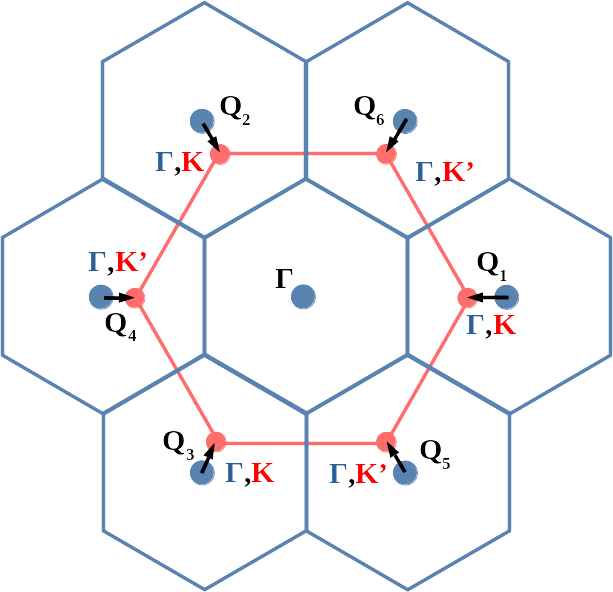}
    \includegraphics[height=1.75in]{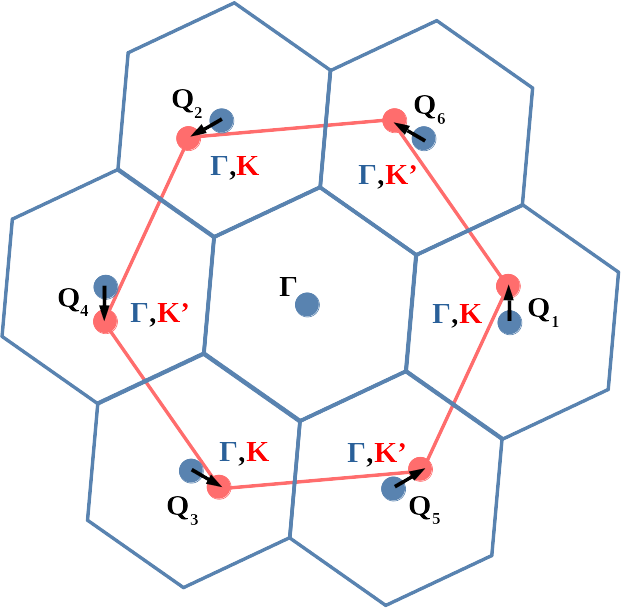}
    \caption{The BZ of graphene (red) arranged on top of the 3D TI BZ (blue), shown both with the lattice mismatch (top) and then with a small relative twist angle while neglecting mismatch (bottom). The momentum differences $Q_i$ are indicated in each case.}
    \label{fig:TIGRQs}
\end{figure}

We consider a 3D TI with a single Dirac cone at $\Gamma$, such as in Bi$_2$Se$_3$. The lattice constants of the TI and graphene differ by a factor of $\sqrt{3}$ up to an approximate $3\%$ lattice mismatch\cite{RossiTriolaOld}; Fig.~\ref{fig:TIGRUntwisted} shows the lattice matching ignoring the lattice mismatch.
The aligned superstructure then folds the $K$ and $K'$ points in graphene onto $\Gamma$ in the TI BZ, with the resulting twisted structure shown in Fig.~\ref{fig:TIGRQs}. We assume the TI has a sixfold rotational symmetry, as well as time-reversal symmetry.
(The sixfold symmetry is not exact: while each layer of Bi$_2$Se$_3$ has a sixfold rotation axis, the layers are stacked such that the axes do not coincide \cite{zhang2009topological}.
Thus, surface states mostly localized on the top layer will exhibit an approximate sixfold rotational symmetry.)
The TI Hamiltonian is given by $v_{TI}k\cdot\sigma$, as in Sec.~\ref{Sec:TITISame}. Graphene's Hamiltonian consists of four Dirac cones including spin and valley; the matrix form is given in Appendix~\ref{Apx:GRHam}.

In the subsequent sections, we will derive a perturbative analytic calculation of the renormalization of the TI Dirac cone that arises from its coupling to graphene.
The analogous calculation for graphene is more complicated due to its spin degeneracy and is derived in Appendix~\ref{Apx:GRCorrections}; when the interlayer hopping is spin-preserving, almost all corrections vanish, with the nonvanishing coefficients only rescaling the spectrum.

To go beyond our perturbative results, we compute the surface band structure, specifying to the most physically interesting case where spin is conserved. 
Our results both validate the first-order perturbative calculation for the TI Dirac cone renormalization and provide insight into the higher-order corrections to the graphene Dirac cone.

\subsection{Corrections to TI Dirac cone\label{Sec:TIGRPert}}
We begin by computing the corrections to the TI Dirac cone to first order in perturbation theory. In so doing, we make use of the corrections to a single Dirac cone derived in Eqs.~\eqref{eq:1DCSelfEn}-\eqref{eq:1DCCoeffs}.
We decompose the interlayer coupling terms $T_Q$, defined in Eq.~(\ref{eq:Topdef}), into $T_{Q,k_0,s}$, where $T_{Q,k_0,s}$ couples each graphene Dirac cone to the TI Dirac cone, with the $k_0$ running over valley and $s$ over spin. At any particular $Q$, only one valley will couple to the TI; without loss of generality, we consider $Q$ corresponding to the $K$ point, so that $T_{Q,K'}=0$. 
The interlayer coupling terms for other values of $Q$ can then be determined by time-reversal symmetry.

When the lattices are aligned, there is a moir\'e pattern that arises from the lattice mismatch. In this case, we can choose an orientation such that $Q=Q_x\hat{x}$. On the other hand, if we ignore the lattice mismatch and consider arranging the graphene layer on the TI surface with a small twist angle, we can take $Q = Q_y\hat{y}$. In the most general case of a lattice mismatch and a small twist angle, $Q$ could point in an arbitrary direction; our formalism covers this case, but we here focus on the two limits.

Regardless of the combination of twist and mismatch, for a spin-preserving coupling, the corresponding $T_{Q,K,s}$ will take one of two forms (depending on $s$), with the rows corresponding to spins in the TI layer and the columns corresponding to sublattice in the graphene layer:
\begin{subequations}\label{eq:TIGRTQ}
\begin{align}
    T_{Q,K,\uparrow}=\begin{bmatrix}t&t\\0&0\end{bmatrix}\\
    \nonumber\text{or}\qquad\quad\\
    T_{Q,K,\downarrow}=\begin{bmatrix}0&0\\t&t\end{bmatrix}
\end{align}
\end{subequations}
For either of these $T_{Q,K}$, $t_{||}=\pm it_{\perp}$, where $t_{||}$ and $t_\perp$ are defined in Eq.~\eqref{eq:TQdef}. Therefore, the direct corrections $C$ and $D$ to the Fermi velocity (computed according to Eq.~\eqref{eq:1DCCoeffs} and inputted into the Fermi velocity in Eq.~\eqref{eq:1DCFermiVel}) both vanish. Consequently, magic angles from spin-preserving hoppings do not exist for any small twist or strain.

By the same logic, purely $z$-spin-flipping hoppings will also not produce magic angles in this case. Unlike the interface between two 3D TIs (Sec.~\ref{Sec:TITI}), we need both $z$-spin-preserving and $z$-spin-flipping hoppings to produce magic angles. 

Another change compared to the interface between two 3D TIs is that here, due to the small lattice mismatch, the coupling to graphene shifts the TI's surface Dirac cone up or down in energy, while the Dirac cones at the interface between two 3D TIs experience no such shift when the interlayer coupling is spin-preserving. (The lack of an energy shift in that case is a consequence of the parameter $B$ in Eq.~(\ref{eq:TITICoeffB}) vanishing for spin-preserving coupling; here, it vanishes for a small twist, but not for lattice mismatch.) In this case, the energy shift follows the formula:
\begin{equation}\label{eq:GrEnShift}
    \Delta E=-\frac{12t^2v_G|Q|}{v_G^2Q^2+12t^2}
\end{equation}
The sign of the energy shift depends on the direction of mismatch, i.e., whether the graphene unit cell or the TI unit cell is larger compared to the perfect supercell matching illustrated in Fig.~\ref{fig:TIGRUntwisted}.

\subsection{Global band structure\label{Sec:TIGRNum}}
\begin{figure}
    \centering
    \includegraphics[width=0.3\textwidth]{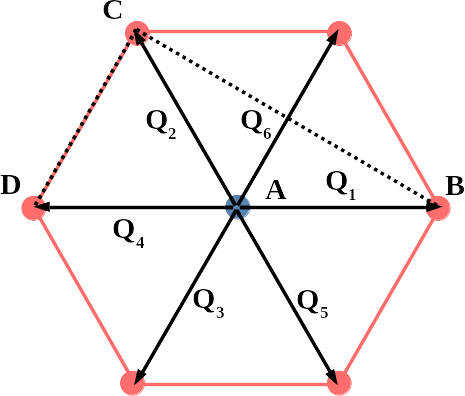}
    \caption{Path along which spectra in Fig.~\ref{fig:TIGRSPHSpec} are plotted in the case of lattice mismatch. For a twist, the figure is rotated 90 degrees counterclockwise to account for the different values for $Q$.}
    \label{fig:TIGRMBZ}
\end{figure}

We now plot the band structure to see how our perturbation theory holds up. We plot all spectra along the path shown in Fig.~\ref{fig:TIGRMBZ}, which is the same path used in \cite{Bistritzer12233}.

\begin{figure*}
    \centering
    \begin{subfigure}[b]{0.37\textwidth}
        \includegraphics[width=\textwidth]{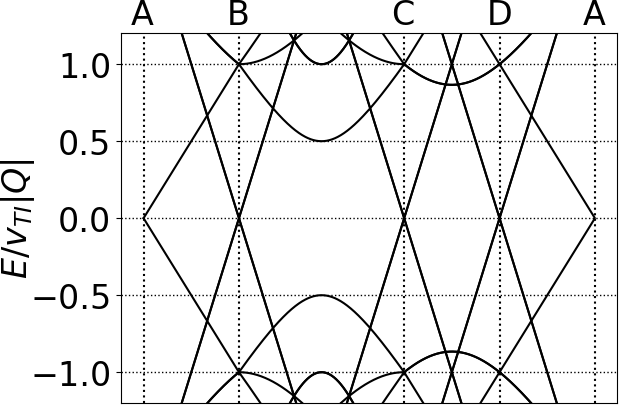}
    \end{subfigure}
    \begin{subfigure}[b]{0.29827140548\textwidth}
        \includegraphics[width=\textwidth]{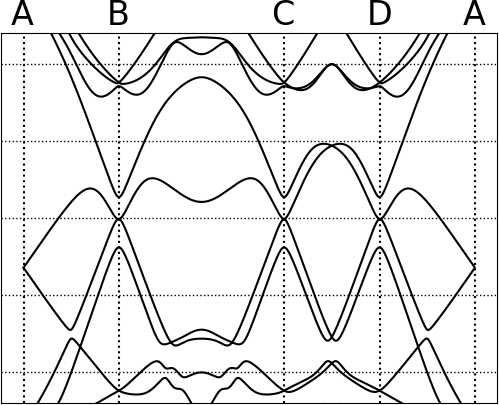}
    \end{subfigure}
    \begin{subfigure}[b]{0.29827140548\textwidth}
        \includegraphics[width=\textwidth]{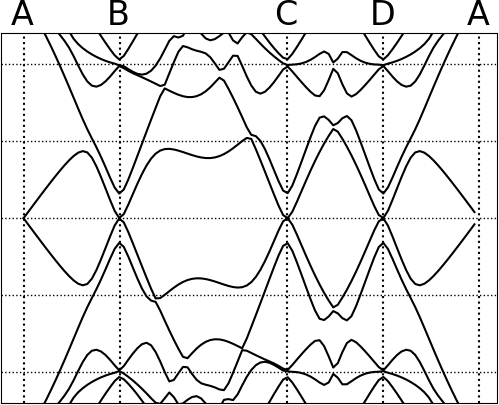}
    \end{subfigure}
    \caption{Spectrum of TI-graphene heterostructure with zero interlayer hopping (left) and spin-preserving sublattice-independent interlayer hopping $t/v_{TI}|Q|=0.25$ for a uniform lattice mismatch without twist (center) and twist without mismatch (right). We assume $v_G=2v_{TI}$ \cite{Rossi_2019}.}
    \label{fig:TIGRSPHSpec}
\end{figure*}

\begin{figure}
    \centering
    \begin{subfigure}[b]{0.35\textwidth}
        \includegraphics[width=\textwidth]{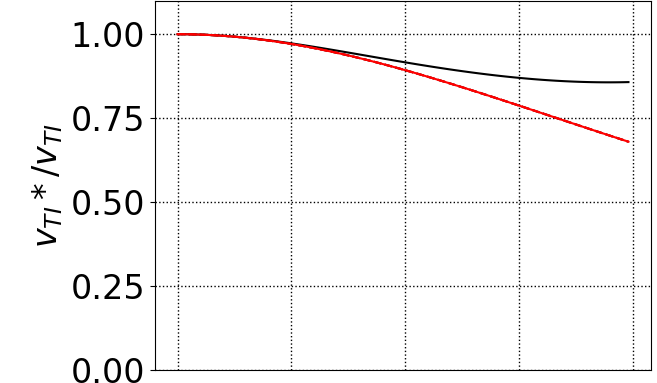}
    \end{subfigure}
    \begin{subfigure}[b]{0.35\textwidth}
        \includegraphics[width=\textwidth]{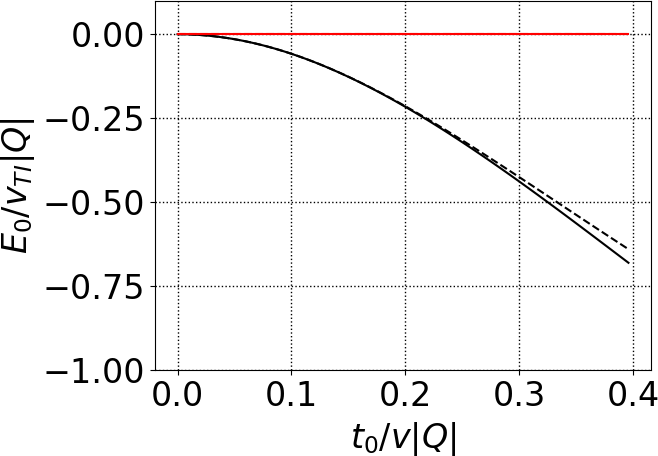}
    \end{subfigure}
    \caption{Renormalized TI Dirac cone velocity $v_{TI}^*/v_{TI}$ (top) and energy shift $E_0/v_{TI}|Q|$ (bottom) as a function of the spin-preserving and sublattice-independent hopping parameter $t/v_{TI}|Q|$ extracted from band structure calculations (solid lines) and computed perturbatively (dashed line, drawn only in the one case where it does not overlap with a solid line), for the case of a small twist angle (red) and small lattice mismatch (black).
    In the twist case, the energy and Fermi velocity perfectly match the perturbative calculations, so those lines are not drawn. In the lattice mismatch case, the Fermi velocity has the same perturbative theory as the twist case (i.e., following the solid red line).}
    \label{fig:TIGRParamPlots}
\end{figure}

We first examine spin-preserving, sublattice-independent coupling (which does not produce magic angles). We present one representative spectrum each for the cases of twist and lattice mismatch in Fig.~\ref{fig:TIGRSPHSpec}, computed with $t/v_{TI}|Q|=0.25$; other values of $t$ look qualitatively similar.
The band structure near the TI Dirac cone at A agrees with our perturbative calculations: in the case of a lattice mismatch, it shows an energy shift whose magnitude and direction matches Eq.~\eqref{eq:GrEnShift}; there is no energy shift in the twist case; and in either case there is only modest velocity renormalization. Fig.~\ref{fig:TIGRParamPlots} illustrates the behavior of Fermi velocity and energy shift, as extracted from band structure, as a function of twist angle.
It shows that the Fermi velocity does not experience nearly the renormalization achieved at the interface between two 3D TIs (Fig.~\ref{fig:TITIvt}). It also shows that at small twist angles and lattice mismatches (large couplings), twisting suppresses Fermi velocity more than mismatch for the same moir\'e lattice size.

The graphene cones at B, C, and D resemble the quadratic bands of untwisted Bernal-stacked bilayer graphene. This is because the TI couples graphene's spin-up and spin-down cones at $K$ (since propagation in the TI layer flips the $z$-component of spin). These cones are not twisted relative to each other, so they couple in much the same way as the Dirac cones in the different layers of bilayer graphene, but with an effective interlayer hopping parameter proportional to $\frac{t^2}{v_{TI}|Q|}$ (instead of $t_{BLG}$).

This correction to the graphene Dirac cones is fourth order in $t$, which agrees with our analytic calculation showing that the only nonvanishing correction to the graphene cone to quadratic order in $t$ is an overall rescaling of the spectrum (details in Appendix~\ref{Apx:GRCorrections}).

Like the TI-TI interface, the spectra in Fig.~\ref{fig:TIGRSPHSpec} remain gapless at all energies.
This is topologically required because the Dirac cone on the 3D TI surface is protected by time reversal symmetry; unlike the TI-TI case discussed in Sec.~\ref{Sec:TITINum}, here the topological protection is strong and not weak (i.e., not dependent on the approximate translation symmetries of the moir\'e lattice) since there is only one TI surface state.

In summary, our calculations predict that arranging graphene on the surface of a 3D TI with a small twist angle is not a promising platform for creating flatter bands, both because no gaps are topologically allowed to open in the spectrum and because under the most physically intuitive conditions, the Dirac cone is only marginally renormalized.

\section{Conclusions}

We have studied moir\'e heterostructures on the surface of a 3D TI. We derived analytic expressions for the leading order corrections to the velocity of the surface Dirac cone induced by
coupling to another lattice-matched Dirac material arranged with a small twist angle. We applied our results to two types of heterostructures: an interface between two TIs and graphene on the surface of a TI. We derived conditions for achieving a ``magic'' angle at which the Fermi velocity vanishes.

One of our main results is that at the interface between two identical 3D TIs arranged with a small twist angle, a spin-flipping interlayer hopping term is a necessary ingredient to achieve a magic angle. The same is true for graphene on a 3D TI. While such a term is symmetry-allowed (enforcing time-reversal and rotational symmetry), it is not clear what physical mechanism would give rise to it. In future work, it would be interesting to perform first-principles calculations for different material combinations (such as graphene on Bi$_2$Se$_3$ or Bi$_2$Te$_3$, or the interface between two 3D TIs) to determine under what conditions the spin-flipping interlayer hopping terms arise.

On the other hand, we found that magic angles are achieved without spin-flipping interlayer hopping at twisted interfaces between two lattice-matched TIs whose Dirac cones have opposite winding number (see Fig.~\ref{fig:SpinTextures}).
At the magic angle, such an interface realizes locally flat bands. Unlike in twisted bilayer graphene, the flat bands are not gapped from the conduction/valence bands due to topological constraints. Nonetheless, such locally flat bands will enhance the density of states at the Dirac point, which is favorable for realizing instabilities to superconducting or quantum anomalous Hall states \cite{fu2008superconducting,santos2010,baum2012magnetic,baum2012density,marchand2012lattice,schmidt2012strong,sitte2013interaction,mendler2015magnetic,wang2021moire}.

This analysis therefore provides two routes for inducing magic angles on the surface of a 3D TI. There are also several potential routes to creating flat bands in such systems that go beyond the setups considered in this paper.

One route would be to consider an interface between two materials with Dirac cones at $\Gamma$. We did not consider this case here because such interfaces would generically gap, and therefore exhibit qualitatively different behavior than other twisted interfaces, although this case has been studied recently for thin slabs \cite{TateishiHirayama}. Whether the resulting bands contain interesting features (e.g., van Hove singularities \cite{wang2021moire}) remains an open question.

Another route would be to reduce the rotational symmetry. We limited ourselves to Dirac cones that would remain isotropic to linear order in the twisted heterostructure, i.e., those that remain a center of three- or four-fold rotation. If one considers Dirac cones with less rotational symmetry, then even where a vanishing Fermi velocity is impossible, it may be possible to achieve a significantly enhanced density of states via anisotropic band flattening \cite{kariyado2019flat, kennes2020one}. In such a setup, one may have ``partial magic angles" where Dirac cone velocity vanishes in one direction. The interacting instabilities of a 3D TI with such a Dirac cone provide a potential scenario for further study.

\section{Acknowledgments}
We are grateful for useful conversations with Shiang Fang, Jed Pixley, Jie Wang, and Justin Wilson and for a careful reading of the manuscript by Sayed Ali Akbar Ghorashi.
This work was partially supported by the Air Force Office of Scientific Research under Grant No. FA9550-20-1-0260.
J.C. acknowledges the support of the Flatiron Institute, a division of the Simons Foundation.

\bibliography{moire}

\begin{thebibliography}{54}%
\makeatletter
\providecommand \@ifxundefined [1]{%
 \@ifx{#1\undefined}
}%
\providecommand \@ifnum [1]{%
 \ifnum #1\expandafter \@firstoftwo
 \else \expandafter \@secondoftwo
 \fi
}%
\providecommand \@ifx [1]{%
 \ifx #1\expandafter \@firstoftwo
 \else \expandafter \@secondoftwo
 \fi
}%
\providecommand \natexlab [1]{#1}%
\providecommand \enquote  [1]{``#1''}%
\providecommand \bibnamefont  [1]{#1}%
\providecommand \bibfnamefont [1]{#1}%
\providecommand \citenamefont [1]{#1}%
\providecommand \href@noop [0]{\@secondoftwo}%
\providecommand \href [0]{\begingroup \@sanitize@url \@href}%
\providecommand \@href[1]{\@@startlink{#1}\@@href}%
\providecommand \@@href[1]{\endgroup#1\@@endlink}%
\providecommand \@sanitize@url [0]{\catcode `\\12\catcode `\$12\catcode
  `\&12\catcode `\#12\catcode `\^12\catcode `\_12\catcode `\%12\relax}%
\providecommand \@@startlink[1]{}%
\providecommand \@@endlink[0]{}%
\providecommand \url  [0]{\begingroup\@sanitize@url \@url }%
\providecommand \@url [1]{\endgroup\@href {#1}{\urlprefix }}%
\providecommand \urlprefix  [0]{URL }%
\providecommand \Eprint [0]{\href }%
\providecommand \doibase [0]{https://doi.org/}%
\providecommand \selectlanguage [0]{\@gobble}%
\providecommand \bibinfo  [0]{\@secondoftwo}%
\providecommand \bibfield  [0]{\@secondoftwo}%
\providecommand \translation [1]{[#1]}%
\providecommand \BibitemOpen [0]{}%
\providecommand \bibitemStop [0]{}%
\providecommand \bibitemNoStop [0]{.\EOS\space}%
\providecommand \EOS [0]{\spacefactor3000\relax}%
\providecommand \BibitemShut  [1]{\csname bibitem#1\endcsname}%
\let\auto@bib@innerbib\@empty
\bibitem [{\citenamefont {Cao}\ \emph {et~al.}(2018{\natexlab{a}})\citenamefont
  {Cao}, \citenamefont {Fatemi}, \citenamefont {Demir}, \citenamefont {Fang},
  \citenamefont {Tomarken}, \citenamefont {Luo}, \citenamefont
  {Sanchez-Yamagishi}, \citenamefont {Watanabe}, \citenamefont {Taniguchi},
  \citenamefont {Kaxiras} \emph {et~al.}}]{cao2018correlated}%
  \BibitemOpen
  \bibfield  {author} {\bibinfo {author} {\bibfnamefont {Y.}~\bibnamefont
  {Cao}}, \bibinfo {author} {\bibfnamefont {V.}~\bibnamefont {Fatemi}},
  \bibinfo {author} {\bibfnamefont {A.}~\bibnamefont {Demir}}, \bibinfo
  {author} {\bibfnamefont {S.}~\bibnamefont {Fang}}, \bibinfo {author}
  {\bibfnamefont {S.~L.}\ \bibnamefont {Tomarken}}, \bibinfo {author}
  {\bibfnamefont {J.~Y.}\ \bibnamefont {Luo}}, \bibinfo {author} {\bibfnamefont
  {J.~D.}\ \bibnamefont {Sanchez-Yamagishi}}, \bibinfo {author} {\bibfnamefont
  {K.}~\bibnamefont {Watanabe}}, \bibinfo {author} {\bibfnamefont
  {T.}~\bibnamefont {Taniguchi}}, \bibinfo {author} {\bibfnamefont
  {E.}~\bibnamefont {Kaxiras}}, \emph {et~al.},\ }\bibfield  {title} {\bibinfo
  {title} {Correlated insulator behaviour at half-filling in magic-angle
  graphene superlattices},\ }\href@noop {} {\bibfield  {journal} {\bibinfo
  {journal} {Nature}\ }\textbf {\bibinfo {volume} {556}},\ \bibinfo {pages}
  {80} (\bibinfo {year} {2018}{\natexlab{a}})}\BibitemShut {NoStop}%
\bibitem [{\citenamefont {Cao}\ \emph {et~al.}(2018{\natexlab{b}})\citenamefont
  {Cao}, \citenamefont {Fatemi}, \citenamefont {Fang}, \citenamefont
  {Watanabe}, \citenamefont {Taniguchi}, \citenamefont {Kaxiras},\ and\
  \citenamefont {Jarillo-Herrero}}]{cao2018unconventional}%
  \BibitemOpen
  \bibfield  {author} {\bibinfo {author} {\bibfnamefont {Y.}~\bibnamefont
  {Cao}}, \bibinfo {author} {\bibfnamefont {V.}~\bibnamefont {Fatemi}},
  \bibinfo {author} {\bibfnamefont {S.}~\bibnamefont {Fang}}, \bibinfo {author}
  {\bibfnamefont {K.}~\bibnamefont {Watanabe}}, \bibinfo {author}
  {\bibfnamefont {T.}~\bibnamefont {Taniguchi}}, \bibinfo {author}
  {\bibfnamefont {E.}~\bibnamefont {Kaxiras}},\ and\ \bibinfo {author}
  {\bibfnamefont {P.}~\bibnamefont {Jarillo-Herrero}},\ }\bibfield  {title}
  {\bibinfo {title} {Unconventional superconductivity in magic-angle graphene
  superlattices},\ }\href@noop {} {\bibfield  {journal} {\bibinfo  {journal}
  {Nature}\ }\textbf {\bibinfo {volume} {556}},\ \bibinfo {pages} {43}
  (\bibinfo {year} {2018}{\natexlab{b}})}\BibitemShut {NoStop}%
\bibitem [{\citenamefont {Sharpe}\ \emph {et~al.}(2019)\citenamefont {Sharpe},
  \citenamefont {Fox}, \citenamefont {Barnard}, \citenamefont {Finney},
  \citenamefont {Watanabe}, \citenamefont {Taniguchi}, \citenamefont
  {Kastner},\ and\ \citenamefont {Goldhaber-Gordon}}]{sharpe2019emergent}%
  \BibitemOpen
  \bibfield  {author} {\bibinfo {author} {\bibfnamefont {A.~L.}\ \bibnamefont
  {Sharpe}}, \bibinfo {author} {\bibfnamefont {E.~J.}\ \bibnamefont {Fox}},
  \bibinfo {author} {\bibfnamefont {A.~W.}\ \bibnamefont {Barnard}}, \bibinfo
  {author} {\bibfnamefont {J.}~\bibnamefont {Finney}}, \bibinfo {author}
  {\bibfnamefont {K.}~\bibnamefont {Watanabe}}, \bibinfo {author}
  {\bibfnamefont {T.}~\bibnamefont {Taniguchi}}, \bibinfo {author}
  {\bibfnamefont {M.}~\bibnamefont {Kastner}},\ and\ \bibinfo {author}
  {\bibfnamefont {D.}~\bibnamefont {Goldhaber-Gordon}},\ }\bibfield  {title}
  {\bibinfo {title} {Emergent ferromagnetism near three-quarters filling in
  twisted bilayer graphene},\ }\href@noop {} {\bibfield  {journal} {\bibinfo
  {journal} {Science}\ }\textbf {\bibinfo {volume} {365}},\ \bibinfo {pages}
  {605} (\bibinfo {year} {2019})}\BibitemShut {NoStop}%
\bibitem [{\citenamefont {Serlin}\ \emph {et~al.}(2020)\citenamefont {Serlin},
  \citenamefont {Tschirhart}, \citenamefont {Polshyn}, \citenamefont {Zhang},
  \citenamefont {Zhu}, \citenamefont {Watanabe}, \citenamefont {Taniguchi},
  \citenamefont {Balents},\ and\ \citenamefont {Young}}]{serlin2020intrinsic}%
  \BibitemOpen
  \bibfield  {author} {\bibinfo {author} {\bibfnamefont {M.}~\bibnamefont
  {Serlin}}, \bibinfo {author} {\bibfnamefont {C.}~\bibnamefont {Tschirhart}},
  \bibinfo {author} {\bibfnamefont {H.}~\bibnamefont {Polshyn}}, \bibinfo
  {author} {\bibfnamefont {Y.}~\bibnamefont {Zhang}}, \bibinfo {author}
  {\bibfnamefont {J.}~\bibnamefont {Zhu}}, \bibinfo {author} {\bibfnamefont
  {K.}~\bibnamefont {Watanabe}}, \bibinfo {author} {\bibfnamefont
  {T.}~\bibnamefont {Taniguchi}}, \bibinfo {author} {\bibfnamefont
  {L.}~\bibnamefont {Balents}},\ and\ \bibinfo {author} {\bibfnamefont
  {A.}~\bibnamefont {Young}},\ }\bibfield  {title} {\bibinfo {title} {Intrinsic
  quantized anomalous hall effect in a moir{\'e} heterostructure},\ }\href@noop
  {} {\bibfield  {journal} {\bibinfo  {journal} {Science}\ }\textbf {\bibinfo
  {volume} {367}},\ \bibinfo {pages} {900} (\bibinfo {year}
  {2020})}\BibitemShut {NoStop}%
\bibitem [{\citenamefont {Bistritzer}\ and\ \citenamefont
  {MacDonald}(2011)}]{Bistritzer12233}%
  \BibitemOpen
  \bibfield  {author} {\bibinfo {author} {\bibfnamefont {R.}~\bibnamefont
  {Bistritzer}}\ and\ \bibinfo {author} {\bibfnamefont {A.~H.}\ \bibnamefont
  {MacDonald}},\ }\bibfield  {title} {\bibinfo {title} {Moir{\'e} bands in
  twisted double-layer graphene},\ }\href
  {https://doi.org/10.1073/pnas.1108174108} {\bibfield  {journal} {\bibinfo
  {journal} {Proceedings of the National Academy of Sciences}\ }\textbf
  {\bibinfo {volume} {108}},\ \bibinfo {pages} {12233} (\bibinfo {year}
  {2011})},\ \Eprint
  {https://arxiv.org/abs/https://www.pnas.org/content/108/30/12233.full.pdf}
  {https://www.pnas.org/content/108/30/12233.full.pdf} \BibitemShut {NoStop}%
\bibitem [{\citenamefont {Lopes~dos Santos}\ \emph
  {et~al.}(2012{\natexlab{a}})\citenamefont {Lopes~dos Santos}, \citenamefont
  {Peres},\ and\ \citenamefont {Castro~Neto}}]{lopes2012continuum}%
  \BibitemOpen
  \bibfield  {author} {\bibinfo {author} {\bibfnamefont {J.~M.~B.}\
  \bibnamefont {Lopes~dos Santos}}, \bibinfo {author} {\bibfnamefont
  {N.~M.~R.}\ \bibnamefont {Peres}},\ and\ \bibinfo {author} {\bibfnamefont
  {A.~H.}\ \bibnamefont {Castro~Neto}},\ }\bibfield  {title} {\bibinfo {title}
  {Continuum model of the twisted graphene bilayer},\ }\href
  {https://doi.org/10.1103/PhysRevB.86.155449} {\bibfield  {journal} {\bibinfo
  {journal} {Phys. Rev. B}\ }\textbf {\bibinfo {volume} {86}},\ \bibinfo
  {pages} {155449} (\bibinfo {year} {2012}{\natexlab{a}})}\BibitemShut
  {NoStop}%
\bibitem [{\citenamefont {Su\'arez~Morell}\ \emph {et~al.}(2010)\citenamefont
  {Su\'arez~Morell}, \citenamefont {Correa}, \citenamefont {Vargas},
  \citenamefont {Pacheco},\ and\ \citenamefont {Barticevic}}]{suarez2010flat}%
  \BibitemOpen
  \bibfield  {author} {\bibinfo {author} {\bibfnamefont {E.}~\bibnamefont
  {Su\'arez~Morell}}, \bibinfo {author} {\bibfnamefont {J.~D.}\ \bibnamefont
  {Correa}}, \bibinfo {author} {\bibfnamefont {P.}~\bibnamefont {Vargas}},
  \bibinfo {author} {\bibfnamefont {M.}~\bibnamefont {Pacheco}},\ and\ \bibinfo
  {author} {\bibfnamefont {Z.}~\bibnamefont {Barticevic}},\ }\bibfield  {title}
  {\bibinfo {title} {Flat bands in slightly twisted bilayer graphene:
  Tight-binding calculations},\ }\href
  {https://doi.org/10.1103/PhysRevB.82.121407} {\bibfield  {journal} {\bibinfo
  {journal} {Phys. Rev. B}\ }\textbf {\bibinfo {volume} {82}},\ \bibinfo
  {pages} {121407(R)} (\bibinfo {year} {2010})}\BibitemShut {NoStop}%
\bibitem [{\citenamefont {Hasan}\ and\ \citenamefont
  {Kane}(2010)}]{hasan2010colloquium}%
  \BibitemOpen
  \bibfield  {author} {\bibinfo {author} {\bibfnamefont {M.~Z.}\ \bibnamefont
  {Hasan}}\ and\ \bibinfo {author} {\bibfnamefont {C.~L.}\ \bibnamefont
  {Kane}},\ }\bibfield  {title} {\bibinfo {title} {Colloquium: topological
  insulators},\ }\href@noop {} {\bibfield  {journal} {\bibinfo  {journal}
  {Reviews of modern physics}\ }\textbf {\bibinfo {volume} {82}},\ \bibinfo
  {pages} {3045} (\bibinfo {year} {2010})}\BibitemShut {NoStop}%
\bibitem [{\citenamefont {Cano}\ \emph {et~al.}(2021)\citenamefont {Cano},
  \citenamefont {Fang}, \citenamefont {Pixley},\ and\ \citenamefont
  {Wilson}}]{cano2021moire}%
  \BibitemOpen
  \bibfield  {author} {\bibinfo {author} {\bibfnamefont {J.}~\bibnamefont
  {Cano}}, \bibinfo {author} {\bibfnamefont {S.}~\bibnamefont {Fang}}, \bibinfo
  {author} {\bibfnamefont {J.~H.}\ \bibnamefont {Pixley}},\ and\ \bibinfo
  {author} {\bibfnamefont {J.~H.}\ \bibnamefont {Wilson}},\ }\bibfield  {title}
  {\bibinfo {title} {Moir\'e superlattice on the surface of a topological
  insulator},\ }\href {https://doi.org/10.1103/PhysRevB.103.155157} {\bibfield
  {journal} {\bibinfo  {journal} {Phys. Rev. B}\ }\textbf {\bibinfo {volume}
  {103}},\ \bibinfo {pages} {155157} (\bibinfo {year} {2021})}\BibitemShut
  {NoStop}%
\bibitem [{\citenamefont {Wang}\ \emph {et~al.}(2021)\citenamefont {Wang},
  \citenamefont {Yuan},\ and\ \citenamefont {Fu}}]{wang2021moire}%
  \BibitemOpen
  \bibfield  {author} {\bibinfo {author} {\bibfnamefont {T.}~\bibnamefont
  {Wang}}, \bibinfo {author} {\bibfnamefont {N.~F.~Q.}\ \bibnamefont {Yuan}},\
  and\ \bibinfo {author} {\bibfnamefont {L.}~\bibnamefont {Fu}},\ }\bibfield
  {title} {\bibinfo {title} {Moir\'e surface states and enhanced
  superconductivity in topological insulators},\ }\href
  {https://doi.org/10.1103/PhysRevX.11.021024} {\bibfield  {journal} {\bibinfo
  {journal} {Phys. Rev. X}\ }\textbf {\bibinfo {volume} {11}},\ \bibinfo
  {pages} {021024} (\bibinfo {year} {2021})}\BibitemShut {NoStop}%
\bibitem [{\citenamefont {Guerci}\ \emph {et~al.}(2022)\citenamefont {Guerci},
  \citenamefont {Wang}, \citenamefont {Pixley},\ and\ \citenamefont
  {Cano}}]{guerci2022designer}%
  \BibitemOpen
  \bibfield  {author} {\bibinfo {author} {\bibfnamefont {D.}~\bibnamefont
  {Guerci}}, \bibinfo {author} {\bibfnamefont {J.}~\bibnamefont {Wang}},
  \bibinfo {author} {\bibfnamefont {J.}~\bibnamefont {Pixley}},\ and\ \bibinfo
  {author} {\bibfnamefont {J.}~\bibnamefont {Cano}},\ }\bibfield  {title}
  {\bibinfo {title} {Designer meron lattice on the surface of a topological
  insulator},\ }\href@noop {} {\bibfield  {journal} {\bibinfo  {journal} {arXiv
  preprint arXiv:2203.04986}\ } (\bibinfo {year} {2022})}\BibitemShut {NoStop}%
\bibitem [{\citenamefont {Fu}\ and\ \citenamefont
  {Kane}(2008)}]{fu2008superconducting}%
  \BibitemOpen
  \bibfield  {author} {\bibinfo {author} {\bibfnamefont {L.}~\bibnamefont
  {Fu}}\ and\ \bibinfo {author} {\bibfnamefont {C.~L.}\ \bibnamefont {Kane}},\
  }\bibfield  {title} {\bibinfo {title} {Superconducting proximity effect and
  majorana fermions at the surface of a topological insulator},\ }\href
  {https://doi.org/10.1103/PhysRevLett.100.096407} {\bibfield  {journal}
  {\bibinfo  {journal} {Phys. Rev. Lett.}\ }\textbf {\bibinfo {volume} {100}},\
  \bibinfo {pages} {096407} (\bibinfo {year} {2008})}\BibitemShut {NoStop}%
\bibitem [{\citenamefont {Santos}\ \emph {et~al.}(2010)\citenamefont {Santos},
  \citenamefont {Neupert}, \citenamefont {Chamon},\ and\ \citenamefont
  {Mudry}}]{santos2010}%
  \BibitemOpen
  \bibfield  {author} {\bibinfo {author} {\bibfnamefont {L.}~\bibnamefont
  {Santos}}, \bibinfo {author} {\bibfnamefont {T.}~\bibnamefont {Neupert}},
  \bibinfo {author} {\bibfnamefont {C.}~\bibnamefont {Chamon}},\ and\ \bibinfo
  {author} {\bibfnamefont {C.}~\bibnamefont {Mudry}},\ }\bibfield  {title}
  {\bibinfo {title} {Superconductivity on the surface of topological insulators
  and in two-dimensional noncentrosymmetric materials},\ }\href
  {https://doi.org/10.1103/PhysRevB.81.184502} {\bibfield  {journal} {\bibinfo
  {journal} {Phys. Rev. B}\ }\textbf {\bibinfo {volume} {81}},\ \bibinfo
  {pages} {184502} (\bibinfo {year} {2010})}\BibitemShut {NoStop}%
\bibitem [{\citenamefont {Baum}\ and\ \citenamefont
  {Stern}(2012{\natexlab{a}})}]{baum2012magnetic}%
  \BibitemOpen
  \bibfield  {author} {\bibinfo {author} {\bibfnamefont {Y.}~\bibnamefont
  {Baum}}\ and\ \bibinfo {author} {\bibfnamefont {A.}~\bibnamefont {Stern}},\
  }\bibfield  {title} {\bibinfo {title} {Magnetic instability on the surface of
  topological insulators},\ }\href {https://doi.org/10.1103/PhysRevB.85.121105}
  {\bibfield  {journal} {\bibinfo  {journal} {Phys. Rev. B}\ }\textbf {\bibinfo
  {volume} {85}},\ \bibinfo {pages} {121105(R)} (\bibinfo {year}
  {2012}{\natexlab{a}})}\BibitemShut {NoStop}%
\bibitem [{\citenamefont {Baum}\ and\ \citenamefont
  {Stern}(2012{\natexlab{b}})}]{baum2012density}%
  \BibitemOpen
  \bibfield  {author} {\bibinfo {author} {\bibfnamefont {Y.}~\bibnamefont
  {Baum}}\ and\ \bibinfo {author} {\bibfnamefont {A.}~\bibnamefont {Stern}},\
  }\bibfield  {title} {\bibinfo {title} {Density-waves instability and a
  skyrmion lattice on the surface of strong topological insulators},\ }\href
  {https://doi.org/10.1103/PhysRevB.86.195116} {\bibfield  {journal} {\bibinfo
  {journal} {Phys. Rev. B}\ }\textbf {\bibinfo {volume} {86}},\ \bibinfo
  {pages} {195116} (\bibinfo {year} {2012}{\natexlab{b}})}\BibitemShut
  {NoStop}%
\bibitem [{\citenamefont {Marchand}\ and\ \citenamefont
  {Franz}(2012)}]{marchand2012lattice}%
  \BibitemOpen
  \bibfield  {author} {\bibinfo {author} {\bibfnamefont {D.~J.~J.}\
  \bibnamefont {Marchand}}\ and\ \bibinfo {author} {\bibfnamefont
  {M.}~\bibnamefont {Franz}},\ }\bibfield  {title} {\bibinfo {title} {Lattice
  model for the surface states of a topological insulator with applications to
  magnetic and exciton instabilities},\ }\href
  {https://doi.org/10.1103/PhysRevB.86.155146} {\bibfield  {journal} {\bibinfo
  {journal} {Phys. Rev. B}\ }\textbf {\bibinfo {volume} {86}},\ \bibinfo
  {pages} {155146} (\bibinfo {year} {2012})}\BibitemShut {NoStop}%
\bibitem [{\citenamefont {Schmidt}(2012)}]{schmidt2012strong}%
  \BibitemOpen
  \bibfield  {author} {\bibinfo {author} {\bibfnamefont {M.~J.}\ \bibnamefont
  {Schmidt}},\ }\bibfield  {title} {\bibinfo {title} {Strong correlations at
  topological insulator surfaces and the breakdown of the bulk-boundary
  correspondence},\ }\href {https://doi.org/10.1103/PhysRevB.86.161110}
  {\bibfield  {journal} {\bibinfo  {journal} {Phys. Rev. B}\ }\textbf {\bibinfo
  {volume} {86}},\ \bibinfo {pages} {161110} (\bibinfo {year}
  {2012})}\BibitemShut {NoStop}%
\bibitem [{\citenamefont {Sitte}\ \emph {et~al.}(2013)\citenamefont {Sitte},
  \citenamefont {Rosch},\ and\ \citenamefont {Fritz}}]{sitte2013interaction}%
  \BibitemOpen
  \bibfield  {author} {\bibinfo {author} {\bibfnamefont {M.}~\bibnamefont
  {Sitte}}, \bibinfo {author} {\bibfnamefont {A.}~\bibnamefont {Rosch}},\ and\
  \bibinfo {author} {\bibfnamefont {L.}~\bibnamefont {Fritz}},\ }\bibfield
  {title} {\bibinfo {title} {Interaction effects on almost flat surface bands
  in topological insulators},\ }\href
  {https://doi.org/10.1103/PhysRevB.88.205107} {\bibfield  {journal} {\bibinfo
  {journal} {Phys. Rev. B}\ }\textbf {\bibinfo {volume} {88}},\ \bibinfo
  {pages} {205107} (\bibinfo {year} {2013})}\BibitemShut {NoStop}%
\bibitem [{\citenamefont {Mendler}\ \emph {et~al.}(2015)\citenamefont
  {Mendler}, \citenamefont {Kotetes},\ and\ \citenamefont
  {Sch\"on}}]{mendler2015magnetic}%
  \BibitemOpen
  \bibfield  {author} {\bibinfo {author} {\bibfnamefont {D.}~\bibnamefont
  {Mendler}}, \bibinfo {author} {\bibfnamefont {P.}~\bibnamefont {Kotetes}},\
  and\ \bibinfo {author} {\bibfnamefont {G.}~\bibnamefont {Sch\"on}},\
  }\bibfield  {title} {\bibinfo {title} {Magnetic order on a topological
  insulator surface with warping and proximity-induced superconductivity},\
  }\href {https://doi.org/10.1103/PhysRevB.91.155405} {\bibfield  {journal}
  {\bibinfo  {journal} {Phys. Rev. B}\ }\textbf {\bibinfo {volume} {91}},\
  \bibinfo {pages} {155405} (\bibinfo {year} {2015})}\BibitemShut {NoStop}%
\bibitem [{\citenamefont {Liu}\ \emph {et~al.}(2021)\citenamefont {Liu},
  \citenamefont {Wang},\ and\ \citenamefont {Wang}}]{liu2021magnetic}%
  \BibitemOpen
  \bibfield  {author} {\bibinfo {author} {\bibfnamefont {Z.}~\bibnamefont
  {Liu}}, \bibinfo {author} {\bibfnamefont {H.}~\bibnamefont {Wang}},\ and\
  \bibinfo {author} {\bibfnamefont {J.}~\bibnamefont {Wang}},\ }\bibfield
  {title} {\bibinfo {title} {Magnetic moir\'e surface states and flat chern
  band in topological insulators},\ }\href@noop {} {\bibfield  {journal}
  {\bibinfo  {journal} {arXiv preprint arXiv:2106.01630}\ } (\bibinfo {year}
  {2021})}\BibitemShut {NoStop}%
\bibitem [{\citenamefont {Chou}\ \emph {et~al.}(2021)\citenamefont {Chou},
  \citenamefont {Cano},\ and\ \citenamefont {Pixley}}]{chou2021band}%
  \BibitemOpen
  \bibfield  {author} {\bibinfo {author} {\bibfnamefont {Y.-Z.}\ \bibnamefont
  {Chou}}, \bibinfo {author} {\bibfnamefont {J.}~\bibnamefont {Cano}},\ and\
  \bibinfo {author} {\bibfnamefont {J.}~\bibnamefont {Pixley}},\ }\bibfield
  {title} {\bibinfo {title} {Band manipulation and spin texture in interacting
  moir\'e helical edges},\ }\href@noop {} {\bibfield  {journal} {\bibinfo
  {journal} {arXiv preprint arXiv:2106.05270}\ } (\bibinfo {year}
  {2021})}\BibitemShut {NoStop}%
\bibitem [{\citenamefont {Bernevig}\ \emph {et~al.}(2021)\citenamefont
  {Bernevig}, \citenamefont {Song}, \citenamefont {Regnault},\ and\
  \citenamefont {Lian}}]{bernevig2020tbg}%
  \BibitemOpen
  \bibfield  {author} {\bibinfo {author} {\bibfnamefont {B.~A.}\ \bibnamefont
  {Bernevig}}, \bibinfo {author} {\bibfnamefont {Z.-D.}\ \bibnamefont {Song}},
  \bibinfo {author} {\bibfnamefont {N.}~\bibnamefont {Regnault}},\ and\
  \bibinfo {author} {\bibfnamefont {B.}~\bibnamefont {Lian}},\ }\bibfield
  {title} {\bibinfo {title} {Twisted bilayer graphene. {III}. interacting
  hamiltonian and exact symmetries},\ }\href
  {https://doi.org/10.1103/PhysRevB.103.205413} {\bibfield  {journal} {\bibinfo
   {journal} {Phys. Rev. B}\ }\textbf {\bibinfo {volume} {103}},\ \bibinfo
  {pages} {205413} (\bibinfo {year} {2021})}\BibitemShut {NoStop}%
\bibitem [{\citenamefont {Rossi}\ and\ \citenamefont
  {Triola}(2019)}]{Rossi_2019}%
  \BibitemOpen
  \bibfield  {author} {\bibinfo {author} {\bibfnamefont {E.}~\bibnamefont
  {Rossi}}\ and\ \bibinfo {author} {\bibfnamefont {C.}~\bibnamefont {Triola}},\
  }\bibfield  {title} {\bibinfo {title} {Van der waals heterostructures with
  spin‐orbit coupling},\ }\href {https://doi.org/10.1002/andp.201900344}
  {\bibfield  {journal} {\bibinfo  {journal} {Annalen der Physik}\ }\textbf
  {\bibinfo {volume} {532}},\ \bibinfo {pages} {1900344} (\bibinfo {year}
  {2019})}\BibitemShut {NoStop}%
\bibitem [{\citenamefont {Jin}\ \emph {et~al.}(2012)\citenamefont {Jin},
  \citenamefont {Im}, \citenamefont {Song},\ and\ \citenamefont
  {Freeman}}]{jin2012multiple}%
  \BibitemOpen
  \bibfield  {author} {\bibinfo {author} {\bibfnamefont {H.}~\bibnamefont
  {Jin}}, \bibinfo {author} {\bibfnamefont {J.}~\bibnamefont {Im}}, \bibinfo
  {author} {\bibfnamefont {J.-H.}\ \bibnamefont {Song}},\ and\ \bibinfo
  {author} {\bibfnamefont {A.~J.}\ \bibnamefont {Freeman}},\ }\bibfield
  {title} {\bibinfo {title} {Multiple {D}irac fermions from a topological
  insulator and graphene superlattice},\ }\href
  {https://doi.org/10.1103/PhysRevB.85.045307} {\bibfield  {journal} {\bibinfo
  {journal} {Phys. Rev. B}\ }\textbf {\bibinfo {volume} {85}},\ \bibinfo
  {pages} {045307} (\bibinfo {year} {2012})}\BibitemShut {NoStop}%
\bibitem [{\citenamefont {Jin}\ and\ \citenamefont
  {Jhi}(2013)}]{jin2013proximity}%
  \BibitemOpen
  \bibfield  {author} {\bibinfo {author} {\bibfnamefont {K.-H.}\ \bibnamefont
  {Jin}}\ and\ \bibinfo {author} {\bibfnamefont {S.-H.}\ \bibnamefont {Jhi}},\
  }\bibfield  {title} {\bibinfo {title} {Proximity-induced giant spin-orbit
  interaction in epitaxial graphene on a topological insulator},\ }\href
  {https://doi.org/10.1103/PhysRevB.87.075442} {\bibfield  {journal} {\bibinfo
  {journal} {Phys. Rev. B}\ }\textbf {\bibinfo {volume} {87}},\ \bibinfo
  {pages} {075442} (\bibinfo {year} {2013})}\BibitemShut {NoStop}%
\bibitem [{\citenamefont {Zhang}\ \emph {et~al.}(2014)\citenamefont {Zhang},
  \citenamefont {Triola},\ and\ \citenamefont {Rossi}}]{RossiTriolaOld}%
  \BibitemOpen
  \bibfield  {author} {\bibinfo {author} {\bibfnamefont {J.}~\bibnamefont
  {Zhang}}, \bibinfo {author} {\bibfnamefont {C.}~\bibnamefont {Triola}},\ and\
  \bibinfo {author} {\bibfnamefont {E.}~\bibnamefont {Rossi}},\ }\bibfield
  {title} {\bibinfo {title} {Proximity effect in
  graphene--topological-insulator heterostructures},\ }\href
  {https://doi.org/10.1103/PhysRevLett.112.096802} {\bibfield  {journal}
  {\bibinfo  {journal} {Phys. Rev. Lett.}\ }\textbf {\bibinfo {volume} {112}},\
  \bibinfo {pages} {096802} (\bibinfo {year} {2014})}\BibitemShut {NoStop}%
\bibitem [{\citenamefont {Cao}\ \emph {et~al.}(2016)\citenamefont {Cao},
  \citenamefont {Zhang}, \citenamefont {Tang}, \citenamefont {Yang},
  \citenamefont {Sofo}, \citenamefont {Duan},\ and\ \citenamefont
  {Liu}}]{cao2016heavy}%
  \BibitemOpen
  \bibfield  {author} {\bibinfo {author} {\bibfnamefont {W.}~\bibnamefont
  {Cao}}, \bibinfo {author} {\bibfnamefont {R.-X.}\ \bibnamefont {Zhang}},
  \bibinfo {author} {\bibfnamefont {P.}~\bibnamefont {Tang}}, \bibinfo {author}
  {\bibfnamefont {G.}~\bibnamefont {Yang}}, \bibinfo {author} {\bibfnamefont
  {J.}~\bibnamefont {Sofo}}, \bibinfo {author} {\bibfnamefont {W.}~\bibnamefont
  {Duan}},\ and\ \bibinfo {author} {\bibfnamefont {C.-X.}\ \bibnamefont
  {Liu}},\ }\bibfield  {title} {\bibinfo {title} {Heavy {D}irac fermions in a
  graphene/topological insulator hetero-junction},\ }\href@noop {} {\bibfield
  {journal} {\bibinfo  {journal} {2D Materials}\ }\textbf {\bibinfo {volume}
  {3}},\ \bibinfo {pages} {034006} (\bibinfo {year} {2016})}\BibitemShut
  {NoStop}%
\bibitem [{\citenamefont {De~Beule}\ \emph {et~al.}(2017)\citenamefont
  {De~Beule}, \citenamefont {Zarenia},\ and\ \citenamefont
  {Partoens}}]{debeule2017transmission}%
  \BibitemOpen
  \bibfield  {author} {\bibinfo {author} {\bibfnamefont {C.}~\bibnamefont
  {De~Beule}}, \bibinfo {author} {\bibfnamefont {M.}~\bibnamefont {Zarenia}},\
  and\ \bibinfo {author} {\bibfnamefont {B.}~\bibnamefont {Partoens}},\
  }\bibfield  {title} {\bibinfo {title} {Transmission in graphene--topological
  insulator heterostructures},\ }\href
  {https://doi.org/10.1103/PhysRevB.95.115424} {\bibfield  {journal} {\bibinfo
  {journal} {Phys. Rev. B}\ }\textbf {\bibinfo {volume} {95}},\ \bibinfo
  {pages} {115424} (\bibinfo {year} {2017})}\BibitemShut {NoStop}%
\bibitem [{\citenamefont {Rodriguez-Vega}\ \emph {et~al.}(2017)\citenamefont
  {Rodriguez-Vega}, \citenamefont {Schwiete}, \citenamefont {Sinova},\ and\
  \citenamefont {Rossi}}]{rodriguez2017giant}%
  \BibitemOpen
  \bibfield  {author} {\bibinfo {author} {\bibfnamefont {M.}~\bibnamefont
  {Rodriguez-Vega}}, \bibinfo {author} {\bibfnamefont {G.}~\bibnamefont
  {Schwiete}}, \bibinfo {author} {\bibfnamefont {J.}~\bibnamefont {Sinova}},\
  and\ \bibinfo {author} {\bibfnamefont {E.}~\bibnamefont {Rossi}},\ }\bibfield
   {title} {\bibinfo {title} {Giant edelstein effect in
  topological-insulator--graphene heterostructures},\ }\href@noop {} {\bibfield
   {journal} {\bibinfo  {journal} {Physical Review B}\ }\textbf {\bibinfo
  {volume} {96}},\ \bibinfo {pages} {235419} (\bibinfo {year}
  {2017})}\BibitemShut {NoStop}%
\bibitem [{\citenamefont {Song}\ \emph {et~al.}(2018)\citenamefont {Song},
  \citenamefont {Soriano}, \citenamefont {Cummings}, \citenamefont {Robles},
  \citenamefont {Ordej{\'o}n},\ and\ \citenamefont {Roche}}]{song2018spin}%
  \BibitemOpen
  \bibfield  {author} {\bibinfo {author} {\bibfnamefont {K.}~\bibnamefont
  {Song}}, \bibinfo {author} {\bibfnamefont {D.}~\bibnamefont {Soriano}},
  \bibinfo {author} {\bibfnamefont {A.~W.}\ \bibnamefont {Cummings}}, \bibinfo
  {author} {\bibfnamefont {R.}~\bibnamefont {Robles}}, \bibinfo {author}
  {\bibfnamefont {P.}~\bibnamefont {Ordej{\'o}n}},\ and\ \bibinfo {author}
  {\bibfnamefont {S.}~\bibnamefont {Roche}},\ }\bibfield  {title} {\bibinfo
  {title} {Spin proximity effects in graphene/topological insulator
  heterostructures},\ }\href@noop {} {\bibfield  {journal} {\bibinfo  {journal}
  {Nano letters}\ }\textbf {\bibinfo {volume} {18}},\ \bibinfo {pages} {2033}
  (\bibinfo {year} {2018})}\BibitemShut {NoStop}%
\bibitem [{\citenamefont {Zollner}\ and\ \citenamefont
  {Fabian}(2020)}]{GRBISE5}%
  \BibitemOpen
  \bibfield  {author} {\bibinfo {author} {\bibfnamefont {K.}~\bibnamefont
  {Zollner}}\ and\ \bibinfo {author} {\bibfnamefont {J.}~\bibnamefont
  {Fabian}},\ }\bibfield  {title} {\bibinfo {title} {Heterostructures of
  graphene and topological insulators {Bi$_2$Se$_3$, Bi$_2$Te$_3$, and
  Sb$_2$Te$_3$}},\ }\href {https://doi.org/10.1002/pssb.202000081} {\bibfield
  {journal} {\bibinfo  {journal} {physica status solidi (b)}\ }\textbf
  {\bibinfo {volume} {258}} (\bibinfo {year} {2020})}\BibitemShut {NoStop}%
\bibitem [{\citenamefont {Zollner}\ and\ \citenamefont
  {Fabian}(2019)}]{PhysRevB.100.165141}%
  \BibitemOpen
  \bibfield  {author} {\bibinfo {author} {\bibfnamefont {K.}~\bibnamefont
  {Zollner}}\ and\ \bibinfo {author} {\bibfnamefont {J.}~\bibnamefont
  {Fabian}},\ }\bibfield  {title} {\bibinfo {title} {Single and bilayer
  graphene on the topological insulator {Bi$_2$Se$_3$}: Electronic and
  spin-orbit properties from first principles},\ }\href
  {https://doi.org/10.1103/PhysRevB.100.165141} {\bibfield  {journal} {\bibinfo
   {journal} {Phys. Rev. B}\ }\textbf {\bibinfo {volume} {100}},\ \bibinfo
  {pages} {165141} (\bibinfo {year} {2019})}\BibitemShut {NoStop}%
\bibitem [{\citenamefont {Dang}\ \emph {et~al.}(2010)\citenamefont {Dang},
  \citenamefont {Peng}, \citenamefont {Li}, \citenamefont {Wang},\ and\
  \citenamefont {Liu}}]{dang2010epitaxial}%
  \BibitemOpen
  \bibfield  {author} {\bibinfo {author} {\bibfnamefont {W.}~\bibnamefont
  {Dang}}, \bibinfo {author} {\bibfnamefont {H.}~\bibnamefont {Peng}}, \bibinfo
  {author} {\bibfnamefont {H.}~\bibnamefont {Li}}, \bibinfo {author}
  {\bibfnamefont {P.}~\bibnamefont {Wang}},\ and\ \bibinfo {author}
  {\bibfnamefont {Z.}~\bibnamefont {Liu}},\ }\bibfield  {title} {\bibinfo
  {title} {Epitaxial heterostructures of ultrathin topological insulator
  nanoplate and graphene},\ }\href@noop {} {\bibfield  {journal} {\bibinfo
  {journal} {Nano letters}\ }\textbf {\bibinfo {volume} {10}},\ \bibinfo
  {pages} {2870} (\bibinfo {year} {2010})}\BibitemShut {NoStop}%
\bibitem [{\citenamefont {Song}\ \emph {et~al.}(2010)\citenamefont {Song},
  \citenamefont {Wang}, \citenamefont {Jiang}, \citenamefont {Zhang},
  \citenamefont {Chang}, \citenamefont {Wang}, \citenamefont {He},
  \citenamefont {Chen}, \citenamefont {Jia}, \citenamefont {Wang} \emph
  {et~al.}}]{song2010topological}%
  \BibitemOpen
  \bibfield  {author} {\bibinfo {author} {\bibfnamefont {C.-L.}\ \bibnamefont
  {Song}}, \bibinfo {author} {\bibfnamefont {Y.-L.}\ \bibnamefont {Wang}},
  \bibinfo {author} {\bibfnamefont {Y.-P.}\ \bibnamefont {Jiang}}, \bibinfo
  {author} {\bibfnamefont {Y.}~\bibnamefont {Zhang}}, \bibinfo {author}
  {\bibfnamefont {C.-Z.}\ \bibnamefont {Chang}}, \bibinfo {author}
  {\bibfnamefont {L.}~\bibnamefont {Wang}}, \bibinfo {author} {\bibfnamefont
  {K.}~\bibnamefont {He}}, \bibinfo {author} {\bibfnamefont {X.}~\bibnamefont
  {Chen}}, \bibinfo {author} {\bibfnamefont {J.-F.}\ \bibnamefont {Jia}},
  \bibinfo {author} {\bibfnamefont {Y.}~\bibnamefont {Wang}}, \emph {et~al.},\
  }\bibfield  {title} {\bibinfo {title} {Topological insulator {Bi$_2$Se$_3$}
  thin films grown on double-layer graphene by molecular beam epitaxy},\
  }\href@noop {} {\bibfield  {journal} {\bibinfo  {journal} {Applied Physics
  Letters}\ }\textbf {\bibinfo {volume} {97}},\ \bibinfo {pages} {143118}
  (\bibinfo {year} {2010})}\BibitemShut {NoStop}%
\bibitem [{\citenamefont {Lee}\ \emph {et~al.}(2015)\citenamefont {Lee},
  \citenamefont {Jin}, \citenamefont {Sung}, \citenamefont {Kim}, \citenamefont
  {Ryu}, \citenamefont {Park}, \citenamefont {Jhi}, \citenamefont {Kim},
  \citenamefont {Kim}, \citenamefont {Yu} \emph {et~al.}}]{lee2015proximity}%
  \BibitemOpen
  \bibfield  {author} {\bibinfo {author} {\bibfnamefont {P.}~\bibnamefont
  {Lee}}, \bibinfo {author} {\bibfnamefont {K.-H.}\ \bibnamefont {Jin}},
  \bibinfo {author} {\bibfnamefont {S.~J.}\ \bibnamefont {Sung}}, \bibinfo
  {author} {\bibfnamefont {J.~G.}\ \bibnamefont {Kim}}, \bibinfo {author}
  {\bibfnamefont {M.-T.}\ \bibnamefont {Ryu}}, \bibinfo {author} {\bibfnamefont
  {H.-M.}\ \bibnamefont {Park}}, \bibinfo {author} {\bibfnamefont {S.-H.}\
  \bibnamefont {Jhi}}, \bibinfo {author} {\bibfnamefont {N.}~\bibnamefont
  {Kim}}, \bibinfo {author} {\bibfnamefont {Y.}~\bibnamefont {Kim}}, \bibinfo
  {author} {\bibfnamefont {S.~U.}\ \bibnamefont {Yu}}, \emph {et~al.},\
  }\bibfield  {title} {\bibinfo {title} {Proximity effect induced electronic
  properties of graphene on {Bi$_2$Te$_2$Se}},\ }\href@noop {} {\bibfield
  {journal} {\bibinfo  {journal} {Acs Nano}\ }\textbf {\bibinfo {volume} {9}},\
  \bibinfo {pages} {10861} (\bibinfo {year} {2015})}\BibitemShut {NoStop}%
\bibitem [{\citenamefont {Steinberg}\ \emph {et~al.}(2015)\citenamefont
  {Steinberg}, \citenamefont {Orona}, \citenamefont {Fatemi}, \citenamefont
  {Sanchez-Yamagishi}, \citenamefont {Watanabe}, \citenamefont {Taniguchi},\
  and\ \citenamefont {Jarillo-Herrero}}]{steinberg2015tunneling}%
  \BibitemOpen
  \bibfield  {author} {\bibinfo {author} {\bibfnamefont {H.}~\bibnamefont
  {Steinberg}}, \bibinfo {author} {\bibfnamefont {L.~A.}\ \bibnamefont
  {Orona}}, \bibinfo {author} {\bibfnamefont {V.}~\bibnamefont {Fatemi}},
  \bibinfo {author} {\bibfnamefont {J.~D.}\ \bibnamefont {Sanchez-Yamagishi}},
  \bibinfo {author} {\bibfnamefont {K.}~\bibnamefont {Watanabe}}, \bibinfo
  {author} {\bibfnamefont {T.}~\bibnamefont {Taniguchi}},\ and\ \bibinfo
  {author} {\bibfnamefont {P.}~\bibnamefont {Jarillo-Herrero}},\ }\bibfield
  {title} {\bibinfo {title} {Tunneling in graphene--topological insulator
  hybrid devices},\ }\href@noop {} {\bibfield  {journal} {\bibinfo  {journal}
  {Physical Review B}\ }\textbf {\bibinfo {volume} {92}},\ \bibinfo {pages}
  {241409(R)} (\bibinfo {year} {2015})}\BibitemShut {NoStop}%
\bibitem [{\citenamefont {Zhang}\ \emph {et~al.}(2016)\citenamefont {Zhang},
  \citenamefont {Yan}, \citenamefont {Wu}, \citenamefont {Yu},\ and\
  \citenamefont {Liao}}]{zhang2016gate}%
  \BibitemOpen
  \bibfield  {author} {\bibinfo {author} {\bibfnamefont {L.}~\bibnamefont
  {Zhang}}, \bibinfo {author} {\bibfnamefont {Y.}~\bibnamefont {Yan}}, \bibinfo
  {author} {\bibfnamefont {H.-C.}\ \bibnamefont {Wu}}, \bibinfo {author}
  {\bibfnamefont {D.}~\bibnamefont {Yu}},\ and\ \bibinfo {author}
  {\bibfnamefont {Z.-M.}\ \bibnamefont {Liao}},\ }\bibfield  {title} {\bibinfo
  {title} {Gate-tunable tunneling resistance in graphene/topological insulator
  vertical junctions},\ }\href@noop {} {\bibfield  {journal} {\bibinfo
  {journal} {ACS nano}\ }\textbf {\bibinfo {volume} {10}},\ \bibinfo {pages}
  {3816} (\bibinfo {year} {2016})}\BibitemShut {NoStop}%
\bibitem [{\citenamefont {Bian}\ \emph {et~al.}(2016)\citenamefont {Bian},
  \citenamefont {Chung}, \citenamefont {Chen}, \citenamefont {Liu},
  \citenamefont {Chang}, \citenamefont {Wu}, \citenamefont {Belopolski},
  \citenamefont {Zheng}, \citenamefont {Xu}, \citenamefont {Sanchez},
  \citenamefont {Alidoust}, \citenamefont {Pierce}, \citenamefont {Quiliams},
  \citenamefont {Barletta}, \citenamefont {Lorcy}, \citenamefont {Avila},
  \citenamefont {Chang}, \citenamefont {Lin}, \citenamefont {Jeng},
  \citenamefont {Asensio}, \citenamefont {Chen},\ and\ \citenamefont
  {Hasan}}]{bian2016experimental}%
  \BibitemOpen
  \bibfield  {author} {\bibinfo {author} {\bibfnamefont {G.}~\bibnamefont
  {Bian}}, \bibinfo {author} {\bibfnamefont {T.-F.}\ \bibnamefont {Chung}},
  \bibinfo {author} {\bibfnamefont {C.}~\bibnamefont {Chen}}, \bibinfo {author}
  {\bibfnamefont {C.}~\bibnamefont {Liu}}, \bibinfo {author} {\bibfnamefont
  {T.-R.}\ \bibnamefont {Chang}}, \bibinfo {author} {\bibfnamefont
  {T.}~\bibnamefont {Wu}}, \bibinfo {author} {\bibfnamefont {I.}~\bibnamefont
  {Belopolski}}, \bibinfo {author} {\bibfnamefont {H.}~\bibnamefont {Zheng}},
  \bibinfo {author} {\bibfnamefont {S.-Y.}\ \bibnamefont {Xu}}, \bibinfo
  {author} {\bibfnamefont {D.~S.}\ \bibnamefont {Sanchez}}, \bibinfo {author}
  {\bibfnamefont {N.}~\bibnamefont {Alidoust}}, \bibinfo {author}
  {\bibfnamefont {J.}~\bibnamefont {Pierce}}, \bibinfo {author} {\bibfnamefont
  {B.}~\bibnamefont {Quiliams}}, \bibinfo {author} {\bibfnamefont {P.~P.}\
  \bibnamefont {Barletta}}, \bibinfo {author} {\bibfnamefont {S.}~\bibnamefont
  {Lorcy}}, \bibinfo {author} {\bibfnamefont {J.}~\bibnamefont {Avila}},
  \bibinfo {author} {\bibfnamefont {G.}~\bibnamefont {Chang}}, \bibinfo
  {author} {\bibfnamefont {H.}~\bibnamefont {Lin}}, \bibinfo {author}
  {\bibfnamefont {H.-T.}\ \bibnamefont {Jeng}}, \bibinfo {author}
  {\bibfnamefont {M.-C.}\ \bibnamefont {Asensio}}, \bibinfo {author}
  {\bibfnamefont {Y.~P.}\ \bibnamefont {Chen}},\ and\ \bibinfo {author}
  {\bibfnamefont {M.~Z.}\ \bibnamefont {Hasan}},\ }\bibfield  {title} {\bibinfo
  {title} {Experimental observation of two massless {D}irac-fermion gases in
  graphene-topological insulator heterostructure},\ }\href@noop {} {\bibfield
  {journal} {\bibinfo  {journal} {2D Materials}\ }\textbf {\bibinfo {volume}
  {3}},\ \bibinfo {pages} {021009} (\bibinfo {year} {2016})}\BibitemShut
  {NoStop}%
\bibitem [{\citenamefont {Vaklinova}\ \emph {et~al.}(2016)\citenamefont
  {Vaklinova}, \citenamefont {Hoyer}, \citenamefont {Burghard},\ and\
  \citenamefont {Kern}}]{vaklinova2016current}%
  \BibitemOpen
  \bibfield  {author} {\bibinfo {author} {\bibfnamefont {K.}~\bibnamefont
  {Vaklinova}}, \bibinfo {author} {\bibfnamefont {A.}~\bibnamefont {Hoyer}},
  \bibinfo {author} {\bibfnamefont {M.}~\bibnamefont {Burghard}},\ and\
  \bibinfo {author} {\bibfnamefont {K.}~\bibnamefont {Kern}},\ }\bibfield
  {title} {\bibinfo {title} {Current-induced spin polarization in topological
  insulator--graphene heterostructures},\ }\href@noop {} {\bibfield  {journal}
  {\bibinfo  {journal} {Nano letters}\ }\textbf {\bibinfo {volume} {16}},\
  \bibinfo {pages} {2595} (\bibinfo {year} {2016})}\BibitemShut {NoStop}%
\bibitem [{\citenamefont {Rajput}\ \emph {et~al.}(2016)\citenamefont {Rajput},
  \citenamefont {Li}, \citenamefont {Weinert},\ and\ \citenamefont
  {Li}}]{rajput2016indirect}%
  \BibitemOpen
  \bibfield  {author} {\bibinfo {author} {\bibfnamefont {S.}~\bibnamefont
  {Rajput}}, \bibinfo {author} {\bibfnamefont {Y.-Y.}\ \bibnamefont {Li}},
  \bibinfo {author} {\bibfnamefont {M.}~\bibnamefont {Weinert}},\ and\ \bibinfo
  {author} {\bibfnamefont {L.}~\bibnamefont {Li}},\ }\bibfield  {title}
  {\bibinfo {title} {Indirect interlayer bonding in graphene--topological
  insulator van der {W}aals heterostructure: Giant spin--orbit splitting of the
  graphene {D}irac states},\ }\href@noop {} {\bibfield  {journal} {\bibinfo
  {journal} {ACS nano}\ }\textbf {\bibinfo {volume} {10}},\ \bibinfo {pages}
  {8450} (\bibinfo {year} {2016})}\BibitemShut {NoStop}%
\bibitem [{\citenamefont {Zhang}\ \emph {et~al.}(2017)\citenamefont {Zhang},
  \citenamefont {Lin}, \citenamefont {Wu}, \citenamefont {Wu}, \citenamefont
  {Huang}, \citenamefont {Chang}, \citenamefont {Ke}, \citenamefont
  {Kurttepeli}, \citenamefont {Tendeloo}, \citenamefont {Xu} \emph
  {et~al.}}]{zhang2017electronic}%
  \BibitemOpen
  \bibfield  {author} {\bibinfo {author} {\bibfnamefont {L.}~\bibnamefont
  {Zhang}}, \bibinfo {author} {\bibfnamefont {B.-C.}\ \bibnamefont {Lin}},
  \bibinfo {author} {\bibfnamefont {Y.-F.}\ \bibnamefont {Wu}}, \bibinfo
  {author} {\bibfnamefont {H.-C.}\ \bibnamefont {Wu}}, \bibinfo {author}
  {\bibfnamefont {T.-W.}\ \bibnamefont {Huang}}, \bibinfo {author}
  {\bibfnamefont {C.-R.}\ \bibnamefont {Chang}}, \bibinfo {author}
  {\bibfnamefont {X.}~\bibnamefont {Ke}}, \bibinfo {author} {\bibfnamefont
  {M.}~\bibnamefont {Kurttepeli}}, \bibinfo {author} {\bibfnamefont {G.~V.}\
  \bibnamefont {Tendeloo}}, \bibinfo {author} {\bibfnamefont {J.}~\bibnamefont
  {Xu}}, \emph {et~al.},\ }\bibfield  {title} {\bibinfo {title} {Electronic
  coupling between graphene and topological insulator induced anomalous
  magnetotransport properties},\ }\href@noop {} {\bibfield  {journal} {\bibinfo
   {journal} {ACS nano}\ }\textbf {\bibinfo {volume} {11}},\ \bibinfo {pages}
  {6277} (\bibinfo {year} {2017})}\BibitemShut {NoStop}%
\bibitem [{\citenamefont {Zalic}\ \emph {et~al.}(2017)\citenamefont {Zalic},
  \citenamefont {Dvir},\ and\ \citenamefont {Steinberg}}]{GRBISE1}%
  \BibitemOpen
  \bibfield  {author} {\bibinfo {author} {\bibfnamefont {A.}~\bibnamefont
  {Zalic}}, \bibinfo {author} {\bibfnamefont {T.}~\bibnamefont {Dvir}},\ and\
  \bibinfo {author} {\bibfnamefont {H.}~\bibnamefont {Steinberg}},\ }\bibfield
  {title} {\bibinfo {title} {High-density carriers at a strongly coupled
  interface between graphene and a three-dimensional topological insulator},\
  }\bibfield  {journal} {\bibinfo  {journal} {Physical Review B}\ }\textbf
  {\bibinfo {volume} {96}},\ \href {https://doi.org/10.1103/physrevb.96.075104}
  {10.1103/physrevb.96.075104} (\bibinfo {year} {2017})\BibitemShut {NoStop}%
\bibitem [{\citenamefont {Khokhriakov}\ \emph {et~al.}(2018)\citenamefont
  {Khokhriakov}, \citenamefont {Cummings}, \citenamefont {Song}, \citenamefont
  {Vila}, \citenamefont {Karpiak}, \citenamefont {Dankert}, \citenamefont
  {Roche},\ and\ \citenamefont {Dash}}]{khokhriakov2018tailoring}%
  \BibitemOpen
  \bibfield  {author} {\bibinfo {author} {\bibfnamefont {D.}~\bibnamefont
  {Khokhriakov}}, \bibinfo {author} {\bibfnamefont {A.~W.}\ \bibnamefont
  {Cummings}}, \bibinfo {author} {\bibfnamefont {K.}~\bibnamefont {Song}},
  \bibinfo {author} {\bibfnamefont {M.}~\bibnamefont {Vila}}, \bibinfo {author}
  {\bibfnamefont {B.}~\bibnamefont {Karpiak}}, \bibinfo {author} {\bibfnamefont
  {A.}~\bibnamefont {Dankert}}, \bibinfo {author} {\bibfnamefont
  {S.}~\bibnamefont {Roche}},\ and\ \bibinfo {author} {\bibfnamefont {S.~P.}\
  \bibnamefont {Dash}},\ }\bibfield  {title} {\bibinfo {title} {Tailoring
  emergent spin phenomena in {Dirac} material heterostructures},\ }\href@noop
  {} {\bibfield  {journal} {\bibinfo  {journal} {Science advances}\ }\textbf
  {\bibinfo {volume} {4}},\ \bibinfo {pages} {eaat9349} (\bibinfo {year}
  {2018})}\BibitemShut {NoStop}%
\bibitem [{\citenamefont {Chae}\ \emph {et~al.}(2019)\citenamefont {Chae},
  \citenamefont {Kang}, \citenamefont {Park}, \citenamefont {Park},
  \citenamefont {Jeong}, \citenamefont {Kim}, \citenamefont {Hong},
  \citenamefont {Kim}, \citenamefont {Kwon}, \citenamefont {Kim} \emph
  {et~al.}}]{GRBISE3}%
  \BibitemOpen
  \bibfield  {author} {\bibinfo {author} {\bibfnamefont {J.}~\bibnamefont
  {Chae}}, \bibinfo {author} {\bibfnamefont {S.-H.}\ \bibnamefont {Kang}},
  \bibinfo {author} {\bibfnamefont {S.~H.}\ \bibnamefont {Park}}, \bibinfo
  {author} {\bibfnamefont {H.}~\bibnamefont {Park}}, \bibinfo {author}
  {\bibfnamefont {K.}~\bibnamefont {Jeong}}, \bibinfo {author} {\bibfnamefont
  {T.~H.}\ \bibnamefont {Kim}}, \bibinfo {author} {\bibfnamefont {S.-B.}\
  \bibnamefont {Hong}}, \bibinfo {author} {\bibfnamefont {K.~S.}\ \bibnamefont
  {Kim}}, \bibinfo {author} {\bibfnamefont {Y.-K.}\ \bibnamefont {Kwon}},
  \bibinfo {author} {\bibfnamefont {J.~W.}\ \bibnamefont {Kim}}, \emph
  {et~al.},\ }\bibfield  {title} {\bibinfo {title} {Closing the surface bandgap
  in thin {Bi$_2$Se$_3$}/graphene heterostructures},\ }\href@noop {} {\bibfield
   {journal} {\bibinfo  {journal} {ACS nano}\ }\textbf {\bibinfo {volume}
  {13}},\ \bibinfo {pages} {3931} (\bibinfo {year} {2019})}\BibitemShut
  {NoStop}%
\bibitem [{\citenamefont {Kariyado}\ and\ \citenamefont
  {Vishwanath}(2019)}]{kariyado2019flat}%
  \BibitemOpen
  \bibfield  {author} {\bibinfo {author} {\bibfnamefont {T.}~\bibnamefont
  {Kariyado}}\ and\ \bibinfo {author} {\bibfnamefont {A.}~\bibnamefont
  {Vishwanath}},\ }\bibfield  {title} {\bibinfo {title} {Flat band in twisted
  bilayer bravais lattices},\ }\href
  {https://doi.org/10.1103/PhysRevResearch.1.033076} {\bibfield  {journal}
  {\bibinfo  {journal} {Phys. Rev. Research}\ }\textbf {\bibinfo {volume}
  {1}},\ \bibinfo {pages} {033076} (\bibinfo {year} {2019})}\BibitemShut
  {NoStop}%
\bibitem [{\citenamefont {Kennes}\ \emph {et~al.}(2020)\citenamefont {Kennes},
  \citenamefont {Xian}, \citenamefont {Claassen},\ and\ \citenamefont
  {Rubio}}]{kennes2020one}%
  \BibitemOpen
  \bibfield  {author} {\bibinfo {author} {\bibfnamefont {D.~M.}\ \bibnamefont
  {Kennes}}, \bibinfo {author} {\bibfnamefont {L.}~\bibnamefont {Xian}},
  \bibinfo {author} {\bibfnamefont {M.}~\bibnamefont {Claassen}},\ and\
  \bibinfo {author} {\bibfnamefont {A.}~\bibnamefont {Rubio}},\ }\bibfield
  {title} {\bibinfo {title} {One-dimensional flat bands in twisted bilayer
  germanium selenide},\ }\href@noop {} {\bibfield  {journal} {\bibinfo
  {journal} {Nature communications}\ }\textbf {\bibinfo {volume} {11}},\
  \bibinfo {pages} {1} (\bibinfo {year} {2020})}\BibitemShut {NoStop}%
\bibitem [{\citenamefont {Luo}\ \emph {et~al.}(2021)\citenamefont {Luo},
  \citenamefont {Xu},\ and\ \citenamefont {Jian}}]{TwistedSquares}%
  \BibitemOpen
  \bibfield  {author} {\bibinfo {author} {\bibfnamefont {Z.-X.}\ \bibnamefont
  {Luo}}, \bibinfo {author} {\bibfnamefont {C.}~\bibnamefont {Xu}},\ and\
  \bibinfo {author} {\bibfnamefont {C.-M.}\ \bibnamefont {Jian}},\ }\bibfield
  {title} {\bibinfo {title} {Magic continuum in a twisted bilayer square
  lattice with staggered flux},\ }\href
  {https://doi.org/10.1103/PhysRevB.104.035136} {\bibfield  {journal} {\bibinfo
   {journal} {Phys. Rev. B}\ }\textbf {\bibinfo {volume} {104}},\ \bibinfo
  {pages} {035136} (\bibinfo {year} {2021})}\BibitemShut {NoStop}%
\bibitem [{\citenamefont {Lopes~dos Santos}\ \emph
  {et~al.}(2012{\natexlab{b}})\citenamefont {Lopes~dos Santos}, \citenamefont
  {Peres},\ and\ \citenamefont {Castro~Neto}}]{Lopes_dos_Santos_2012}%
  \BibitemOpen
  \bibfield  {author} {\bibinfo {author} {\bibfnamefont {J.~M.~B.}\
  \bibnamefont {Lopes~dos Santos}}, \bibinfo {author} {\bibfnamefont
  {N.~M.~R.}\ \bibnamefont {Peres}},\ and\ \bibinfo {author} {\bibfnamefont
  {A.~H.}\ \bibnamefont {Castro~Neto}},\ }\bibfield  {title} {\bibinfo {title}
  {Continuum model of the twisted graphene bilayer},\ }\bibfield  {journal}
  {\bibinfo  {journal} {Physical Review B}\ }\textbf {\bibinfo {volume} {86}},\
  \href {https://doi.org/10.1103/physrevb.86.155449}
  {10.1103/physrevb.86.155449} (\bibinfo {year}
  {2012}{\natexlab{b}})\BibitemShut {NoStop}%
\bibitem [{\citenamefont {Song}\ \emph {et~al.}(2021)\citenamefont {Song},
  \citenamefont {Lian}, \citenamefont {Regnault},\ and\ \citenamefont
  {Bernevig}}]{BernevigII}%
  \BibitemOpen
  \bibfield  {author} {\bibinfo {author} {\bibfnamefont {Z.-D.}\ \bibnamefont
  {Song}}, \bibinfo {author} {\bibfnamefont {B.}~\bibnamefont {Lian}}, \bibinfo
  {author} {\bibfnamefont {N.}~\bibnamefont {Regnault}},\ and\ \bibinfo
  {author} {\bibfnamefont {B.~A.}\ \bibnamefont {Bernevig}},\ }\bibfield
  {title} {\bibinfo {title} {Twisted bilayer graphene. {II}. stable symmetry
  anomaly},\ }\bibfield  {journal} {\bibinfo  {journal} {Physical Review B}\
  }\textbf {\bibinfo {volume} {103}},\ \href
  {https://doi.org/10.1103/physrevb.103.205412} {10.1103/physrevb.103.205412}
  (\bibinfo {year} {2021})\BibitemShut {NoStop}%
\bibitem [{\citenamefont {Zhang}\ \emph {et~al.}(2009)\citenamefont {Zhang},
  \citenamefont {Liu}, \citenamefont {Qi}, \citenamefont {Dai}, \citenamefont
  {Fang},\ and\ \citenamefont {Zhang}}]{zhang2009topological}%
  \BibitemOpen
  \bibfield  {author} {\bibinfo {author} {\bibfnamefont {H.}~\bibnamefont
  {Zhang}}, \bibinfo {author} {\bibfnamefont {C.-X.}\ \bibnamefont {Liu}},
  \bibinfo {author} {\bibfnamefont {X.-L.}\ \bibnamefont {Qi}}, \bibinfo
  {author} {\bibfnamefont {X.}~\bibnamefont {Dai}}, \bibinfo {author}
  {\bibfnamefont {Z.}~\bibnamefont {Fang}},\ and\ \bibinfo {author}
  {\bibfnamefont {S.-C.}\ \bibnamefont {Zhang}},\ }\bibfield  {title} {\bibinfo
  {title} {Topological insulators in {Bi$_2$Se$_3$}, {Bi$_2$Te$_3$} and
  {Sb$_2$Te$_3$} with a single {Dirac} cone on the surface},\ }\href@noop {}
  {\bibfield  {journal} {\bibinfo  {journal} {Nature physics}\ }\textbf
  {\bibinfo {volume} {5}},\ \bibinfo {pages} {438} (\bibinfo {year}
  {2009})}\BibitemShut {NoStop}%
\bibitem [{\citenamefont {Xia}\ \emph {et~al.}(2009)\citenamefont {Xia},
  \citenamefont {Qian}, \citenamefont {Hsieh}, \citenamefont {Wray},
  \citenamefont {Pal}, \citenamefont {Lin}, \citenamefont {Bansil},
  \citenamefont {Grauer}, \citenamefont {Hor}, \citenamefont {Cava} \emph
  {et~al.}}]{xia2009observation}%
  \BibitemOpen
  \bibfield  {author} {\bibinfo {author} {\bibfnamefont {Y.}~\bibnamefont
  {Xia}}, \bibinfo {author} {\bibfnamefont {D.}~\bibnamefont {Qian}}, \bibinfo
  {author} {\bibfnamefont {D.}~\bibnamefont {Hsieh}}, \bibinfo {author}
  {\bibfnamefont {L.}~\bibnamefont {Wray}}, \bibinfo {author} {\bibfnamefont
  {A.}~\bibnamefont {Pal}}, \bibinfo {author} {\bibfnamefont {H.}~\bibnamefont
  {Lin}}, \bibinfo {author} {\bibfnamefont {A.}~\bibnamefont {Bansil}},
  \bibinfo {author} {\bibfnamefont {D.}~\bibnamefont {Grauer}}, \bibinfo
  {author} {\bibfnamefont {Y.~S.}\ \bibnamefont {Hor}}, \bibinfo {author}
  {\bibfnamefont {R.~J.}\ \bibnamefont {Cava}}, \emph {et~al.},\ }\bibfield
  {title} {\bibinfo {title} {Observation of a large-gap topological-insulator
  class with a single {D}irac cone on the surface},\ }\href@noop {} {\bibfield
  {journal} {\bibinfo  {journal} {Nature physics}\ }\textbf {\bibinfo {volume}
  {5}},\ \bibinfo {pages} {398} (\bibinfo {year} {2009})}\BibitemShut {NoStop}%
\bibitem [{\citenamefont {Tateishi}\ and\ \citenamefont
  {Hirayama}(2021)}]{TateishiHirayama}%
  \BibitemOpen
  \bibfield  {author} {\bibinfo {author} {\bibfnamefont {I.}~\bibnamefont
  {Tateishi}}\ and\ \bibinfo {author} {\bibfnamefont {M.}~\bibnamefont
  {Hirayama}},\ }\href {https://doi.org/10.48550/ARXIV.2112.13770} {\bibinfo
  {title} {Quantum spin hall effect from multi-scale band inversion in twisted
  bilayer bi$_2$(te$_{1-x}$se$_x$)$_3$}} (\bibinfo {year} {2021})\BibitemShut
  {NoStop}%
\bibitem [{\citenamefont {Takagi}(1924)}]{takagi1924algebraic}%
  \BibitemOpen
  \bibfield  {author} {\bibinfo {author} {\bibfnamefont {T.}~\bibnamefont
  {Takagi}},\ }\bibfield  {title} {\bibinfo {title} {On an algebraic problem
  reluted to an analytic theorem of {C}arath{\'e}odory and {F}ej{\'e}r and on
  an allied theorem of {L}andau},\ }in\ \href@noop {} {\emph {\bibinfo
  {booktitle} {Japanese journal of mathematics: transactions and abstracts}}},\
  Vol.~\bibinfo {volume} {1}\ (\bibinfo {organization} {The Mathematical
  Society of Japan},\ \bibinfo {year} {1924})\ pp.\ \bibinfo {pages}
  {83--93}\BibitemShut {NoStop}%
\bibitem [{\citenamefont {Autonne}(1915)}]{autonne1915matrices}%
  \BibitemOpen
  \bibfield  {author} {\bibinfo {author} {\bibfnamefont {L.}~\bibnamefont
  {Autonne}},\ }\href@noop {} {\emph {\bibinfo {title} {Sur les matrices
  hypohermitiennes et sur les matrices unitaires}}},\ Vol.~\bibinfo {volume}
  {38}\ (\bibinfo  {publisher} {A. Rey},\ \bibinfo {year} {1915})\BibitemShut
  {NoStop}%
\end{thebibliography}%

\appendix
\section{General Hamiltonians for Dirac Materials\label{Apx:CanHamForm}}
In this appendix, we discuss the Dirac Hamiltonian in a single layer. In Sec.~\ref{Apx:vkdotsigma}, we discuss our basis choice for the Dirac Hamiltonian. In Sec.~\ref{Apx:TRS} we consider the representations of rotational and time-reversal symmetry on this Hamiltonian, and show that, after a further change of basis, time-reversal symmetry can be brought to a canonical form of $i\sigma_y K$. Finally, in Sec.~\ref{Apx:GRHam} we elaborate on the details and symmetry representation of the graphene Hamiltonian, along with explicit matrices that perform the aforementioned changes of basis.

\subsection{Dirac Hamiltonians\label{Apx:vkdotsigma}}
Let us begin by considering a two-state Hamiltonian in two dimensions which identically vanishes in energy at $k=0$. 
The most general expansion to linear order in $k$ can be expressed as:
\begin{equation}
    H(k)= (a\cdot k)\I+k\cdot b\cdot\sigma+(c\cdot k)\sigma_z
\end{equation}

If the spectrum is isotropic, then the vectors $a$ and $c$ must vanish, while the tensor $b$ is constrained to the trace and antisymmetric parts (which commute with rotations). Then the Hamiltonian can be written as a sum of two terms:
\begin{equation}
    H(k) = b_{1s}(\bar k\cdot\sigma)+b_{1a}(\bar k\times\sigma)
\end{equation}

We can then perform a unitary transformation described by the matrix:
\begin{equation}
    \exp(i\arctan(b_{1a}/b_{1s})\sigma_z/2)
\end{equation}
This will transform our Hamiltonian into:
\begin{equation}
    \sqrt{b_{1s}^2+b_{1a}^2}(\bar k\cdot\sigma)
\end{equation}
Thus, an isotropic Dirac cone at $k=0$ can generically be written as $v\bar k \cdot \sigma$ in the appropriate basis.

A Dirac cone not at the origin will instead take the form $v(k-k_0)\cdot\sigma$. For this reason, we now define $\bar{k}$ as the difference in momentum space between $k$ and the center of the Dirac cone $k_0$:
\begin{equation}\label{kbardef}
    \bar{k}=k-k_0
\end{equation}

This leads to the more general assertion that any isotropic, gapless Dirac cone at charge-neutrality can have an effective Hamiltonian written in the form $H_D(\bar{k})$, with $H_D$ as defined in Eq.~\eqref{eq:DiracHam}.

More generally, for a Hamiltonian with multiple Dirac cones indexed by the Dirac points $k_{0,i}$, we can do the same operations for each cone in terms of $\bar{k}_i:=k-k_{0,i}$.

At the level of the perturbative small-angle calculation performed here, we can drop the $i$ sub-index, and regard $\bar{k}$ as a formal parameter rather than a point in momentum space per se. This is computationally convenient because it allows for valley-mixing basis changes (which we will use in the following section); accordingly, we will use $\bar{k}$ without sub-index in the rest of this paper's exposition. (To extract the physical meaning of the terms we compute, the basis change must be reversed to disentangle the valleys.)

Therefore, under the assumptions we make in this paper, if we index our Dirac cones by $i$ and $j$, we can always write a single-layer Hamiltonian as:
\begin{equation}
    H(\bar{k})=H_D(\bar k)\delta_{ij}.\label{eq:GenDiracHam}
\end{equation}

\subsection{Time-reversal and rotational symmetry\label{Apx:TRS}}
We now also assume that our materials have both time-reversal symmetry $\mathcal{T}$ and in-plane $n$-fold rotational symmetry $C_n$. Note both of these operations leave each layer invariant, so we can discuss them before coupling two layers; however, in materials with multiple Dirac cones, the symmetries do not have to leave each Dirac cone invariant - the Dirac cones may be mixed under the symmetry transformations.

Time-reversal symmetry can be expressed as the condition that:
\begin{equation}
    \mathcal{T}H_{\bar k}\mathcal{T}^{-1}=H_{-\bar k}
\end{equation}

Similarly, if we denote by $\mathfrak{R}$ the usual action of $C_n$ on the plane, then the rotation symmetry requires:
\begin{equation}
    C_nH_{\bar k}C_n^\dagger=H_{\mathfrak{R}\bar k}
\end{equation}

We now derive the matrix forms of $\mathcal{T}$ and $C_n$, which will be implemented by anti-unitary and unitary matrices, respectively, when acting on our Dirac Hamiltonian in Eq.~\eqref{eq:GenDiracHam}. Given that $\mathcal{T}$ is antiunitary, it can generally be written as:
\begin{equation}\label{eq:genTRS}
    \mathcal{T}=\tilde T_{ij}(i\sigma_y\mathcal{K}),
\end{equation}
where $\tilde T$ is a matrix in $ij$ space (trivial in $\sigma$ space) whose structure will depend on the specific problem at hand. Since $\mathcal{T}$ is anti-unitary and satisfies $\mathcal{T}^2=-1$, it must be that $\tilde T\tilde T^\dagger=\tilde T\tilde T^*=1$, and therefore that $\tilde T$ is complex-symmetric.

The rotation matrix can be written as:
\begin{equation}\label{eq:genRot}
    C_n=\tilde U_{ij}\exp(i\pi\sigma_z/n),
\end{equation}
Similar to time-reversal symmetry, there is a matrix $\tilde U$ which acts only in $ij$ space and will depend on the specific problem. $\tilde U$ must be unitary and have the property that $\tilde U^n=1$.

To simplify our calculations of twisted multi-Dirac band structures (Appendix~\ref{Apx:SelfEnCalcDetails}), we would like to simplify the form of $\tilde{T}$ and $\tilde{U}$ as much as possible. 
As these unitary transformations should also preserve the form of our Hamiltonian, we constrain ourselves to only operating on the $i$ and $j$ indices (rather than on those of $H_D(\bar{k})$).

Since $\tilde T$ is complex-symmetric, we can perform an Autonne-Takagi factorization \cite{takagi1924algebraic, autonne1915matrices} to diagonalize it into $\delta_{ij}$. (Note the condition that $\mathcal{T}^2=-1$ plays a vital role in this decomposition by ensuring the symmetry of $\tilde T$; therefore, this decomposition does not directly apply to spinless models.)

This change of basis yields new rotation matrices $U_{ij}$ from our previous $\tilde U_{ij}$, such that the rotation in the new basis is $U_{ij}\exp(i\pi\sigma_z/n)$. These matrices $U$ cannot generally be simplified by a further basis transformation while preserving the properties established thus far. (We still have the ability to act on the $ij$ indices with a real orthogonal matrix without disturbing either our Hamiltonian or the form of $\mathcal{T}$, but this is of comparatively limited use.)

In some cases, it is beneficial to diagonalize in the basis of $\mathcal{T}':=C_2\mathcal{T}$, instead of $\mathcal{T}$. This will most often the case if the model has a subset of Dirac cones (decoupled from the rest) invariant under $\mathcal{T}'$ but not $\mathcal{T}$, which occurs, for instance, in twisted bilayer graphene, where the $K$ and $K'$ Dirac cones are decoupled from each other.

In this circumstance, $\mathcal{T}'$ takes the matrix form:
\begin{equation}
    \mathcal{T}'=\tilde T_{ij}'(\sigma_x\mathcal{K})
\end{equation}

The condition that $\mathcal{T}'^2=+1$ results in $T_{ij}'$ also being complex-symmetric, and therefore the same Autonne-Takagi factorization applies to reduce this to $\delta_{ij}$. However, this diagonalization cannot (generally) be performed simultaneously with the previous one, so one basis or the other should be chosen for the computation on a case-by-case basis.

\subsection{Effective graphene Hamiltonian and symmetry representations\label{Apx:GRHam}}
We now present the application of the above methodology to graphene as a concrete example. In the process, we develop a model of ``half-graphene" which will serve an effective simpler stepping stone in our calculations.

\subsubsection{Graphene fundamentals}
We can write an effective graphene Hamiltonian as four Dirac cones coupled to sublattice, indexed by spin and valley:
\begin{equation}
    H_G=\begin{bmatrix}vk\cdot\tau&0&0&0\\0&vk\cdot\tau&0&0\\0&0&-vk\cdot\tau^*&0\\0&0&0&-vk\cdot\tau^*\end{bmatrix}
\end{equation}

We will here use $\tau$ to refer to the sublattice degree of freedom, $\sigma$ to refer to the spin degree of freedom, and $\mu$ to refer to the valley degree of freedom. Therefore, the above Hamiltonian can be equivalently written as:
\begin{equation}
    H_G=v[k_x\tau_x\mu_z+k_y\tau_y\mu_0]\sigma_0
\end{equation}

In this basis, time-reversal symmetry will be written as:
\begin{equation}
    \mathcal{T}=i\sigma_y\mu_x\mathcal{K}
\end{equation}

For rotations, it will be computationally convenient to separately consider $C_2$ and $C_3$ (rather than $C_6$). These take the form:
\begin{equation}
    C_2=i\sigma_z\tau_x\mu_x
\end{equation}
\begin{equation}
    C_3=-\exp(i\pi\sigma_z/3)\exp(i\pi\tau_z\mu_z/3)
\end{equation}

\subsubsection{Half-graphene model}
A computationally simpler model would have only one Dirac cone each at $K$ and $K'$. Considering a spinless model of graphene will not work if we want to preserve time-reversal symmetry with $\mathcal{T}^2=-1$. However, what we can do is write a Hamiltonian for a material like graphene with spin-valley locking. In other words, we take a material with Dirac cones coupled to sublattice at $K$ and $K'$, but take a single Dirac cone at $K$ with one spin and one at $K'$ with the opposite spin. This setup breaks $C_2$ symmetry, but preserves both $C_3$ and time-reversal.

The effective Hamiltonian of such a model, before any basis change, could be written by picking out the first and fourth cones of our previous graphene model:
\begin{equation}
    H_{HG}=\begin{bmatrix}vk\cdot\tau&0\\0&-vk\cdot\tau^*\end{bmatrix}
\end{equation}

We can therefore inherit our symmetry representations from that model as well. Let $\sigma$ denote the combined spin-valley degree of freedom. Then, time reversal takes the form:
\begin{equation}
    \mathcal{T}=i\sigma_y\mathcal{K}
\end{equation}
Similarly, threefold rotations take the form:
\begin{equation}
    C_3=-\exp(i\pi\sigma_z(\tau_z+\tau_0)/3)
\end{equation}

\subsubsection{Basis change in half-graphene\label{Apx:HGCanHam}}
We now take our simplified model of half-graphene and put it into the standard form with a Hamiltonian of $(vk\cdot\tau)\sigma_0$ with $\mathcal{T}=i\sigma_y\mathcal{K}$.

First, we transform the Hamiltonian to our standard form with a $\sigma_y$ transformation on the second Dirac cone only. Time-reversal then takes the form $(i\tau_y\mathcal{K})\sigma_x$. We can easily transform $\sigma_x$ into $\sigma_z$ with a rotation by $e^{i\pi\sigma_y/4}$, after which we transform the second Dirac cone by a factor of $i$ to eliminate the minus sign (exploiting antiunitarity to eliminate the cone-dependent phase in $\mathcal{T}$).

Combining these transformations together yields the matrix:
\begin{equation}\label{eq:HGBasisChange}
    \left(\frac{1+i}{2}\right)\begin{bmatrix}i\tau_0&-\tau_y\\-\tau_0&i\tau_y\end{bmatrix}
\end{equation}

Due to this complicated transformation, the $\tau$ and $\sigma$ matrices in this basis refer to some combination of the original degrees of freedom in a non-obvious way. (I.e.: we use these symbols to refer to the indices in the matrix structure rather than their original degrees of freedom.)

With our Hamiltonian and time-reversal symmetry both in standard form, only the $C_3$ rotation operator remains nontrivial. That the operator is a symmetry of the Hamiltonian determines how it acts on $\tau$ space, so what remains is to determine how it acts on $\sigma$ space, which is precisely specifying the matrix $U$ in Eq.~\eqref{eq:genRot}). Computing $U$ in this case yields a standard rotation matrix: 
\begin{equation}\label{eq:HGU3}
    U_{3,HG}=\exp(i\pi\sigma_y/3)
\end{equation}

\subsubsection{Basis change in graphene\label{Apx:GRCanHam}}
We now perform an analogous series of steps in our full graphene model. There is some additional freedom in our basis change here that we exploit to make our rotational symmetries as simple as possible. In the end, our basis transformation is:
\begin{equation}
    \left(\frac{1+i}{2}\right)\begin{bmatrix}i\tau_0&0&0&-\tau_y\\-\tau_0&0&0&i\tau_y\\0&i\tau_0&\tau_y&0\\0&\tau_0&i\tau_y&0\end{bmatrix}
\end{equation}

With our basis transformations performed, we can now write the rotations in the new basis. Again, we take Eq.~\eqref{eq:genRot} and specify the $U$ matrices defined therein, this time both for $C_3$ and $C_2$ symmetry:
\begin{equation}\label{eq:GrU3}
    U_3=\exp(i\pi\sigma_y/3)\mu_0
\end{equation}
\begin{equation}\label{eq:GrU2}
    U_2=\sigma_0\mu_x
\end{equation}

That these expressions factor is precisely the motivation for working with $C_3$ and $C_2$ rather than $C_6$: $C_3$ now acts trivially on the $\mu$ indices and $C_2$ acts trivially on the $\sigma$ indices, whereas $C_6$ acts nontrivially on both indices simultaneously.

Moreover, we can now see the exact sense in which our half-graphene model was, indeed, half of graphene: our full model of graphene can be written as two $C_2$-related copies of our half-graphene model. This will allow us to work out the details of the $C_3$ symmetry in our half-graphene model in Sec.~\ref{Apx:HGCorrections}, then add the details of $C_2$ symmetry in the full graphene model in Sec.~\ref{Apx:GRCorrections}.

\section{Interlayer Couplings\label{Apx:CoupDer}}

In this appendix, we derive the coupling between two layers stacked with a relative twist angle. We begin with a general derivation of interlayer couplings in a tight-binding model given the lattice vectors of each layer and then specialize to the case of small twist angle about a commensurate crystal structure. We then specialize to the case of coupled Dirac cones and incorporate the constraints of time-reversal and rotational symmetry.

\subsection{General two-layer tight-binding couplings\label{Apx:GenCoup}}

We begin by considering the interlayer coupling between two layers with different lattices, with no constraints on the Hamiltonian of each layer.

Letting $I,J$ index all non-momentum degrees of freedom (sublattice, orbital, spin, etc., but not valley, since it is included in the integral over $k$) of layer 1 and $I',J'$ of layer 2, the two-layer Hamiltonian can generally be written:
\begin{widetext}
\begin{multline}
    H=\sum_{I,J}\int_{\text{BZ}\ 1} d^2k\psi_{1,I,k}^\dagger H_1^{IJ}(k)\psi_{1,J,k}+\sum_{K,L}\int_{\text{BZ}\ 2} d^2k'\psi_{2,I',k'}^\dagger H_2^{I'J'}(k')\psi_{2,J',k'}\\
    +\left[\sum_{I,J'}\int_{\text{BZ}\ 1}d^2k\int_{\text{BZ}\ 2}d^2k'\left(\psi_{1,I,k}^\dagger T^{IJ'}(k,k')\psi_{2,J',k'}\right)+h.c.\right]
\end{multline}
\end{widetext}

We now simplify $T(k,k')$ by imposing some minimal assumptions. We start by Fourier transforming $T^{IJ'}(k,k')$:
\begin{equation}
    T^{IJ'}(k,k')=\sum_{R,R'}e^{-ik\cdot (R+r_I)}e^{ik'\cdot (R'+r_{J'})}T^{IJ'}(R,R')
\end{equation}

In the above equation, $R$ and $R'$ denote the centers of the unit cells of the two layers, whereupon $r_I$ and $r_J'$ denote the location of sublattices $I$ and $J'$ relative to the centers of their respective unit cells.

We then make a standard moir\'e tight-binding assumption that the crystal locally has the translation invariance of the original lattice, i.e.:
\begin{equation}
    T^{IJ'}(R,R')=t^{IJ'}(R+r_I-R'-r_{J'})
\end{equation}

Combining this equation with the previous one and Fourier transforming again allows us to simplify significantly:
\begin{widetext}
\begin{equation}\begin{split}
T^{IJ'}(k,k')=&\sum_{R,R'}e^{-ik\cdot (R+r_I)}e^{ik'\cdot (R'+r_{J'})}t^{IJ'}(R+r_I-R'-r_{J'})\\
=&\sum_{R,R'}e^{-ik\cdot (R+r_I)}e^{ik'\cdot (R'+r_{J'})}\sum_{k''}e^{ik''\cdot(R+r_I-R'-r_{J'})}t^{IJ'}_{k''}\\
=&\sum_{R,R'}\sum_{k''}e^{-i(k-k'')\cdot R}e^{i(k'-k'')\cdot R'}e^{i(k''-k)\cdot r_I}e^{-i(k''-k')\cdot r_{J'}}t^{IJ'}_{k''}\\
=&\sum_{G,G'}\sum_{k''}\delta_{k+G,k''}\delta_{k'+G',k''}e^{i(k''-k)\cdot r_I}e^{-i(k''-k')\cdot r_{J'}}t^{IJ'}_{k''}\\
=&\sum_{G,G'}\delta_{k+G,k'+G'}e^{i(G\cdot r_I-G'\cdot r_{J'})}t^{IJ'}_{k+G}
\end{split}
\label{eq:Tkk1}
\end{equation}
\end{widetext}

In the above expressions, $k''$ is allowed to run over all of momentum space, whereas $G$ and $G'$ are the reciprocal lattice vectors of the two layers.

This final expression relates the coupling between two materials of arbitrarily mismatched lattices to the Fourier transform of the tight-binding interlayer hopping $t^{IJ}(r)$.

\subsection{Small-angle twisting\label{Apx:TwistCoup}}

We now look specifically at the case where the two layers are twisted with a small angle relative to a commensurate cell and, further, that the low-energy physics is well described by a $k\cdot p$ expansion about a particular point $k_0$ in the BZ. (The derivation proceeds similarly for a small lattice mismatch instead of a small twist angle.)

Making the substitutions $k\rightarrow k_0 + \bar k$, $k' = k_0' + \bar k'$, Eq.~\eqref{eq:Tkk1} becomes:
\begin{equation}
   T^{IJ'}(k,k') =  \sum_{G,G'}\delta_{\bar k+k_0+G,\bar k'+k_0'+G'}e^{i(G\cdot r_I-G'\cdot r_{J'})}t^{IJ'}_{\bar k+k_0+G}
\end{equation}

From here, we explicitly separate out twist angle dependence: we split $k_0$ as $k_0=\hat k_0+\delta k_0$, where $\hat k_0$ is twist-angle-independent and $|\delta k_0|\sim\theta|k_0|$. Therefore, for small twist angle, $\delta k_0$ will be small. (As an example, in graphene, $\hat k_0$ would be the position of the $K$ point before twisting, and $\delta k_0$ would be its deviation after twisting.) We do a similar decomposition with $G$, $k_0'$, and $G'$.

We now suppose that the interlayer spacing is much greater than the interatomic spacing, which motivates an assumption that $t$ decays slowly in position space and therefore quickly in momentum space \cite{Bistritzer12233}.
Then, the smallest values for $\bar{k}+k_0+G$ (i.e., within the first few BZ) dominate. 
Combined with our assumption for small $\bar{k}$, this justifies the following decomposition:
\begin{equation}
    \delta_{\bar k+k_0+G,\bar k'+k_0'+G'}\simeq \delta_{\hat k_0+\hat G,\hat k_0'+\hat G'}\delta_{\bar k+\delta k_0+\delta G,\bar k'+\delta k_0'+\delta G'}
\end{equation}

The first term is $\theta$-independent: it implies that cones will only couple after twisting if they fold onto the same point in the BZ before twisting, allowing us a simple way to determine what valleys do or do not couple. The second term then specifies the twist angle dependence, as it contains the $\theta$-dependent terms.

To simplify, we explicitly input the rotation by declaring $\delta G'=M\hat G'$, $\delta k_0'=M\hat k_0'$, and $\delta G=\delta k_0=0$, where $M$ is the difference between the rotation matrix and the identity.
Then, with an approximation in $t$ to lowest order in $\bar{k}$, we end up with the following expression for $T^{IJ'}(k,k')$:
\begin{equation}
    \sum_{\hat G,\hat G'}\underbrace{\delta_{\hat k_0+\hat G,\hat k_0'+\hat G'}}_\text{Valley matching}\underbrace{\delta_{\bar k,\bar k'+M(\hat k_0'+\hat G')}}_{\theta/Q\text{-dependence}}\underbrace{e^{i(\hat G\cdot \hat r_I-\hat G'\cdot \hat r_{J'})}t^{IJ'}_{\hat k_0+\hat G}}_\text{Coupling strength}
    \label{eq:TIJkkp}
\end{equation}
where $\hat r_{I,J'}$ denote the positions of the sublattice degrees of freedom before being twisted.
Finally, we define $Q:=\bar k'-\bar k$ and express this interlayer hopping as $T_Q$, giving the notation from the main text in Eq.~\eqref{eq:Topdef}. 

\subsection{Symmetry constraints on $T_Q$\label{Apx:SymmTq}}

We now derive the effect of symmetry on the interlayer coupling terms. In so doing, we explicitly assume the Dirac Hamiltonian is written in the basis discussed in Appendix~\ref{Apx:CanHamForm}, including the specified form of the symmetry operators.

For simplicity of expression, we explicitly break the $T_Q$ into a set of 2-by-2 matrices $T_{Q,ij'}$. Note that $ij'$ used here are not the same as $IJ'$ in the previous section: $IJ'$ run over all non-valley degrees of freedom, whereas $ij'$ run over time-reversal-invariant Dirac cones (which may include valley degrees of freedom).

Time-reversal symmetry relates $Q$ to $-Q$, and gives the relation:
\begin{equation}
    \sigma^yT^*_{ik,-Q}\sigma^y=T_{ik,Q}\label{eq:TRSonT}
\end{equation}

For in-plane rotations, we return to the $U$ matrices defined following Eq.~\eqref{eq:genRot}, which may be different for each layer.
Call the matrix for one layer $U$ and for the other $U'$. Implementing the rotation in momentum space by $Q \mapsto \mathfrak{R}Q$, the resulting constraint is:
\begin{equation}
    e^{i\theta\sigma_z/2}[U_{ij}T_{jk',\mathfrak{R}Q}U'^\dagger_{k'l'}]e^{-i\theta\sigma_z/2}=T_{il',Q}\label{eq:RotonT}
\end{equation}

Notice that the constraints imposed by time-reversal and rotation symmetry in Eqs.~\eqref{eq:TRSonT} and \eqref{eq:RotonT} relate the interlayer coupling at different $Q$. Thus, once $T_Q$ is known for one choice of $Q$, it is determined for all other $Q$ related by time-reversal or rotation symmetry. The exception to this is the combined symmetry operation $C_2\mathcal{T}$, which leaves $Q$ invariant and (due to antilinearity) acts as a reality constraint on each individual $T_Q$.

In the case of identical materials, there may also be layer-interchanging rotations, like $C_{2x}$ or $C_{2y}$. These follow a similar formula to the in-plane rotations, except instead of relating $T$ to $T$, they relate $T$ to $T^\dagger$. Letting $\phi$ denote the angle between the axis of rotation and the $x$-axis, define:
\begin{equation}
    \sigma_\phi=\cos(\phi)\sigma_x+\sin(\phi)\sigma_y
\end{equation}

Then, letting $\mathfrak{L}Q$ denote the action of the rotation on $Q$ and $V$ a material-dependent transformation matrix representing the action of the symmetry on the $ij$ indices:
\begin{equation}
   \sigma_\phi[V_{ij}T_{jk',-\mathfrak{L}Q}^\dagger V^\dagger_{k'l'}]\sigma_\phi=T_{il',Q}\label{eq:ILRonT}
\end{equation}

There is a subtlety here in the formula involving $T_{-\mathfrak{L}Q}$ instead of $T_{\mathfrak{L}Q}$. 
Specifically, $Q$ is defined as the difference in momentum from layer 1 to layer 2 (see below Eq.~\eqref{eq:TIJkkp}). When the two layers are exchanged, the meaning of $Q$ has to change accordingly, which swaps $Q\rightarrow -Q$.

In practice, it is only necessary to consider one out-of-plane rotation symmetry, since the others are generated by products with the in-plane rotations. When considering a particular $Q$, it is prudent to consider an out-of-plane rotation that preserves that $Q$ (as done in Sec.~\ref{Sec:TITISame}), so that the out-of-plane rotation imposes a constraint on $T_Q$, rather than relating two distinct terms $T_Q$ and $T_{-\mathfrak{L}Q}$. This then (being an antilinear constraint on a single $Q$, akin to $C_2\mathcal{T}$) imposes a reality constraint on individual $T_Q$. A material with both $C_2\mathcal{T}$ and such an out-of-plane symmetry may have conflicting reality constraints that can cause certain interlayer hopping terms to vanish, as again can be seen in Sec.~\ref{Sec:TITISame}.

These symmetries are the only ones that we expect to arise in generic models, since reflection symmetries are (generically) broken in twisted bilayers and internal symmetries are model-specific. Without further constraints, we expect all interlayer hopping terms that satisfy Eqs.~\eqref{eq:TRSonT}, \eqref{eq:RotonT} and \eqref{eq:ILRonT} to arise, although symmetry does not constrain their amplitudes.

\section{Calculations of Self-Energy\label{Apx:SelfEnCalcDetails}}
In the following appendix, we compute the self-energy of a Dirac cone in one layer of material that results from hopping into the other layer.
We show that the self-energy takes the general form given by Eq.~\eqref{eq:1DCSelfEn} and derive the coefficients defined in Eq.~\eqref{eq:1DCCoeffs}. We also derive the more general results that arise when there is more than one Dirac cone in each layer.

We begin by discussing the general case in the special basis described in Appendix~\ref{Apx:CanHamForm} (in particular parts \ref{Apx:vkdotsigma} and \ref{Apx:TRS}). We then discuss the extent to which this calculation requires this change of basis and the extent to which it can be ignored.

With the general case worked out, we proceed to compute the quadratic corrections to graphene (and our simplified half-graphene model) described in \ref{Apx:GRHam}. 

\subsection{General calculation\label{Apx:SelfEnCalc}}
We first calculate the self-energy of a Dirac cone in layer 1 due to its coupling to a Dirac cone in layer 2; the self-energy of the Dirac cone in layer 2 from hopping into layer 1 is identical up to the definition of $Q$ (which is the momentum shift in going from layer 1 to layer 2, not the converse) and an interchange of the roles of $T$ and $T^\dagger$.
The calculation of self-energy, to lowest-order in perturbation theory, proceeds as follows:
\begin{widetext}
\begin{align}
    \Sigma_{ij}(\bar k,\omega)=&\sum_{Q,l'} T_{Q,il'}[\omega-v(\bar k+Q)\cdot\sigma]^{-1}T_{Q,jl'}^\dagger\label{eq:gen2PSE}\\
    \simeq&-\sum_{Q,l'}\frac{1}{v^2Q^4}T_{Q,il'}[Q^2\omega+v(Q^2(\bar k\cdot\sigma)-2(Q\cdot \bar k)(Q\cdot\sigma)+Q^2(Q\cdot\sigma))]T_{Q,jl'}^\dagger\label{eq:LinExpSE}\\
    =&-\sum_{Q,l'}\frac{1}{2v^2Q^4}\left\{\left(T_{Q,il'}T^\dagger_{Q,jl'}+T_{-Q,il'}T^\dagger_{-Q,jl'}\right)Q^2\omega+vQ^2\left(T_{Q,il'}(Q\cdot\sigma)T^\dagger_{Q,jl'}-T_{-Q,il'}(Q\cdot\sigma)T^\dagger_{-Q,jl'}\right)\right.\nonumber\\
    &\qquad\left.+v\left[T_{Q,il'}\left(\bar k\cdot(Q^2\I-2Q\otimes Q)\cdot\sigma\right)T^\dagger_{Q,jl'}+T_{-Q,il'}\left(\bar k\cdot(Q^2\I-2Q\otimes Q)\cdot\sigma\right)T^\dagger_{-Q,jl'}\right]\right\}\label{eq:SEQmQ}\\
    =&-\sum_{Q,l'}\frac{1}{2v^2Q^4}\left\{\left(T_{Q,il'}T^\dagger_{Q,jl'}+\sigma_y(T_{Q,il'}T^\dagger_{Q,jl'})^*\sigma_y\right)Q^2\omega+vQ^2\left(T_{Q,il'}(Q\cdot\sigma)T^\dagger_{Q,jl'}+\sigma_y(T_{Q,il'}(Q\cdot\sigma)T^\dagger_{Q,jl'})^*\sigma_y\right)\right.\nonumber\\
    &\qquad\left.+v\left(T_{Q,il'}\left(\bar k\cdot(Q^2\I-2Q\otimes Q)\cdot\sigma\right)T^\dagger_{Q,jl'}-\sigma_y\left(T_{Q,il'}\left(\bar k\cdot(Q^2\I-2Q\otimes Q)\cdot\sigma\right)T^\dagger_{Q,jl'}\right)^*\sigma_y\right)\right\}\label{eq:SETRS}
\end{align}
\end{widetext}
Let us now explain each equality:
Eq.~\eqref{eq:LinExpSE} is derived by expanding Eq.~\eqref{eq:gen2PSE} to linear order in the dimensionless parameters $\omega/v|Q|$ and $|\bar k|/|Q|$. Eq.~\eqref{eq:SEQmQ} is derived by averaging the terms $T_Q$ and $T_{-Q}$, and Eq.~\eqref{eq:SETRS}
relates $T_{-Q,il'}$ to $T_{Q,il'}$ by the constraint imposed by time reversal symmetry in Eq.~\eqref{eq:TRSonT}.

The above equation for the self-energy expanded to linear order in $\bar k$ and $\omega$ can be decomposed as:
\begin{equation}\label{eq:fullygenSE}
    \Sigma_{ij}(\bar k,\omega)=\mathcal{A}_{ij}\omega+\mathcal{B}_{ij}+\bar k\cdot\mathcal{V}_{ij},
\end{equation}
where the coefficients can be further decomposed as:
\begin{subequations}\label{eq:fullygencoeffs}
\begin{align}
    \mathcal{A}_{ij}=&A_{ij}\sigma_0+iA'_{ij}\sigma_z+iA''_{ij}\cdot\sigma\\
    \mathcal{B}_{ij}=&B_{ij}\sigma_0+iB'_{ij}\sigma_z+iB''_{ij}\cdot\sigma\\
    \mathcal{V}_{ij}=&V_{ij}\sigma_z+iV'_{ij}\sigma_0+M_{ij}\cdot\sigma
\end{align}
\end{subequations}
In the above equations, aside from the $ij$ indices, some of the coefficients have implicit indices implied by the dot products, i.e., $A$, $A'$, $B$, and $B'$ are scalars; $A''$, $B''$, $V$, and $V'$ are two-component vectors; and $M$ is a rank-2 tensor.

The factors of $i$ have been chosen so that all are real due to time-reversal symmetry. 
Since the self-energy is Hermitian, all unprimed coefficients are symmetric in the $ij$ indices, whereas all primed or double-primed indices are antisymmetric. (In particular, the antisymmetric terms vanish on the $ij$ diagonal, and therefore do not appear in the case of a single-Dirac cone, for which $i=j=1$).

We rewrite the above equations in terms of traces over the $T$ matrices using the following identities for 2x2 Hermitian matrices $H$:
\begin{subequations}
\begin{align}
    H+\sigma_yH^*\sigma_y=&\Tr[H]\sigma_0\\
    H-\sigma_yH^*\sigma_y=&\Tr[H\sigma^\mu]\sigma_\mu
\end{align}
\end{subequations}
By applying these identities to Eq.~\eqref{eq:SETRS} and isolating the coefficients defined in Eqs~\eqref{eq:fullygenSE}-\eqref{eq:fullygencoeffs}, we find the following explicit expressions for the individual coefficients defined in Eq.~(\ref{eq:fullygencoeffs}):
\begin{subequations}\label{eq:coeffsA}
\begin{align}
\label{eq:coeffA}A_{ij}&=\sum_{Q,i'}\frac{1}{2v^2Q^2}\Re\Tr[T_{Q,ii'}T_{Q,ji'}^\dagger]\\
\label{eq:coeffAp}A_{ij}'&=\sum_{Q,i'}\frac{1}{2v^2Q^2}\Im\Tr[T_{Q,ii'}T_{Q,ji'}^\dagger\sigma^z]\\
\label{eq:coeffAdp}A_{ij}''^{\mu}&=\sum_{Q,i'}\frac{1}{2v^2Q^2}\Im\Tr[T_{Q,ii'}T_{Q,ji'}^\dagger\sigma^\mu]
\end{align}
\end{subequations}

\begin{subequations}\label{eq:coeffsB}
\begin{align}
\label{eq:coeffB}B_{ij}&=\sum_{Q,i'}\frac{1}{2vQ^2}\Re\Tr[T_{Q,ii'}(Q\cdot\sigma)T_{Q,ji'}^\dagger]\\
\label{eq:coeffBp}B_{ij}'&=\sum_{Q,i'}\frac{1}{2vQ^2}\Im\Tr[T_{Q,ii'}(Q\cdot\sigma)T_{Q,ji'}^\dagger\sigma^z]\\
\label{eq:coeffBdp}B_{ij}''^\mu&=\sum_{Q,i'}\frac{1}{2vQ^2}\Im\Tr[T_{Q,ii'}(Q\cdot\sigma)T_{Q,ji'}^\dagger\sigma^\mu]
\end{align}
\end{subequations}

\begin{subequations}\label{eq:coeffsV}
\begin{align}
\label{eq:coeffV}V_{ij}^{\mu}&=\sum_{Q,i'}\frac{1}{2vQ^2}\left(\delta^\mu_{\ \lambda}-2\frac{Q^\mu Q_\lambda}{Q^2}\right)\Re\Tr[T_{Q,ii'}\sigma^\lambda T_{Q,ji'}^\dagger\sigma^z]\\
\label{eq:coeffVp}V_{ij}'^{\mu}&=\sum_{Q,i'}\frac{1}{2vQ^2}\left(\delta^\mu_{\ \lambda}-2\frac{Q^\mu Q_\lambda}{Q^2}\right)\Im\Tr[T_{Q,ii'}\sigma^\lambda T_{Q,ji'}^\dagger]
\end{align}
\end{subequations}

\begin{equation}
    \label{eq:coeffM}M_{ij}^{\mu\nu}=\sum_{Q,i'}\frac{1}{2vQ^2}\left(\delta^\mu_{\ \lambda}-2\frac{Q^\mu Q_\lambda}{Q^2}\right)\Re\Tr[T_{Q,ii'}\sigma^\lambda T_{Q,ji'}^\dagger\sigma^\nu]
\end{equation}

So far, we have only imposed time-reversal symmetry (in the form noted in Sec.~\ref{Apx:TRS}, using the constraints derived in Sec.~\ref{Apx:SymmTq}). To further constrain the form of the self-energy, we should impose rotational symmetry. While we cannot further simplify without knowing the specific form of the rotation matrices $U$ (defined in Appendix~\ref{Apx:TRS}), which are material-dependent, we now outline the procedure for simplifying further.

The core aspect of the simplification is relating the terms in the sums over $Q$ in Eqs~\eqref{eq:coeffsA}-\eqref{eq:coeffM} using Eq.~\eqref{eq:genRot}. Writing, for instance, Eq.~\eqref{eq:coeffA} as $A_{ij}=\sum_Q A_{ij}(Q)$ and noting $A_{il}(\mathfrak{R}Q)=U_{ij}A_{jk}(Q)U_{kl}^\dagger$, it follows that $A_{ij}$ can be expressed in terms of only a single representative $Q_0$ (of each symmetry-related-set of $Q$), i.e., we need only compute $A_{ij}(Q_0)$. Summing over different $Q$ is then equivalent to picking out the rotation-invariant part of $A_{ij}(Q_0)$ (under the representation of rotations given by the $U$ matrices) and multiplying by an appropriate symmetry factor.

The same procedure can be applied to the rest of Eqs.~\eqref{eq:coeffsA}-\eqref{eq:coeffM} very similarly. 
The only modification required is incorporating the $\mu$ and $\nu$ indices, where present: $A''(Q)$, etc., pick up a (vector) transformation on those indices in addition to the $U$ matrices that transform the $ij$ indices. This makes the decomposition into irreps different, but the philosophy is the same: find the rotation-invariant part of those matrices and multiply by an appropriate symmetry factor.

It is often useful to decompose $M$ into the irreps of the $\mu\nu$ indices (before considering the $ij$ indices): the trace $C$, the antisymmetric part $D$, and the symmetric trace-free part $N$, i.e.,
\begin{equation}\label{eq:Mdecomp}
    M^{\mu\nu}_{ij}=C_{ij}\delta^{\mu\nu}+D_{ij}\epsilon^{\mu\nu}+N_{ij}^{\mu\nu}
\end{equation}

We can be more concrete in the specific case where there is only one Dirac cone. In this case, all primed coefficients vanish as discussed below Eq.~(\ref{eq:fullygencoeffs}). In addition, $U$ is simply a phase factor. Thus, we only need consider the rotational transformations of $A$, $B$, $V$, and $M$, all of which are specified by $U$ being a phase: $A(Q)$ and $B(Q)$ are identical for all $Q$; $V(Q)$ transforms like a vector; and $M(Q)$ transforms like a two-index tensor.

In the sum over $Q$, therefore, the $V(Q)$ will cancel for any rotational symmetry (including $C_2$): denoting the rotation operation on a vector by $\mathfrak{R}$, our above calculation reveals that $V(\mathfrak{R}^n Q)=\mathfrak{R^n}V(Q)$, and so the sum of rotation-related $Q$ terms vanishes ($\sum_n V(\mathfrak{R}^n Q)=0$).

A similar but more complicated representation theory argument reveals that $M$ is constrained to only its trace and antisymmetric parts $C$ and $D$ (i.e., $N$ vanishes). This yields the single-cone result given in Eq.~\eqref{eq:1DCSelfEn}, with coefficients as given in Eq.~\eqref{eq:1DCCoeffs}.

\subsection{$U(N)$ symmetry\label{Apx:UnSym}}

The above calculation, in principle, only holds in the special basis in which both materials have a $k\cdot\sigma$ Hamiltonian and time reversal takes the form $i\sigma_y K$. This requires a change of basis that is inconvenient for developing intuition in materials where the cones do not naturally come in such a form, such as graphene (where time-reversal does not naturally flip the internal degree of freedom of a Dirac cone).

However, using our final results \eqref{eq:coeffsA}--\eqref{eq:coeffM}, 
we find that the expression for self-energy of the first layer is invariant under a change of basis in the second layer that acts only on the $i' j'$ indices, since they are summed over. Thus, our result is invariant under the change of basis indicated in \ref{Apx:TRS} (for the second layer), although not under the change of basis described in Appendix~\ref{Apx:vkdotsigma}.

In this paper, this invariance allows us to avoid changing the graphene basis in Sec.~\ref{Sec:TIGR}; otherwise, we would be forced to make all the changes of basis highlighted in Appendix~\ref{Apx:GRHam}. Instead, we only need transform the Hamiltonian to a $vk\cdot\sigma$ form (as described in Appendix~\ref{Apx:vkdotsigma}), which is trivial for the purposes of our calculation.

However, note the full basis change on the first layer (the one for which we are calculating the self-energy) is still necessary: a unitary change of basis on the $ij$ indices will not necessarily leave $\Sigma_{ij}$ invariant. Hence, for the computation of the self-energy of the graphene cones Appendices~\ref{Apx:HGCorrections} and \ref{Apx:GRCorrections}, the complex changes of basis are still necessary.

\subsection{Half-graphene corrections\label{Apx:HGCorrections}}
We now apply the above computational method to the case of the corrections to graphene from tunneling into the Dirac cone on the surface of a TI, assuming for simplicity that only the smallest set of $Q$s contribute nonnegligibly to the sums in Eqs.~\eqref{eq:coeffsA}-\eqref{eq:coeffM}. We begin with the simpler half-graphene model presented in Sec.~\ref{Apx:GRHam}.

For each of the terms presented in Eqs.~\eqref{eq:coeffsA}-\eqref{eq:coeffM}, we can divide the contributions into rotational irreps (where the action of the rotations are defined by the action of $U$ on the $ij$ indices combined with the usual action on any vector indices $\mu\nu$). For the trivial irreps, the summands for different (symmetry-related) $Q$s will all be identical, and hence we can simply take a representative $Q$ and multiply by a symmetry factor of $6$. For nontrivial irreps, the summands will cancel.

Consider first the corrections from $A$, $A'$, $B$, $B'$, $C$, and $D$ (as defined in Eqs. \eqref{eq:fullygencoeffs} and \eqref{eq:Mdecomp}). For these terms, since there are no vector indices, any irreps with $ij$ indices that transform nontrivially under $C_3$ will vanish in the sum over $Q$, which means (using the representation of rotational symmetry given in Eq.~\eqref{eq:HGU3}) these terms can only contribute $\sigma_0$ or $\sigma_y$ terms; the pair $(\sigma_x,\sigma_z)$ transform like a vector instead. By symmetry in $ij$, the unprimed terms contribute $\sigma_0$ and the primed $\sigma_y$. Therefore, the contributions from these terms to the self-energy in Eq.~(\ref{eq:fullygenSE}) can be written as:
$$(A_0\omega+B_0)\tau_0\sigma_0+(A_0'\omega+B_0')\tau_z\sigma_y+C_0(\bar k\cdot\tau)\sigma_0+D_0(\bar k\times\tau)\sigma_0$$

Next, consider the terms with vector indices: $A''^\mu_{ij}$, $B''^\mu_{ij}$, $V^\mu_{ij}$, and $V'^\mu_{ij}$, (as defined in eq. \eqref{eq:fullygencoeffs}). Again, we find the irreps that transform trivially. As $\mu$ is a vector index, another vector index from the $ij$ is required, and such an index can be found in the vector combination $(\sigma^x,\sigma^z)$. However, since these matrices are symmetric in $ij$, the only term they can contribute to is $V$, and so $A''=B''=V'=0$.

The surviving term, $V$, we have now decomposed as $V^\mu_{ij}=V^\mu_a\sigma^a_{ij}$ where $a=x,z$ and $\mu = x,y$. The rotationally-invariant terms are those that contract the $\mu$ and $a$ indices via dot and cross products, as follows:
for any matrix $S^a_\mu$, the rotation-invariant parts are the contractions of $V^\mu_a$ with $S_\mu^a$ and $\epsilon_\mu^{\ \nu} S_\nu^a$. 

Therefore, the contribution to the self-energy from $V$ decomposes as:
$$(\bar k\cdot V_{ij})= V_1(\bar k_x\sigma^x_{ij}+\bar k_y\sigma^z_{ij})+V_2(\bar k_x\sigma^z_{ij}-\bar k_y\sigma^x_{ij})$$
This choice of $V_1$ and $V_2$ is, of course, not unique, but the invariant subspace will be the same for all choices of decomposition.

We can similarly decompose our last coefficient, $N^{\mu\nu}_{ij}$: the $\mu\nu$ indices are a combination of $\alpha^x$ and $\alpha^z$, which transform as a vector, and therefore we must take a corresponding combination of $\sigma^x$ and $\sigma^z$, as before. Therefore, we can express our solution as:
$$N^{\mu\nu}_{ij}=N_1(\alpha_x^{\mu \nu}\sigma^x_{ij}+\alpha_z^{\mu \nu}\sigma^z_{ij})+N_2(\alpha_x^{\mu\nu}\sigma^z_{ij}-\alpha_z^{\mu\nu}\sigma^x_{ij})$$

Concatenating these all together, we find the corrections to the self-energy for our half-graphene model (in the special basis of Appendix~\ref{Apx:HGCanHam}) can be written in terms of ten coefficients:
\begin{equation}
\begin{split}
    \Sigma_{HG}(\bar k,\omega)=&(A_0\omega+B_0)\tau_0\sigma_0+(A_0'\omega+B_0')\tau_z\sigma_y\\
    +&C_0\sigma_0(\bar k\cdot\tau)+D_0\sigma_0(\bar k\times\tau)\\
    +&\left[V_1(\bar k_x\sigma^x+\bar k_y\sigma^z)+V_2(\bar k_x\sigma^z-\bar k_y\sigma^x)\right]\tau_z\\
    +&(N_1\sigma^x+N_2\sigma^z)(\bar k_x\tau_y+\bar k_y\tau_x)\\
    +&(N_1\sigma^z-N_2\sigma^x)(\bar k_x\tau_x-\bar k_y\tau_y)
\end{split}
\label{Eq:HGSE}
\end{equation}

In the original basis, the self-energy takes the form:
\begin{equation}
\begin{split}
    \Sigma_{HG}(\bar k,\omega)=&(A_0\omega+B_0)\tau_0\sigma_0+(A_0'\omega+B_0')\tau_z\sigma_0\\
    +&C_0\begin{bmatrix}\bar k\cdot\tau&0\\0&-\bar k\cdot\tau^*\end{bmatrix}+D_0\begin{bmatrix}\bar k\times\tau&0\\0&-\bar k\times\tau^*\end{bmatrix}\\
    +&\left[V_1(\bar k_x\sigma^y+\bar k_y\sigma^x)+V_2(\bar k_x\sigma^x-\bar k_y\sigma^y)\right]\tau_x\\
    +&(\tau_0-\tau_z)(N_1(k\cdot\sigma)-N_2(k\times\sigma))
\end{split}
\label{Eq:HGSE2}
\end{equation}
Some of these terms may vanish to lowest order in perturbation theory; Eq.~(\ref{Eq:HGSE}) is the most general form constrained by symmetry.

For spin-preserving sublattice-independent hopping with amplitude $t$, we can explicitly compute these terms by by multiplying the term from $Q=|Q|\hat{x}$ (for twist) or $Q=|Q|\hat{x}$ (for lattice mismatch) by a symmetry factor of six.
We find that (in either case) the only nonvanishing coefficient in the self-energy is:
\begin{equation}
    A_0=\frac{3t^2}{v^2Q^2}
\end{equation}

Hence, in this half-graphene model with spin-preserving sublattice-independent hopping to the TI layer, the effect of the lowest-order self-energy corrections is an overall energy rescaling by a factor of $1+A_0$.

\subsubsection{Corrections to TI cone from spin-preserving hopping}
For sake of completeness, we here include the self-energy corrections to the TI Dirac cone with spin-preserving hopping in this model. Similar to graphene (as discussed in Sec.~\ref{Sec:TIGRPert}), we can split $T_Q$ into $T_{Q,K}$ and $T_{Q,K'}$ (with no extra spin index in this case). Taking a $Q$ for which $T_{Q,K}$ is nonvanishing, we find that it takes one of the forms:

\begin{subequations}\label{eq:TIHGTQ}
\begin{align}
    T_{Q,K,\uparrow}=\begin{bmatrix}a&b\\0&0\end{bmatrix}\\
    \nonumber\text{or}\qquad\quad\\
    T_{Q,K,\downarrow}=\begin{bmatrix}0&0\\a&b\end{bmatrix}
\end{align}
\end{subequations}

All other $T_Q$ are symmetry-related, as in the graphene case. Note unlike graphene, the two coefficients $a$ and $b$ can be different due to the lack of $C_2$ symmetry.

Like graphene, the $C$ and $D$ coefficients vanish, but the $A$ and $B$ coefficients do not. As a consequence of these coefficients, the energy shift is: 
\begin{equation}\label{eq:HGEnShift}
    \Delta E=-\frac{6abv_{HG}|Q|}{v_{HG}^2Q^2+3(a^2+b^2)}
\end{equation}

\subsection{Full graphene corrections\label{Apx:GRCorrections}}

We now work with our full model of graphene, using the half-graphene case as a starting point. We work in the basis described in Appendix~\ref{Apx:GRCanHam}.

We first decompose the indices $ij$ into sub-indices for the $\mu$ and $\sigma$ degrees of freedom. For example, $A_{ij}=A_{pa,qb}$, where the $\mu$ indices are $\mu_{pq}$ and the $\sigma$ indices are $\sigma_{ab}$.
As discussed below Eqs.~(\ref{eq:GrU3}) and (\ref{eq:GrU2}),
the unitary rotation matrix $U_3$ only acts nontrivially on the $ab$ indices, while $U_2$ only acts nontrivially on the $pq$ indices, which allows us to straightforwardly separate the two symmetries (rather than working with a more complicated $U_6=U_2U_3^{-1}$ matrix). 

As we did for the half-graphene case, we begin by considering the scalar coefficients $A$, $B$, $C$, $D$, $A'$, and $B'$. We now filter down to the subrepresentations trivial under both $C_3$ and $C_2$. $C_3$ implies each coefficient must commute with $\sigma_y$ and $C_2$ implies it must commute with $\tau_x$, by similar logic to that used in Appendix~\ref{Apx:HGCorrections}. Considering these in the context of the symmetry properties of the different matrices yields the following contribution to the self-energy:
\begin{multline*}
    [(A_1\mu_0+A_2\mu_x)\sigma_0\tau_0+(A_1'\mu_0+A_2'\mu_x)\sigma_y\tau_z]\omega\\
    +(B_1\mu_0+B_2\mu_x)\sigma_0\tau_0+(B_1'\mu_0+B_2'\mu_x)\sigma_0\tau_z\\
    +(C_1\mu_0+C_2\mu_x)\sigma_0(k\cdot\tau)+(D_1\mu_0+D_2\mu_x)\sigma_0(k\times\tau)
\end{multline*}
We next consider the vector parts $A''$, $B''$, $V$, and $V'$. The vector index must pair with a combination of $ab$ and $pq$ indices which transforms like a vector under both $C_3$ and $C_2$. We use $(\sigma_x,\sigma_z)$ as a vectorlike pair as before for our $ab$ indices, and either $\mu_y$ or $\mu_z$ in our $pq$ indices (to anticommute with $\mu_x$ for $C_2$). Combining this with the knowledge that $V$ is symmetric and the others are antisymmetric in $ij$, the resulting allowed terms are:
\begin{multline*}
    [A_1''(\sigma_x\tau_x+\sigma_z\tau_y)+A_2''(\sigma_x\tau_y-\sigma_z\tau_x)]\mu_y\omega\\
    +[B_1''(\sigma_x\tau_x+\sigma_z\tau_y)+B_2''(\sigma_x\tau_y-\sigma_z\tau_x)]\mu_y\\
    +[V_1'(\bar k_x\sigma_x+\bar k_y\sigma_z)+V_2'(\bar k_x\sigma_z-\bar k_y\sigma_x)]\tau_0\mu_y\\
    +[V_1(\bar k_x\sigma_x+\bar k_y\sigma_z)+V_2(\bar k_x\sigma_z-\bar k_y\sigma_x)]\tau_z\mu_z
\end{multline*}

Finally, we consider the two-index $N$ matrix, which again is decomposable into $\alpha_x$ and $\alpha_z$. These transform like a vector under $C_3$ but a scalar under $C_2$, resulting in the allowed terms:
\begin{multline*}
     N=(N_1\mu_0+N_2\mu_x)(\alpha_x\sigma_x+\alpha_z\sigma_z)\\
     +(N_3\mu_0+N_4\mu_x)(\alpha_x\sigma_z-\alpha_z\sigma_x)
\end{multline*}

Therefore, the full self-energy of this system, in the special basis presented in Appendix~\ref{Apx:GRCanHam}, is:

\begin{equation}
\begin{split}
     \Sigma_G&(\bar k,\omega)=[(A_1\mu_0+A_2\mu_x)\sigma_0\tau_0+(A_1'\mu_0+A_2'\mu_x)\sigma_y\tau_z]\omega\\
    +&(B_1\mu_0+B_2\mu_x)\sigma_0\tau_0+(B_1'\mu_0+B_2'\mu_x)\sigma_0\tau_z\\
    +&(C_1\mu_0+C_2\mu_x)\sigma_0(k\cdot\tau)+(D_1\mu_0+D_2\mu_x)\sigma_0(k\times\tau)\\
    +&[A_1''(\sigma_x\tau_x+\sigma_z\tau_y)+A_2''(\sigma_x\tau_y-\sigma_z\tau_x)]\mu_y\omega\\
    +&[B_1''(\sigma_x\tau_x+\sigma_z\tau_y)+B_2''(\sigma_x\tau_y-\sigma_z\tau_x)]\mu_y\\
    +&[V_1'(\bar k_x\sigma_x+\bar k_y\sigma_z)+V_2'(\bar k_x\sigma_z-\bar k_y\sigma_x)]\tau_0\mu_y\\
    +&[V_1(\bar k_x\sigma_x+\bar k_y\sigma_z)+V_2(\bar k_x\sigma_z-\bar k_y\sigma_x)]\tau_z\mu_z\\
    +&(N_1\mu_0+N_2\mu_x)[(k_x\tau_y+k_y\tau_x)\sigma_x+(k_x\tau_x-k_y\tau_y)\sigma_z]\\
    +&(N_3\mu_0+N_4\mu_x)[(k_x\tau_y+k_y\tau_x)\sigma_z-(k_x\tau_x-k_y\tau_y)\sigma_x]
\end{split}
\end{equation}

This 24-coefficient equation is the fully general self-energy for all possible varieties of coupling. Some terms may vanish to leading order in perturbation theory.

For the special case of spin-preserving sublattice-independent coupling, which is the most physically intuitive tunneling term,
the only nonvanishing term to first order (both for lattice mismatch and for twist) is $A_0=\frac{3t_0^2}{v^2Q^2}$, and hence (like the half-graphene case) the practical effect of the lowest-order corrections is a rescaling of the self-energy.

\section{Extensions\label{Apx:Extensions}}
We now discuss straightforward but nontrivial extensions of the self-energy computed in Appendix~\ref{Apx:SelfEnCalc}.

\subsection{$C_2\mathcal{T}$ symmetry\label{Apx:TC2SelfEn}}

In a material that lacks $\mathcal{T}$ symmetry but possesses $C_2\mathcal{T}$ symmetry, it is more convenient to diagonalize in the basis of $C_2\mathcal{T}$ than $\mathcal{T}$. This is also the case in twisted bilayer graphene, where each valley is invariant under $C_2\mathcal{T}$.
Fortunately, the calculation of the self-energy with $C_2\mathcal{T}$ symmetry is methodologically similar to the calculation with $\mathcal{T}$ symmetry shown in Sec.~\ref{Apx:SelfEnCalc}, although, in general, more terms appear.

Specifically, the calculation is identical up to Eq.~\eqref{eq:LinExpSE}, but then instead of averaging over $T_Q$ and $T_{-Q}$ and using $\mathcal{T}$ symmetry to relate them, we use the fact that $T_Q=\frac{1}{2}(T_Q+(C_2\mathcal{T})T_Q(C_2\mathcal{T})^{-1})$, at which point the calculation proceeds similarly.

If we use Autonne-Takagi factorization on $C_2 \mathcal{T}$ as discussed in Sec.~\ref{Apx:TRS}, simplifying the operator to $\sigma_x \mathcal{K}$, then we instead find a self-energy of: 

\begin{widetext}
\begin{align}
    \Sigma_{ij}(\bar k,\omega)=&\sum_{Q,l'} T_{Q,il'}[\omega-v(\bar k+Q)\cdot\sigma]^{-1}T_{Q,jl'}^\dagger\label{eq:gen2PSE2}\\
    \simeq&-\sum_{Q,l'}\frac{1}{v^2Q^4}T_{Q,il'}[Q^2\omega+v(Q^2(\bar k\cdot\sigma)-2(Q\cdot \bar k)(Q\cdot\sigma)+Q^2(Q\cdot\sigma))]T_{Q,jl'}^\dagger\label{eq:LinExpSE2}\\
    =&-\sum_{Q,il'}\frac{1}{2v^2Q^4}\left\{\left(T_{Q,il'}T^\dagger_{Q,il'}+\sigma_x(T_{Q,il'}T^\dagger_{Q,il'})^*\sigma_x\right)Q^2\omega+vQ^2\left(T_{Q,il'}(Q\cdot\sigma)T^\dagger_{Q,il'}+\sigma_x(T_{Q,il'}(Q\cdot\sigma)T^\dagger_{Q,il'})^*\sigma_x\right)\right.\nonumber\\
    &\qquad\left.+v\left(T_{Q,il'}\left(\bar k\cdot(Q^2\I-2Q\otimes Q)\cdot\sigma\right)T^\dagger_{Q,il'}+\sigma_x\left(T_{Q,il'}\left(\bar k\cdot(Q^2\I-2Q\otimes Q)\cdot\sigma\right)T^\dagger_{Q,il'}\right)^*\sigma_x\right)\right\}\label{eq:SEC2Tsum}
\end{align}
\end{widetext}

Then the self-energy takes the same form as Eq.~\eqref{eq:fullygenSE}, except the symmetry decomposition of the matrices in Eq.~\eqref{eq:fullygencoeffs} are replaced with:
\begin{subequations}\label{eq:fullygencoeffsC2T}
\begin{align}
    \mathcal{A}_{ij}=&A_{ij}\sigma_0+iA'_{ij}\sigma_z+A''_{ij}\cdot\sigma\\
    \mathcal{B}_{ij}=&B_{ij}\sigma_0+iB'_{ij}\sigma_z+B''_{ij}\cdot\sigma\\
    \mathcal{V}_{ij}=&V_{ij}\sigma_0+iV'_{ij}\sigma_z+M_{ij}\cdot\sigma
\end{align}
\end{subequations}

In this case (contrary to the terms in Eq.~(\ref{eq:fullygencoeffs})), unprimed \textit{and} double-primed terms are real symmetric matrices in $ij$, whereas (only) single-primed terms are antisymmetric in those indices. (Also note the corresponding interchange of $V$ and $V'$, since $C_2\mathcal{T}$ imposes different constraints.)

The equations for these coefficients are generally the same as Eqs.~\eqref{eq:coeffsA}-\eqref{eq:coeffM}, except that real and imaginary parts of the trace are changed.

For a single-cone material, the self-energy simplifies to:
\begin{multline}
    \Sigma^{C_2\mathcal{T}}(\bar k,\omega)\simeq (A_0\I+A_1\cdot\sigma)\omega\\
    +(B_0\I+B_1\cdot\sigma)
    +\bar k\cdot M\cdot\sigma+(\bar k\cdot \hat V)\I\label{eq:C2TSE}
\end{multline}

We have not yet applied rotational symmetry. 
In a generic model, we naively expect more terms allowed by $C_2\mathcal{T}$ (without $\mathcal{T}$) than by $\mathcal{T}$ (without $C_2\mathcal{T}$): the $ij$-symmetric (i.e., unprimed) parts of \eqref{eq:fullygencoeffs} allow two scalars, one vector, and a tensor; here we have three vectors instead of one (and none of the three vectors in the $C_2\mathcal{T}$ care are the same as the vector in the $\mathcal{T}$ case).

However, when rotation symmetry is present, Eq.~(\ref{eq:C2TSE}) again simplifies to Eq.~\eqref{eq:1DCSelfEn}. In particular, any rotational symmetry will cause the $A_1$, $B_1$, and $V$ terms to vanish (and a rotation of order greater than $2$ will cause the non-scalar portion of $M$ to vanish), in exactly the same fashion as the $\mathcal{T}$-based calculation.
Then, the remaining coefficients ($A_0$, $B_0$, and the remaining parts of $M$) are precisely analogous to the coefficients in \eqref{eq:1DCSelfEn}. Thus, we find a similar condition on magic angles.

As an example of where this would be useful, consider (a spinless model of) TBLG. While TBLG can in principle be accounted for by a $\mathcal{T}$-based transformation, it is much simpler to use this $C_2\mathcal{T}$-based result because $C_2\mathcal{T}$ leaves each valley invariant. The ultimate consequence of this is the the analogy between AB hopping in TBLG and spin-flipping hopping in our TI-TI model discussed in Sec.~\ref{Sec:TITI}.

\subsection{More general Dirac Hamiltonians\label{Apx:GapChem}}
We now offer a few comments on extending to a broader class of Dirac Hamiltonians, including Dirac cones not at charge neutrality, gapped Dirac cones, and anisotropic Dirac cones. In all cases, the calculation is straightforward to adapt, but it is worth highlighting the extent to which these alternative Hamiltonians may yield qualitative changes.

For Dirac cones not at charge neutrality, we find that the coefficients $A$ and $B$ mix, i.e., if  $A_0$ and $B_0$ are the coefficients at charge neutrality, then away from charge neutrality $A$ and $B$ are both linear combinations of $A_0$ and $B_0$.

For Dirac cones with mass terms, several new contributions to $V$ and $M$ appear (potentially yielding interesting new physics). 
To compute the self-energy, it is most convenient to choose a different basis than the one outlined in Sec.~\ref{Apx:CanHamForm}: one would rather have the Hamiltonian (with mass term) in the form $v(k\cdot\sigma)\tau_0+m\sigma_z\tau_z$ with time-reversal symmetry in the form $\mathcal{T}=i\tau_x\sigma_y\mathcal{K}$.

This can always be done: performing the sequence of steps through \eqref{Apx:TRS} puts the mass term as $im\sigma_z\hat{A}_{ij}$, for $\hat A$ a real antisymmetric matrix. This $A_{ij}$ can always be remapped to a block-diagonal $\tau_y$ while preserving the form of time-reversal symmetry (by a Youla decomposition), at which point we have our Hamiltonian in the form $v(k\cdot\sigma)\tau_0+m\sigma_z\tau_y$, with $\mathcal{T}=i\sigma_y\tau_0\mathcal{K}$. From here, a quarter-rotation about $\tau_x$ will put both matrices in the above-specified form.

Finally, anisotropies also produce additional contributions to $M$ and $V$, which may lead to different kinds of behavior. For example, anisotropy may lead to ``partial magic angles," wherein one direction of a Dirac cone has a vanishing Fermi velocity and the other does not (as discussed in \cite{kariyado2019flat,kennes2020one}).

\end{document}